# Extraordinary magnetometry
# - a review on extraordinary magnetoresistance

*Thierry Désiré Pomar[1,†], Ricci Erlandsen[1,†], Bowen Zhou[2], Leonid Iliushyn[2], Rasmus Bjørk[1], Dennis Valbjørn Christensen[1*]*

*[1] Department of Energy Conversion and Storage, Technical University of Denmark, 2800 Kgs. Lyngby, Denmark*
*[2] Department of Physics, Technical University of Denmark, 2800 Kgs. Lyngby, Denmark*
*[†] These authors contributed equally to this manuscript*
*[*] dechr@dtu.dk*

**Abstract**:
Extraordinary magnetoresistance (EMR) is a geometric magnetoresistance effect occurring in hybrid devices consisting of a high-mobility material joined by a metal. The change in resistance can exceed $10^7$% at room temperature when a magnetic field of 5 T is applied. Magnetic field sensors based on EMR hold the potential formeasuring weak magnetic fields with an unprecedented sensitivity, yet, to date this potential is largely unmet. In this work, we provide an extensive review of the current state-of-the-art in EMR sensors with a focus on the hybrid device geometries, the constituent material properties and applications of EMR. We present a direct comparison of the best devices in literature across magnetoresistance, sensitivity and noise equivalent field for different materials and geometric designs. The compilation of studies collected in this review illustrates the extremely rich possibilities for tuning the magnetoresistive behavior varying the device geometry and material properties. In addition, we aim to improve the understanding of the EMR effect and its interplay with geometry and material properties. Finally, we discuss recent trends in the field and future perspectives for EMR.

# Table of Contents

# Table 1 – Overview of Experimental EMR Devices

| Dim | Device Material | Shunt Material | Device Geometry | Sensitivity (Ω/T) | $B_{NEF}$ (nT/$\sqrt{Hz}$) or SNR (dB) | MR (%) | μ ($cm^2$/Vs) | n | B (T) | Rc | I | T (K) | Size | Year & Ref. |
|---|---|---|---|---|---|---|---|---|---|---|---|---|---|---|
| 3D | **InSb** (Te-doped) | Ti/Pt/Au | vdP Disc | R · 24 $T^{-1}$<br>(α = 13/16, 300K) | | $10^6$ % (5T) | 45,500 | $2.6 \cdot 10^{16}$ $cm^{-3}$ | -5 to 5 | | | 295 | 1 mm diameter | 2000 [1] |
| 3D | **InSb** (Te-doped) | Ti/Pt/Au | vdP Disc (off-centered) | | | 2,500%<br>(0.1T) | 40,200 | $2.11 \cdot 10^{16}$ $cm^{-3}$ | -0.1 to 0.1 | | | 295 | N/A | 2001 [2] |
| 2D | **InSb**/InAlSb | Au | Bar<br>(IVVI) | 528 Ω/T<br>(0.2T) | SNR = 43 | 35%<br>(0.05T) | 23,000 | $2.7 \cdot 10^{11}$ $cm^{-2}$ | 0 to 5 | | 3.6 μA | 295 | 600 nm x 30 nm<br>(semiconductor) | 2003 [3] |
| 3D | **InSb** | Ti/Au | Corbino Disc | 800 %/T<br>(0.9T) | | 4685%<br>(8T) | 38,000 | $2.0 \cdot 10^{16}$ $cm^{-3}$ | 0 to 8 | | | 295 | 100 μm diameter | 2005 [4] |
| 2D | **InSb**/InAlSb | Au | Bar<br>(IVVI) | 585 Ω/T<br>(0.2T) | $B_{NEF}$ = 21<br>(calc. 0.2T)<br>$B_{NEF}$ = 2,200<br>(calc. 0.2T) | | 23,000 | $5 \cdot 10^{11}$ $cm^{-2}$ | | | 74 μA<br><br>2.2 μA | 300 | 1μm x 1μm x 25μm<br>(semiconductor)<br>35nm x 30nm x 25nm | 2006 [5] |
| 3D | **InSb** | Ti/Au | vdP Disc | 500 Ω/T<br>(α = 0.2) | | 9,900%<br>(α=0.7, 5T) | 12,230 | $4.2 \cdot 10^{16}$ $cm^{-3}$ | -5 to 5 | | | 295 | 100 μm diameter | 2009 [6] |
| 3D | **InSb**<br>(flash evaporated) | Au | vdP Square | | | 4% (0.9T) | | | -0.9 to 0.9 | | $<1 \cdot 10^{-7}$ $\Omega cm^{-2}$ | 295 | 3 mm x 3 mm | 2010 [7] |
| 3D | **InSb** | Ag | Planar<br>(Bar - IVIV) | | | 22,500%<br>(1T) | 21,300 | $8.0 \cdot 10^{16}$ $cm^{-3}$ | -1 to 1 | | 10 mA | 295 | 10 mm x 0.65 mm<br>(semiconductor) | 2012 [8] |
| 3D | **InSb** on glass<br>(flash evaporated)<br><br>**InAs** on glass<br>(flash evaporated) | Au | vdP Disc | | | 50%<br>(InSb, 0.8T)<br><br>3%<br>(InAs,0.8T) | | | -0.8 to 0.8 | | | 295 | 18 mm diameter | 2016 [9] |
| 3D | **InAs** (Si-doped) | Ti/Au | Bar<br>(2 terminal) | 85 Ω/T<br>(0.1T, 300K)<br><br>562 Ω/T<br>(0.26T, 75K) | | 61%<br>(1T, 300K)<br><br>176%<br>(1T, 100K) | 8,160<br>(300K)<br><br>25,000<br>(80 K) | $5.6 \cdot 10^{16}$ $cm^{-3}$ | -1 to 1 | $1 \cdot 10^{-7}$ $\Omega cm^2$ | 100 μA | 5 - 300 | 300 μm x 20 μm<br>(semiconductor) | 2012 [10] |
| 3D | **InAs** | Ti/Au | Bar<br>(3 terminal) | 2 Ω/T<br>(0.01T) | | 50% (1T) | 8,160 | $5.6 \cdot 10^{16}$ $cm^{-3}$ | -1 to 1 | $1 \cdot 10^{-7}$ $\Omega cm^2$ | 100 μA | 295 | 100 μm x 10 μm<br>(semiconductor) | 2012 [11] |
| 3D | **InAs** | Ti/Au | Bar<br>(2 terminal) | 440 Ω/T<br>(0.5T, 100μA) | | 80% (1T) | 8,200 | $5.6 \cdot 10^{16}$ $cm^{-3}$ | -1 to 1 | | 100 μA | 295 | 6 μm x 0.15 μm<br>(semiconductor) | 2013 [12] |
| 3D | **InAs** (Si-doped) | Ti/Au | Corbino Disc<br><br>vdP Disc<br><br>Bar<br>(2 terminal)<br><br>Bar<br>(4 terminal IVVI) | 2 Ω/T<br>(1T, 300K)<br><br>N/A<br><br>160 Ω/T<br>(1T, 300K)<br><br>3 Ω/T<br>(1T, 300K) | $B_{NEF}$ = 20<br>(calc. 1T, 300K)<br><br>N/A<br><br>$B_{NEF}$ = 12<br>(calc. 1T, 300K)<br><br>$B_{NEF}$ = 190<br>(calc. 1T, 300K) | 40%<br>(1T, 300K)<br><br>900%<br>(1T, 300K)<br><br>90%<br>(1T, 300K)<br>250%<br>(1T, 300K) | 8,160<br>(300K)<br><br>25,000<br>(75 K) | $5.6 \cdot 10^{16}$ $cm^{-3}$<br>(300K) | -1 to 1 | $10^{-7}$ $\Omega cm^2$ | 100 μA | 5 - 300 | 160 μm diameter<br><br>160 μm diameter<br><br>1000 μm x 40 μm<br>(semiconductor)<br><br>1000 μm x 40 μm<br>(semiconductor) | 2013 [13] |
| 2D | **InAs** | Co gate<br><br>Fe gate | vdP Square | 681 Ω/T<br>(0.3T, 4.2K)<br><br>78 Ω/T<br>(0.3T, 295K) | | 12,000%<br>(1T, 4.2K)<br><br>800%<br>(1T, 295K) | 193,800<br>(4.2K)<br><br>31,000<br>(295K) | $9.46 \cdot 10^{11}$ $cm^{-2}$<br>(4.2K)<br><br>$2.28 \cdot 10^{11}$ $cm^{-2}$<br>(295K) | -1 to 1 | | | 4.2<br><br>295 | 8 x 8 mm device<br>2 x 8 mm gate | 2005 [14] |
| 2D | **InAs**/InGaAs | Au | Bar<br>(IVVI) | 1000 Ω/T<br>(4.2K) and<br>900 Ω/T<br>(300K) | | 115,000%<br>(1T, 4.2 K)<br>and<br>1,900%<br>(1T, 295 K) | 62,000<br>(dark) and<br>149,000<br>(light) | $2.85 \cdot 10^{11}$ $cm^{-2}$<br>(dark) and<br>$5.7 \cdot 10^{11}$ $cm^{-2}$<br>(light) | -8 to 8 | $2 \cdot 10^{-8}$ $\Omega cm^{-2}$ | | 4.2 | 200 μm x 12 μm<br>(semiconductor) | 2002 [15] |
| 2D | **InAs**/InGaAs | Au | Bar<br>(IVVI) | 315 Ω/T<br>(0T, 4.2K) | $B_{NEF}$ = 1.5<br>(0T, 4.2K) and<br>$B_{NEF}$ = 1.3<br>(0.8T, 4.2K) | 1,200%<br>(1T) | 150,000<br>(4.2K, light)<br>and<br>21,000<br>(295K) | $6 \cdot 10^{11}$ $cm^{-2}$<br>(4.2K, light) and<br>$3.5 \cdot 10^{11}$ $cm^{-2}$<br>(295K) | -1 to 1 | | 7.5mA | 4.2<br><br>295 | 200 μm x 10 μm<br>(semiconductor)<br>20 μm x 8 μm<br>(semiconductor) | 2004 [16] |
| 2D | **InAs**/InGaAs | Au | Bar<br>(IVVI) | | | 1,100%<br>(1T) | 167,000<br>(4.2K) | $5.7 \cdot 10^{11}$ $cm^{-2}$<br>(4.2K) | -1 to 1 | $3 \cdot 10^{-8}$ $\Omega cm^{-2}$ | | 4.2 | 200 μm x 20 μm<br>(semiconductor) | 2004 [17] |
| 2D | **InAs**/AlSb | Metal | Bar<br>(IVIV) | 500 Ω/T | | 300%<br>(2T, 5K) | 7,000<br>(300K) and<br>28,000<br>(<50K) | $5 \cdot 10^{11}$ $cm^{-2}$ | -2 to 2 | $2.6 \cdot 10^{-7}$ $\Omega cm^{-2}$ | 100 μA | 5<br><br>300 | 9 μm x 0.45 μm<br>100 or 300 nm tabs<br>(semiconductor) | 2006 [18] |
| 2D | **InAs**/AlSb | Ta/Au | Bar (IVIV)<br><br>Tabless Bar | 67 Ω/T<br>(45mT, 1mA) | | | | $2 \cdot 10^{12}$ $cm^{-2}$<br>(7 nm deep)<br><br>$5 \cdot 10^{11}$ $cm^{-2}$<br>(18 nm deep) | | | 200 μA Bar<br><br>1 mA Tabless | 300 | 2 μm x 0.1 μm<br>Bar (semiconductor)<br><br>500 nm x 100 nm<br>Tabless (semiconductor) | 2009 [19] |
| 2D | **InGaAs**/AlGaAs | Ni/Ge/Au | Bar<br>(2 terminal) | 488 Ω/T<br>(1T) | $B_{NEF}$ = 0.1<br>(calc. 0.1T)<br>$B_{NEF}$ = 0.012<br>(calc. 1T) | 17.3% (1T) | 6,040 | $1.2 \cdot 10^{12}$ $cm^{-2}$ | -1 to 1 | $7.2 \cdot 10^{-5}$ $\Omega cm^{-2}$ | 100 mA | 295 | 100 μm x 20 μm<br>(semiconductor) | 2016 [20] |
| 3D | **Si** (doped) | $TiSi_2$ | Bar (IVVI)<br><br>Bar (IVIV) | 0.17 Ω/T<br>(IVVI)<br><br>0.46 Ω/T<br>(IVIV) | | 3.5%<br>(10T, IVVI)<br><br>15.3%<br>(10T, IVIV) | 65 | $6.4 \cdot 10^{19}$ $cm^{-3}$ | -10 to 10 | $8.5 \cdot 10^{-7}$ $\Omega cm^{-2}$ | 50 μA | 300 | 17.3 μm x 1 μm<br>(semiconductor) | 2006 [21] |
| 2D | **Graphene** | Ta/Au | Bar<br>(IVIV) | 1000 Ω/T | SNR = 17 | 10% (1T) | 2,500 | $10^{12}$ $cm^{-2}$ | -0.5 to 0.5 | $3.7 \cdot 10^{-6}$ $\Omega cm^{-2}$ | 150 μA | 295 | 1.2 μm x 0.15 μm<br>(graphene) | 2010 [22,23] |
| 2D | **Graphene** | Pd<br><br>Ti/Au | vdP Disc | 1000 Ω/T<br>(2 terminal) | | 50,000%<br>(9T, Pd) | 5,000 | $1.6 \cdot 10^{12}$ $cm^{-2}$ | -9 to 9 | | | 295 | 8 μm diameter | 2011 [24] |
| 2D | **Graphene**<br>(CVD) | Ti/Au | vdP Disc | 150 Ω/T | SNR = 27 | 600% (12T) | 1,350 – 2,500 | | 0 to 12 | | | 4.2 | N/A | 2011 [25] |
| 2D | **Graphene**<br>(CVD)<br><br>**Graphene**<br>(Bilayer - CVD)<br><br>**$Bi_2Se_3$** | Pd/Au | Bar<br>(Planar - IVIV) | | | 80% (0.6T)<br><br>600%<br>(0.6T)<br><br>2% (0.6T) | 1,200<br><br>5,000<br><br>50 | $9 \cdot 10^{12}$ $cm^{-2}$<br><br>$1 \cdot 10^{13}$ $cm^{-2}$<br><br>N/A | -0.6 to 0.6 | | | 295 | 6.5 mm x 0.15 mm<br>(Graphene)<br><br>180 μm x 30 μm<br>($Bi_2Se_3$) | 2019 [26] |
| 2D | hBN/**Graphene/**<br>hBN | Ti/Al | vdP Disc | 30,000 Ω/T (1T) | | $10^7$ % (8T) | | $6.7 \cdot 10^{10}$ $cm^{-2}$ $V^{-1}$ $\cdot V_g$ | -8 to 8 | | | 295 | 9 μm diameter | 2020 [27] |
| 2D | **Graphene**<br>(Suspended) | Ti/Au | Corbino Disc | 12,000 Ω/T<br>(0.15T, 4K) | $B_{NEF}$ = 60<br>(0.15T, 4K) | 100%<br>(0.15T, 27K) | 100,000 | $3 \cdot 10^{10}$ $cm^{-2}$ | -0.15 to 0.15 | | 10 μA | 4<br><br>27 | 2.25 μm diameter | 2021 [28] |

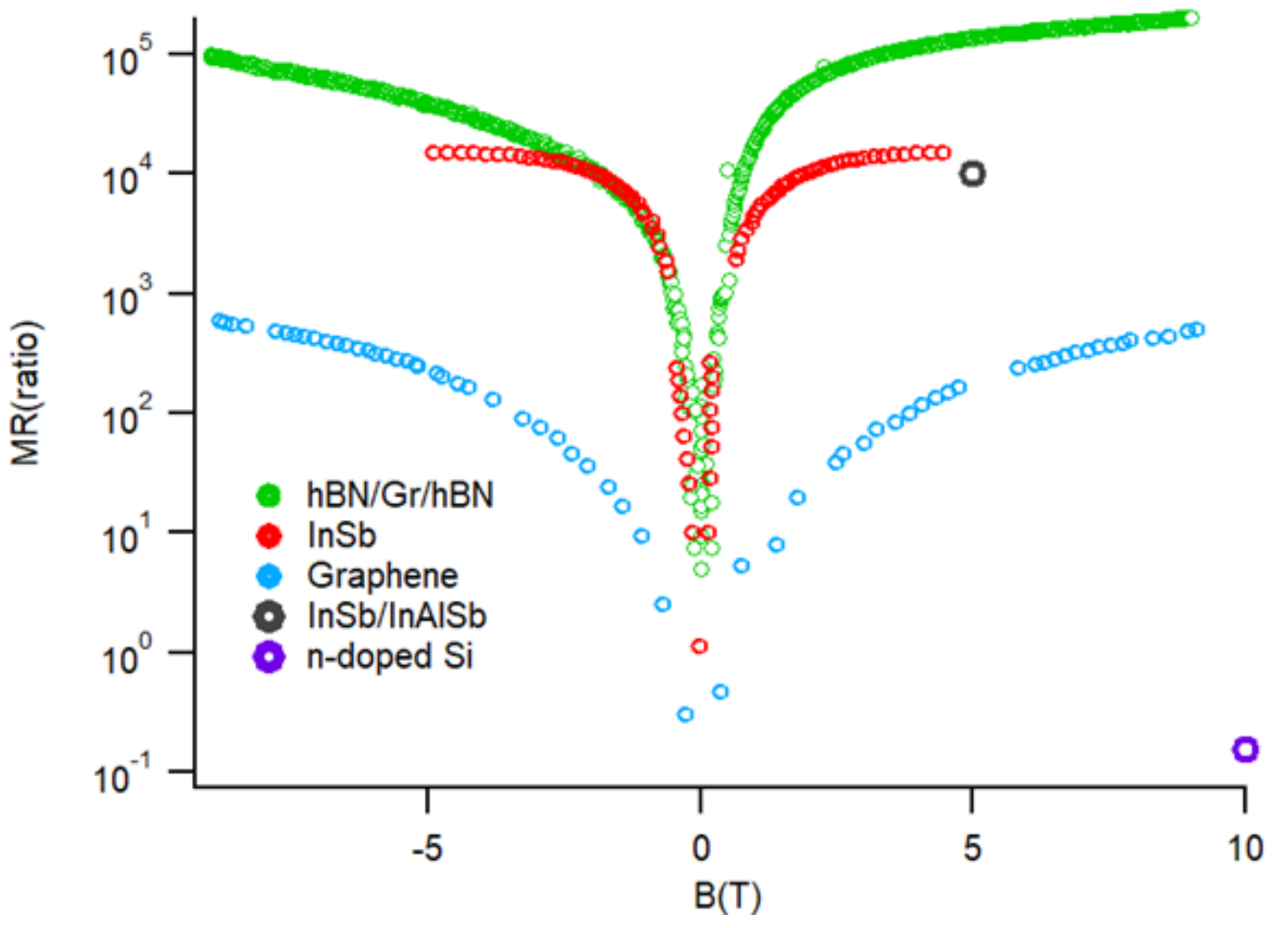

*Figure xx: The best magnetoresistance (MR) values reached in EMR devices made by different materials*

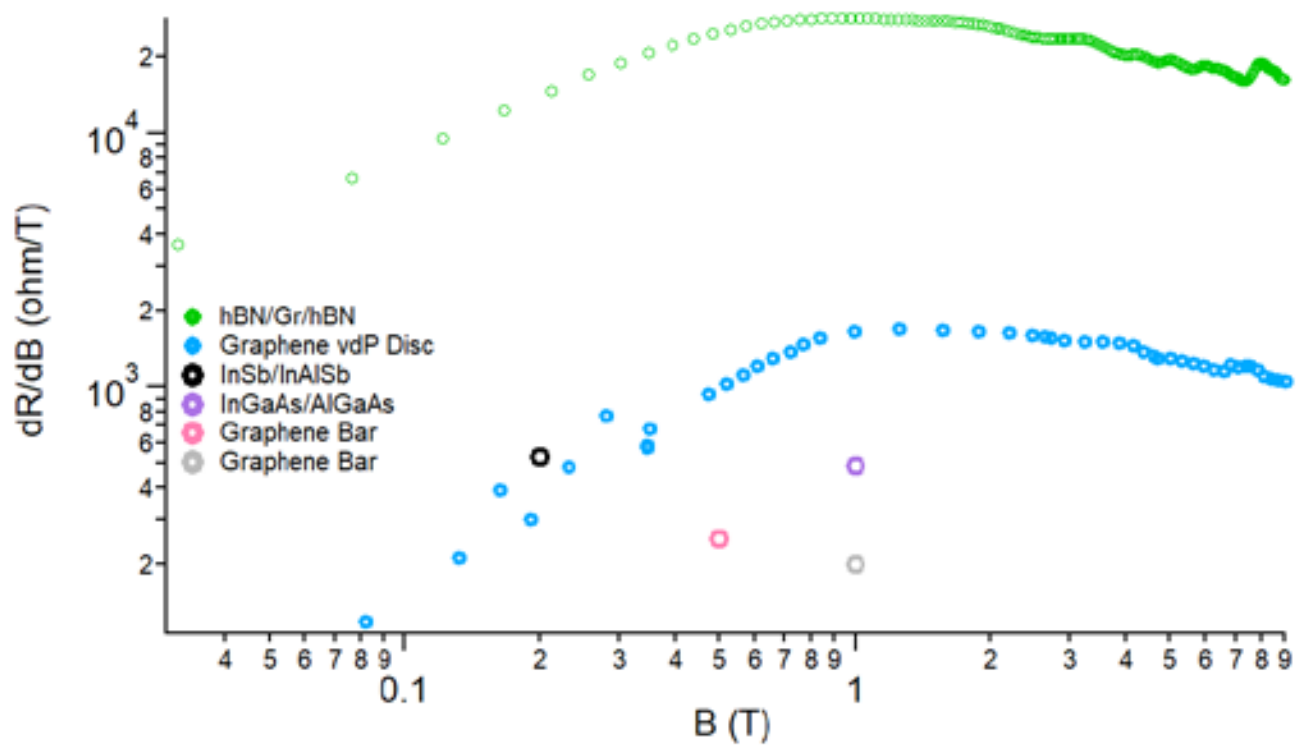

*Figure xx: The best sensitivity (dR/dB) values in EMR devices made by different materials.*

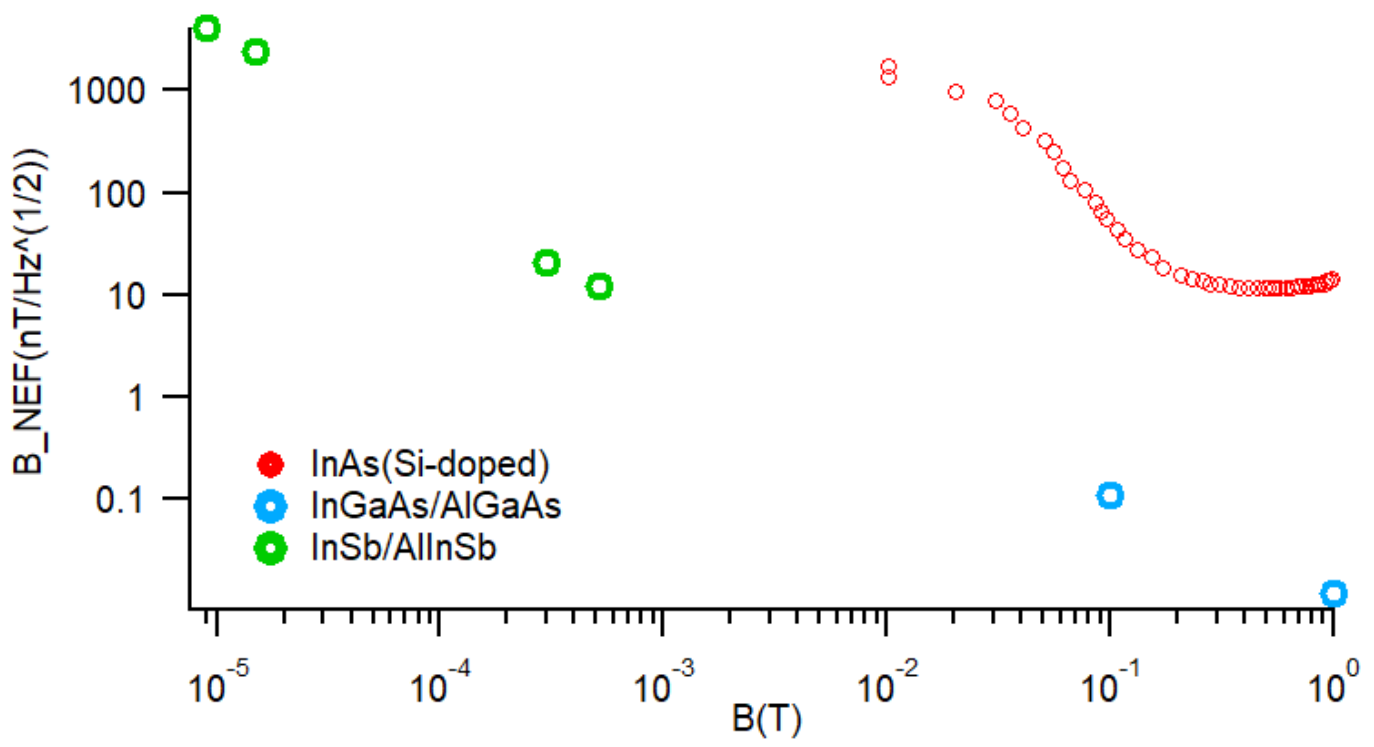

*Figure 1.1: The best values of the a) magnetoresistance, b) the sensitivity (dR/dB) and c) the noise-equivalent field ($B_{NEF}$) in extraordinary magnetoresistive devices made by different materials.*

# 1. Extraordinary Magnetoresistance

## 1.1. Introduction

Magnetic fields are a fundamental part of nature and play a role in numerous physical processes and key technologies, ranging from the simple compass to the complex magnetic resonance imaging systems or tokamak fusion chambers. It is of essential scientific interest to detect and measure magnetic fields with as high an accuracy as possible to better understand nature as well as to optimize technologies.

Magnetoresistive sensors have been used to detect magnetic fields for decades through magnetic field-induced changes to the sensor resistance. The development of magnetic hard drives in particular has been a key area of application for this technology as the detection of small magnetic fields using the the readhead is crucial.[29,30] Numerous other areas ranging from the detection of biological brain activity to space applications have also been explored.[31] The highlight work on magnetoresistive sensors was the awarding of the 2007 Nobel prize in physics to Albert Fert and Peter Grünber for the discovery of the giant magnetoresistance (GMR), but several other classes of magnetoresistances are actively researched including colossal magnetoresistance, tunnel magnetoresistance, and extreme magnetoresistive materials.

Magnetoresistance can have both an intrinsic and geometric contribution. The intrinsic contribution stems from changes to material properties such as the magnetization, electronic band structure, and mobility which are induced by a magnetic field. In contrast, the geometric contribution depends on the design of the device, including the placement of contacts and geometries of the constituent materials. By using geometric magnetoresistance and combining several materials into a hybrid device, it is possible to geometrically enhance the magnetoresistance by orders of magnitude – such devices are known as extraordinary magnetoresistance (EMR) sensors. EMR devices hold the potential for turning the large magnetoresistance into sensitive magnetometers that can measure weak magnetic fields at room temperature using simple measurements of the electrical device resistance. EMR sensors further combines sensing capabilities in either two- and four-terminal measurement modes at room temperature with a device design that do not contain any magnetic elements, providing advantages of conventional magnetoresisive and Hall sensors. Despite some progress towards weak-field magnetometry using EMR sensors, the potential is to date largely unmet. Besides the decline in commercial prospects as read-heads for ultrahigh-density magnetic storage became obsolete with the emergence of solid state harddrives, a few key challenges of realizing EMR sensors also exist including complex device fabrication and the need for performing optimization within an overwhelmingly large parameter space. In this work, we review and discuss this field with a focus on elucidating the impact that the hybrid device geometry and constituent materials have on determining the magnetoresistive performance as well as highlighting the key findings for use in the development of the next generation of EMR sensors. The review serve as a more comprehensive and up-to-date account of the field compared to the previous review on EMR published in 20XX (ref). The main comparison of the experimentally realized EMR devices presented in the literature is given in Table 1 and with Figure XX presenting the state-of-the-art devices with respect to magnetoresistance, sensitivity, and noise equivalent field. Each of these are discussed substantially during the course of this review.

The review is structured as follows: First, the physics of the EMR effect is discussed, a historical perspective on the discovery of the EMR effect is provided and key metrics for describing EMR are given. Following this, EMR devices in literature are reviewed in three major sections with respect to their geometry, material parameters, and application in magnetometery, respectively. In the section on geometry, we describe both the main device geometries used in EMR as well as aspects on 3-dimensional inclusions and contact placement. With regards to material parameters, besides the carrier density and mobility of the constituent materials, effects such as contact resistance and material specific phenomena are discussed. Finally, with regards to magnetometers, fabrication techniques as well as the concept of noise equivalent fields are discussed and reviewed. Ending the review, we conclude on the present state-of-the-art, trends in the field, and directions of future research.

## 1.2. Fundamentals of Extraordinary Magnetoresistance

Magnetoresistance is a property of some material systems where the electrical resistance changes in response to a magnetic field. The types of phenomena which produce a magnetoresistive effect can be separated into two general categories: intrinsic magnetoresistance which originates from the properties of a material, and geometric magnetoresistance which results from the interaction between charge carriers and the device geometry. The extraordinary magnetoresistance effect is a special case of geometric magnetoresistance which can be observed in certain composite materials where the matrix material has a high carrier mobility and the second phase is much more electrically conductive than the matrix.

EMR was first described by Solin et al. in 2000 where devices were presented that showed very large changes in the measured resistance induced by a magnetic field; up to a 1,500,000% increase in the resistance when the magnetic flux density was increased from 0 to 5 T.[1] The effect can be understood by considering a high mobility semiconductor with an embedded metal disk as show in Figure 1.1.[32]

An unbound charged particle moving in an electromagnetic field experiences a force contribution parallel to the electric field ($\boldsymbol{E}$) and a Lorentz force perpendicular to its velocity ($\boldsymbol{v}$) and the magnetic field ($\boldsymbol{B}$) according to:

$$\boldsymbol{F} = e(\boldsymbol{E} + \boldsymbol{v} \times \boldsymbol{B})$$

where $e$ is the charge of the particle

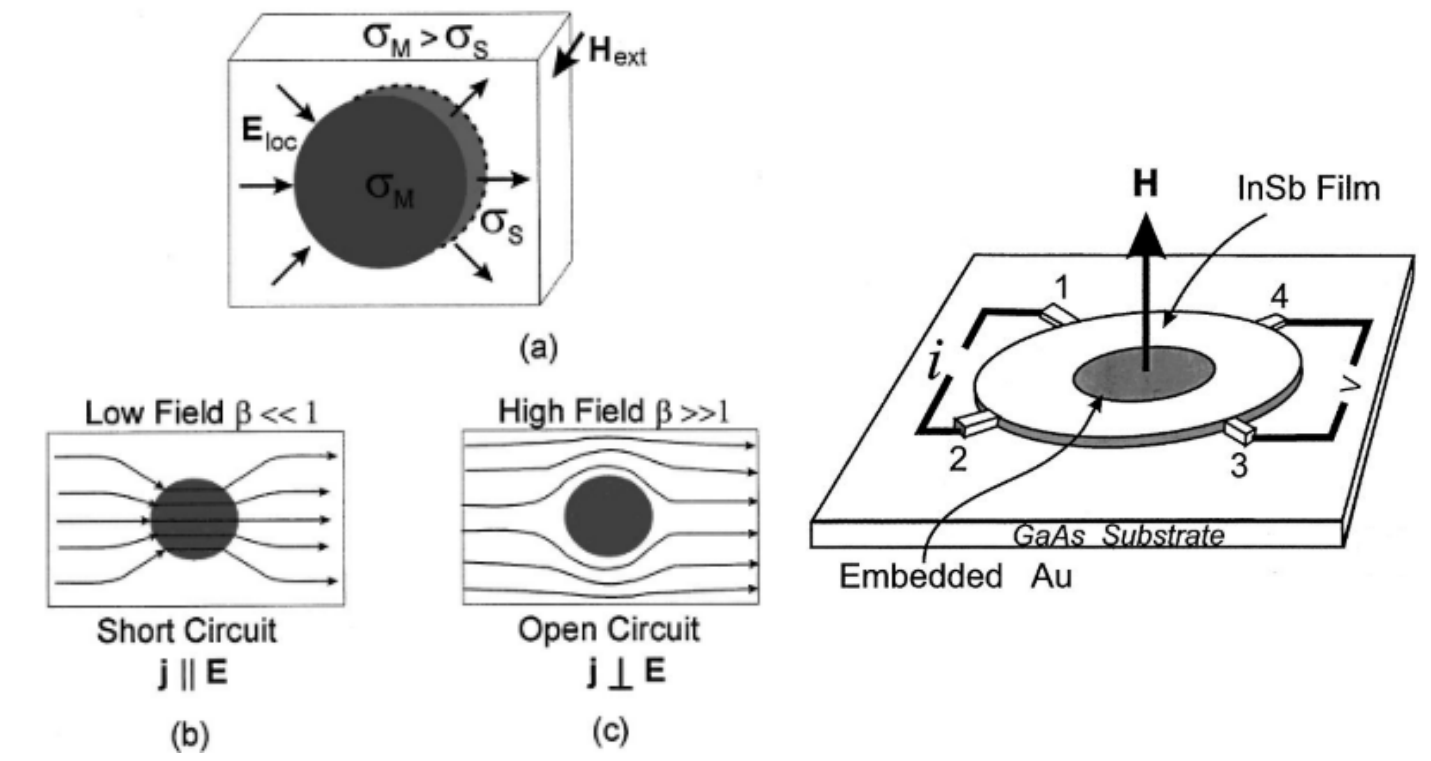

*Figure 1.1: Orientation of the local electrical field in a semiconductor material with a metal inhomogeneity (a). Current paths through the composite material in the low (b) and high magnetic field regimes (c). Schematic of a typical EMR device in the vdP geometry (right).*[32]

In the case of a semiconductor/metal hybrid system where the metal is much more conductive than the semiconductor, the metal can be considered an equipotential volume. In this case, the local electrical field around the semiconductor/metal boundary is oriented perpendicular to the interface at all points (see Figure 1.1a). This leads the current into the metal in the absence of a magnetic field and produces a low zero-field resistance (Figure 1.1b). If a magnetic field is applied however, charge carriers approaching the boundary will experience a perpendicular deflection due to the Lorentz force, resulting in a force component tangential to the semiconductor/metal interface. As the magnetic field strength or the charge velocity is increased, the deflection becomes stronger and more of the current is forced around the conductive material and instead travels through the higher resistive semiconductor (Figure 1.1c). The magnitude of the deflection is given by the Hall angle, which approaches 90° in the high field limit and corresponds to a total expulsion of the current from the conductive material. Thus, the current deflection increases the resistance across the device as a function of the applied field, leading to a positive magnetoresistance.

The constitutive relation between the current, $\boldsymbol{j}$, and the electric field, $\boldsymbol{E}$, in the presence of a magnetic field can be expressed through the following equation:

$$\boldsymbol{j} = \boldsymbol{\sigma}(\boldsymbol{B})\boldsymbol{E}$$

$\boldsymbol{\sigma}$ is the magnetoconductivity tensor which is given by:

$$\boldsymbol{\sigma}(\boldsymbol{B}) = \frac{\sigma_0}{1+|\beta|^2}\begin{bmatrix} (1+\beta_x^2) & (-\beta_z+\beta_y\beta_x) & (\beta_y+\beta_z\beta_x) \\ (\beta_z+\beta_y\beta_x) & (1+\beta_y^2) & (-\beta_x+\beta_y\beta_z) \\ (-\beta_y+\beta_z\beta_x) & (\beta_x+\beta_y\beta_z) & (1+\beta_z^2) \end{bmatrix}$$

Here, $\beta$ is the unitless magnetic field:

$$\beta_i = \mu B_i$$

and $\sigma_0$ is the intrinsic conductivity of the material in the absence of a magnetic field:

$$\sigma_0 = \frac{ne^2\tau}{m^*} = ne\mu$$

where $n$ is the density of charge carriers, $\tau$ is the momentum relaxation time between scattering events, $m^*$ is the effective mass of the charge carriers, and $\mu$ is the carrier mobility. Here we consider a material with only one conduction band and an isotropic mobility. For a thin film structure with a magnetic field perpendicular to the plane of the device ($\boldsymbol{B} = B\hat{\boldsymbol{z}}$) the system can be reduced to two dimensions with a simplermagnetoconductivity tensor:

$$\sigma_{ij}(B) = \frac{\sigma_0}{1+\beta^2}\begin{bmatrix} 1 & -\beta & 0 \\ \beta & 1 & 0 \\ 0 & 0 & 1 \end{bmatrix}$$

The off-diagonal elements of the tensor describe the deflection of the current and depend on the unitless magnetic field $\beta = \mu B$. A strong deflection can therefore be achieved either by increasing the carrier mobility or the magnetic flux density. High mobility materials are preferred for EMR applications for this reason, as they require lower magnetic fields to achieve the deflection necessary to force the current around conducting inhomogeneities.[32–34]

## 1.3. History of Extraordinary Magnetoresistance

Extraordinary magnetoresistance can be better understood if contextualized within the greater historical framework out of which it arose. The origins of EMR began in an unlikely place: in the research laboratories of the infamous chemical and biotechnological company Monsanto. In the late 1960s Monsanto was also a producer of raw semiconductor materials, including GaAs and GaAsP, which later established its own high volume production lines of optical electronic devices and became the first company to mass produce light emitting diodes.

In addition to developing GaAs materials for diodes, the company also explored their use for lasing applications. F. V. Williams, one of Monsanto's researchers in GaAs lasers,[35–37] noticed that in some GaAs samples grown under metal-rich conditions the measured electron mobility was anomalously high. Unable to explain this phenomenon, Williams presented this problem in a private correspondence to C. M. Wolfe at MIT.[38]

Wolfe set about trying to understand the observed behavior and posited that it could be explained by the presence of inhomogeneities,[39] particularly if the inhomogeneous region was significantly more conductive than the surrounding bulk. In 1971, Wolfe modeled a van der Pauw disc with two concentric sections where the outer section represented the bulk semiconductor and the inner section was comprised of an inhomogeneity with higher conductivity (see inset of Figure 1.1).[40] A set of analytical expressions was derived for the apparent resistivity, $\rho_{app}$, and Hall coefficient, $R_{H,app}$, of a van der Pauw disc with an out of plane magnetic field in the limit where the conductivity of the inhomogeneity, $\sigma_0$, is much higher than that of the bulk, $\sigma$, ($\sigma_0/\sigma \rightarrow \infty$). Here, the Hall coefficient was found by sourcing current between diagonally arranged leads while deducing the transverse voltage drop whereas the resistivity was found by sourcing current between adjacent leads and measuring the voltage drop at the opposing leads.

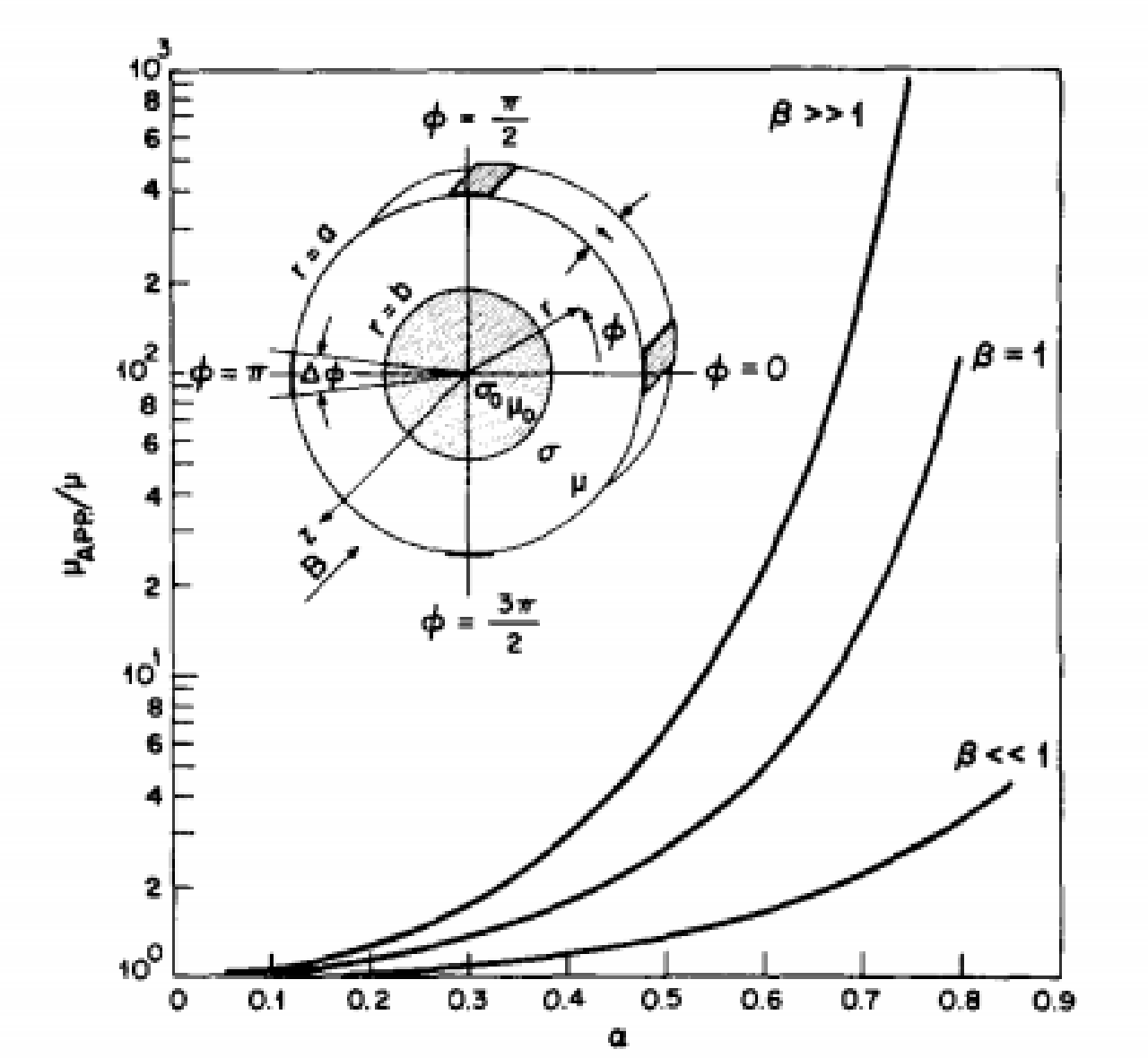

*Figure 1.2: Ratio between the apparent and nominal mobility as a function of the relative size of the conductive inhomogeneity for different regimes of the magnetic field strength.*[40]

The apparent mobility, $\mu_{app}$, of the device is then given by $R_{H,app}/\rho_{app}$. Plotting the ratio between the apparent mobility and bulk mobility as a function of the ratio of the inner and outer radii, α, at different magnetic field strengths yields the set of curves shown in Figure 1.2. Wolfe observes that in the low field limit, β << 1, as the size of the inhomogeneity increases the resistivity and Hall constant both decrease. However, the resistivity decreases faster than the Hall constant and therefore results in a higher apparent mobility. In the high field limit, β >> 1, the Hall constant approaches $\mu/\sigma$ which is equal to that of a homogenous sample as there is no current flow through the inhomogeneity. The observed behavior is thus a geometric effect, produced a current redistribution around the conductive inhomogeneity in the presence of a magnetic field.

To demonstrate this effect, inhomogeneities were intentionally created in thin epitaxial layers of GaAs grown on semi-insulating substrates in 19XX (ref). Conducting inclusions were simulated by alloying Ga into parts of the exposed surface of the epitaxial layers. The optical phonon limited carrier mobility of pure GaAs is calculated to be around 8,000 $cm^2V^{-1}s^{-1}$ at 300 K while the lattice limited mobility at 77 K is approximately 240,000 $cm^2V^{-1}s^{-1}$ (ref). Samples with conducting inhomogeneities were measured to have apparent carrier mobilities that exceeded the theoretical maximum lattice-limited mobility. The apparent mobility of the samples ranged from 7,400 to 24,000 $cm^2V^{-1}s^{-1}$ at 300 K and between 150,000 and 740,000 $cm^2V^{-1}s^{-1}$ at 77 K, demonstrating that it is possible for this mechanism to produce the observed anomalous behavior.[40]

Wolfe then further elaborated his theory in a subsequent paper published the following year.[38] The effect of material parameters on the behavior of the system were detailed by plotting the dependence of the derived expressions on various independent variables. By mapping these relationships, several features became apparent: 1) Relatively small differences in the conductivity of the medium can significantly affect the mobility measurement; 2) Even small values of α and $\sigma_0/\sigma$ have an appreciable effect on the Hall coefficient; and 3) The apparent mobility can be larger than the real mobility by up to several orders of magnitude depending on the various parameters.

Additional experimental results were obtained to further buttress the theory (ref) and as a result of these investigations Wolfe recommended against using the mobility of a semiconductor as a figure of merit unless the homogeneity of the sample can be determined first. Inhomogeneities can appear in semiconductor materials through various mechanisms: for example, due to precipitates or metallic inclusions formed in samples grown under metal-rich conditions, variations in the doping level of doped materials, or through the uptake of impurities at different rates through the different crystallographic faces of polycrystalline materials. It was also suggested that the level of homogeneity could be estimated by measuring the Hall coefficient as a function of an applied magnetic field (ref).

Wolfe's work was found particularly relevant 26 years later in 1998 by T. Thio and S. A. Solin, who at the time were working as researchers for the computer manufacturer NEC. In one of their research efforts, Thio and Solin explored the use of $Hg_{1-x}Cd_xTe$ as a magnetic sensor for hard disc read heads. To assess the performance of the material for magnetic sensing, heavily doped, lightly doped, and undoped samples of $Hg_{1-x}Cd_xTe$ were manufactured into both Hall bar and Corbino disc geometries (ref). The data produced by the Hall bar samples suggested that the doped samples perform worse for applications requiring magnetoresistive behavior. However, for the Corbino samples it was found that for the doped samples the electron mobility was roughly 400% higher than expected. Thio and Solin theorized that the observed behavior may be due to the presence of microscopic inhomogeneities in the doped samples, possibly due to separation into dopant rich and dopant poor phases.[41]

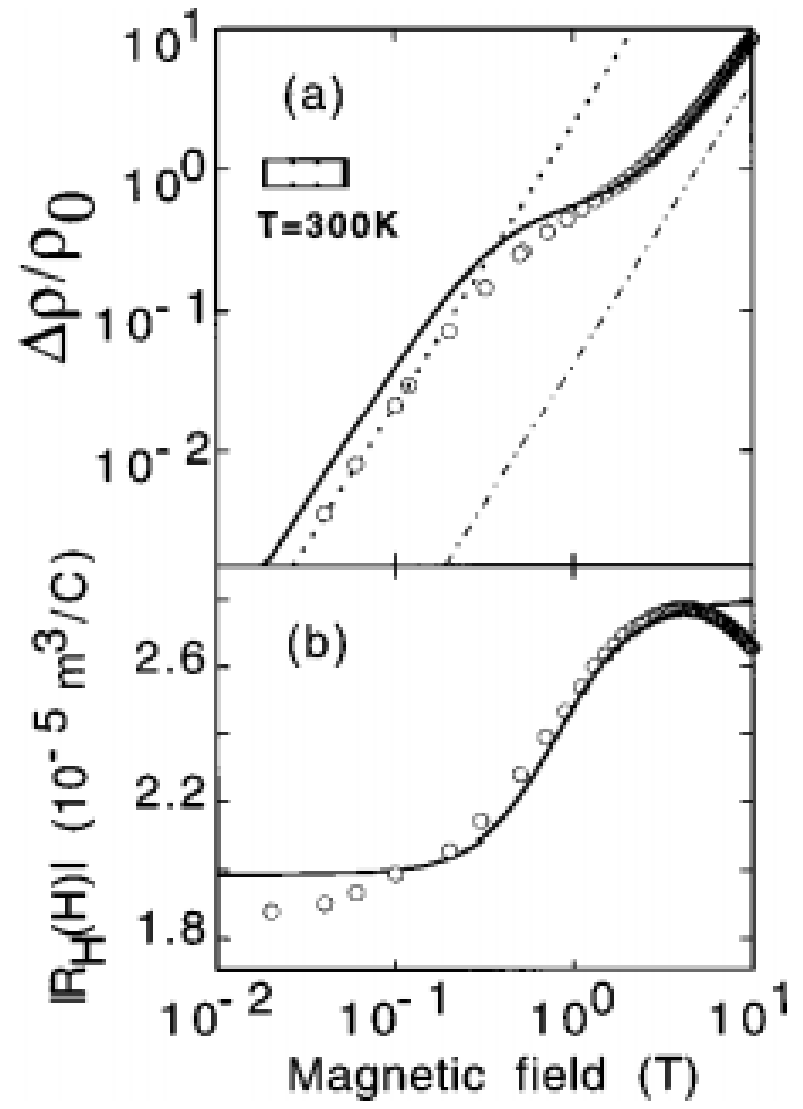

*Figure 1.3: Magnetoresistance (top) and Hall coefficient (bottom) data for $Hg_{1-x}Cd_xTe$ samples as a function of magnetic field.*[42]

In a second paper published that same year, they argued that the effect could be explained by the Wolfe model, if it was augmented to include physical magnetoresistance.[42] Hall bar devices were fabricated by growing films of $Hg_{1-x}Cd_xTe$ using molecular beam epitaxy with $x \approx 0.10$ and a compositional fluctuations around $\Delta x \pm$ 1.5%. The size of the inhomogeneities was estimated to be between 30-220 nm in diameter. An interesting behavior could be observed in the magnetoresistance, $\Delta\rho/\rho_0 = (\rho(B) - \rho_0)/\rho_0$ and Hall coefficient data, as shown in Figure 1.3. At both low and high fields the MR followed a quadratic dependence, but the curvature was 30 times higher at low fields than at high fields, with a cross-over point around 0.4 T. For intrinsic semiconductors, $(n = p)$, the MR is expected to increase quadratically as a function of field with a curvature of $\mu_e \mu_h B^2$, yet here the low field conductivity was much higher than this value.

Another anomaly was observed in the Hall coefficient which at zero-field was 30% lower than at high fields, a similar result to what Wolfe observed in his experiments. Thio and Solin explain these behaviors by referring to the Wolfe model. They posited that in the absence of a magnetic field the current flows preferentially through the low resistance inhomogeneity, reducing the apparent resistivity of the sample. At high fields, the Hall angle approaches 90°, forcing the current around the inhomogeneity and reducing the conductance as the cross-sectional area that the current can flow through is decreased. The cross-over field occurs at $B = 1/\mu$, where $\mu$ is the mobility of the bulk semiconductor.

They point out that Wolfe's model explicitly ignores the case where semiconductors possess intrinsic physical magnetoresistance but then derive another set of equations which can capture this effect. The amended expression for the apparent resistivity fit the data well, lending validity to their approach. Thio and Solin concluded the paper by suggesting that the geometric enhancement of magnetoresistance at low fields could have important applications in the development of magnetic field sensors.

## 1.4. The Discovery of Extraordinary Magnetoresistance

Two years later, in 2000, Solin and Thio published results which became the foundational text for the field of extraordinary magnetoresistance. In it, the authors described the first sensor which made use of the aforementioned effect to measure magnetic fields. The sensor utilized the geometry which was proposed in the original Wolfe paper[40]; a vdP disc with four equally spaced contacts and a circular metal inclusion (see inset in Figure 1.4).

Metal organic vapor phase epitaxy was used to grow the semiconductor material for the devices. First, a 200 nm buffer layer of InSb was deposited on a GaAs substrate. A 1.3 µm thick active layer of Te-doped InSb was then grown with a mobility and carrier density of 45,000 $cm^2V^{-1}s^{-1}$ and $2.6\times10^{16}$ $cm^{-3}$, respectively. This was followed by a 50 nm InSb contacting layer, and the structure was terminated with a 200 nm thick passivating layer of $Si_3N_4$. The lattice mismatch between the GaAs substrate and InSb creates a high degree of disorder in InSb thin films which drastically reduces the mobility in the buffer layer. Band bending at the InSb/$Si_3N_4$ interface also reduces the mobility and depletes the number of carriers in the contacting layer. As such, neither layer represents a parallel conduction channel and current only runs through the active layer. Reactive ion etching was used to define the shape of the device and Ti/Pt/Au layers were successively deposited to form the metal inclusion and metal contacts of the device. Devices were produced with varying metal filling factors defined as the ratio of the inner to outer diameter ($\alpha = r_i/r_o$). The magnetoresistance measured in a four-terminal mode transitioned from being very weak in absence of gold inclusion ($\alpha = 0$) to very strong as the filling factor increased (see Figure 1.4).For the InSb used in this experiment, the optimal value of the filling factor for high field magnetoresistance was found to be 13/16, but shifted to 12/16 in the low field regime. For $\alpha$ = 12/16 the resistance of the device increased by 83% at 0.05T and 400% at 0.1 T. In the high field regime, the device with $\alpha$ = 13/16 showed incredibly large magnetoresistances of 8,100% at 0.25 T, 42,000% at 1 T, and 1,500,000% at 5 T.

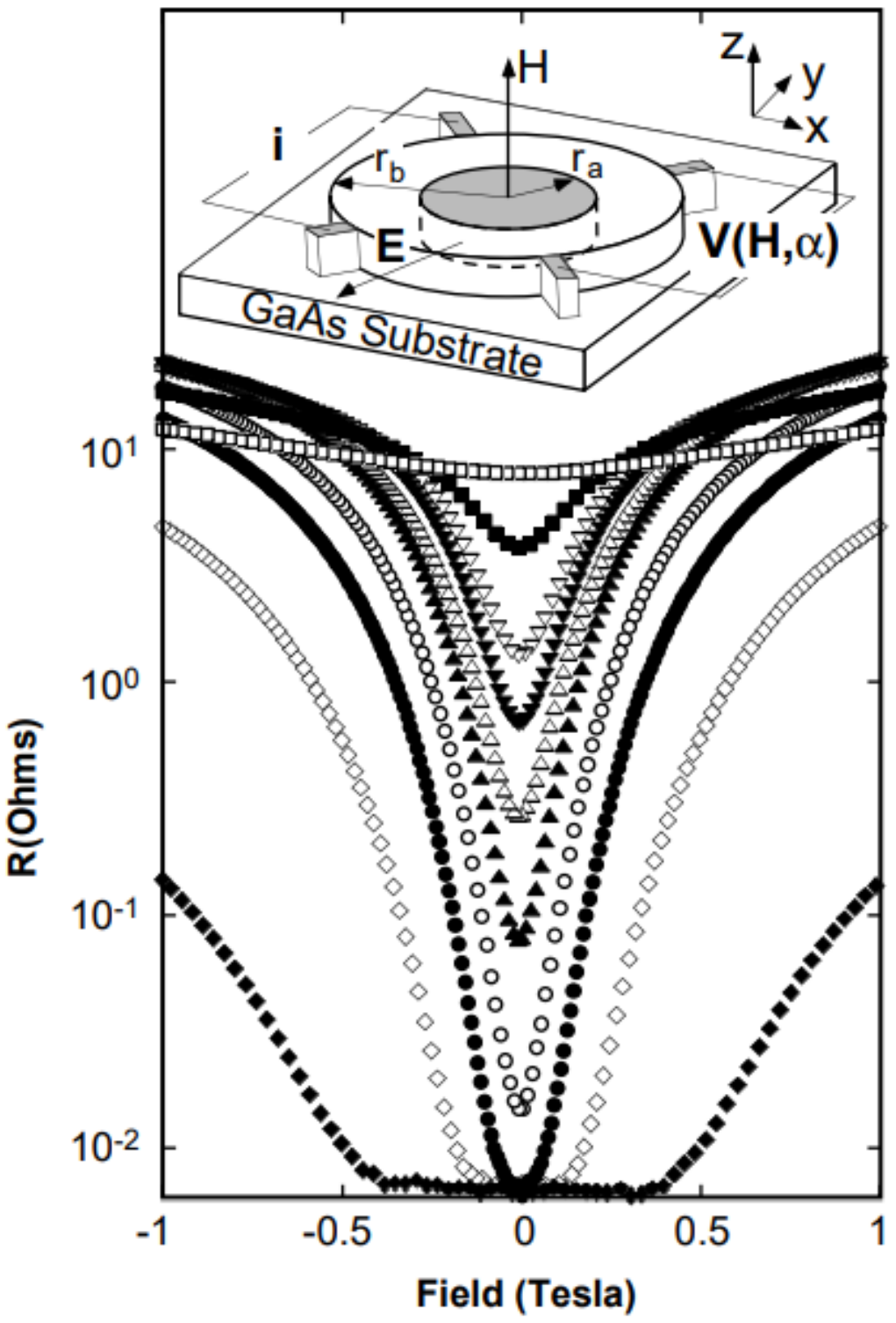

*Figure 1.4 Resistance as a function of magnetic field in InSb/metal hybrid EMR devices with different values of the metal filling factor α = $r_i/r_o$ = 0/16 (□), 6/16 (■), 8/16 (▽), 9/16 (▼), 10/16 (△), 11/16 (▲), 12/16 (○), 13/16 (●), 14/16 (◇), and 15/16 (◆).*[1]

## 1.5. Analytical and Numerical Modeling

The physics of the EMR devices described above is theoretically well-established and well suited for both analytical and numerical analysis. For concentric circular EMR devices with contacts placed symmetrically as in the case of Figure 1.4, the equations reduce to a solvable Laplace equation on a circular geometry. For this the electrical potential between the two voltage contacts on the periphery of the device can be found as function of the magnetic flux density $B$ by writing up the solution as an infinite series.[43] The solution is:

$$V_i(\alpha, \gamma, \eta, \beta, \sigma_0, \Delta, t, \theta) = \frac{1+\beta^2}{\sigma_0} \frac{I}{\pi t \Delta} \sum_{n=1}^{\infty} \frac{1}{n^2} \frac{1}{J^2+K^2} \{[(JU - KW)(1 - \alpha^{2n}\gamma) - \alpha^{2n}\eta(KU + JW)]cos(n\theta) + [(KU + JW) \times (1 - \alpha^{2n}\gamma) + \alpha^{2n}\eta(JU - KW)]sin(n\theta)\}$$

where

$$J = 1 + \alpha^{2n}\gamma + \beta\alpha^{2n}\eta$$

$$K = \beta + \alpha^{2n}\eta - \beta\alpha^{2n}\gamma$$

$$U = sin(n\pi/2 + n\Delta/2) - sin(n\pi/2 - n\Delta/2) - 2sin(n\Delta/2)$$

$$W = cos(n\pi/2 - n\Delta/2) - cos(n\pi/2 + n\Delta/2)$$

$$\gamma = [(\omega_0^2 - \omega^2) + (\omega_0\beta_0 - \omega\beta)^2]/[(\omega_0 + \omega)^2 + (\omega_0\beta_0 - \omega\beta)^2]$$

$$\eta = [2\omega(\omega_0\beta_0 - \omega\beta)]/[(\omega_0 + \omega)^2 + (\omega_0\beta_0 - \omega\beta)^2]$$

Here, $\beta = \mu B$, $\beta_0 = \mu_0 B$, $\omega = \sigma/(1+\beta^2)$, $\omega_0 = \sigma_0/(1+{\beta_0}^2)$, the contact width is $\Delta$ for all contacts, $\theta$ is the angular placement of the contact, $I$ is the current, $t$ is the thickness of the semiconductor, $\mu$ and $\sigma$ are the mobility and the electrical conductivity of the semiconductor, while $\mu_0$ and $\sigma_0$ are the mobility and the electrical conductivity of the metal. The analytical model shows a good agreement with EMR device experiments for symmetrical devices as displayed in Figure XX, in particular since no free fitting parameters is used to calculated the magnetoresistance.[2,32,44] An analytical solution also exists for an asymmetric circular device where the inner metal inclusion is displaced to the side[2] as well as for the bar-shaped device discussed in the next chapter (ref)

Numerically, the EMR governing equations can be solved in steady state using both finite difference and finite element methods.[32,43] Finite element solutions are particularly well-suited due to its flexibility with meshing geometrical features. The EMR finite element model used to model the concentric circular EMR device matches the analytically expression very well as well as displaying a good agreement with experimental data in the range from –1 to 1 T as observed in Figure XX.[32,43] Due to this agreement and because the EMR equations are relatively easy to implement in standard finite element software, numerical studies of the EMR effect have been performed in numerous articles, as will be detailed throughout this manuscript.

### 1.6. Figures of Merit

The studies on extraordinary magnetoresistance are not compressed to a single universal figure of merit, but a variety of metrics are typically used. The four most commonly used metrics include:

1) The field-dependent four-terminal resistance (*R*):

$$R(B) = \frac{\Delta V}{I}$$

where the current (*I*) between two contacts is typical fixed and the resulting field-dependent electrostatic potential drop ($\Delta V$) is measured between two voltage contacts.

2) The magnetoresistance (*MR*) which generally defined as:

$$MR(B) = \frac{R(B) - R(B = 0\,T)}{R(B = 0\,T)}$$

More conservative measures of the magnetoresistance combining both two- and four-terminal resistances have also been used.[23]

3) The sensitivity as defined by the change of resistance around a bias magnetic field:

$$Sensitivity = \left[\frac{dR}{dB}\right]_{B_{bias}}$$

Sensitivity is an important metric that relates directly to the voltage signal, $V_{signal} = I \cdot \left[\frac{dR}{dB}\right]_{B_{bias}} \Delta B$, generated in EMR magnetometers as they are subjected to a magnetic field signal $\Delta B = B - B_{bias}$. In some cases, the sensitivity is scaled with a factor of $1/R$ or $1/\sqrt{R}$ to account for effects such the effect that a change in the EMR device resistance has on the noise.

4) The two key metrics often used when EMR is applied to magnetometery are the signal-to-noise ratio and the noise-equivalent field. These are described in detail in Section 4 where the application of EMR for magnetometry is discussed.

These metrics are used to describe the performance of EMR devices throughout this review.

# 2. Geometry

The EMR effect depends on physically deflecting charge carriers and thereby changing the route the current must take as it travels from source to drain. Therefore, along with the properties of the constituent materials, the device geometry plays a major role in the performance of EMR devices, for instance going from a negligible MR for $\alpha = 0$ to MR(1T) = 42,000% for $\alpha = 13/16$ in Figure 1.4. In this chapter we will review the various designs which have been studied and how various geometric factors affect their performance. There is almost unlimited degrees of freedom for designing these devices as one can manipulate virtually any geometric part of the structure. However, the geometric variations can essentially be generalized to one of three types: 1) changes to the overall shape of the device, e.g. circular vs. bar-type outer device boundary as displayed in Figure Figure 2.1a and b, respectively, 2) changes to the internal shape of the boundary between the constituent materials, e.g. as in the transition from concentric circular device to the asymmetric circular device in Figure 2.1a and c, respectively, or 3) changes to the locations and number of the electrical contacts such as *IVVI* configuration displayed in Figure 2.1b to an *IVIV* configuration. The most common EMR geometries are the concentric circular device and the bar-type device. Optimizations have led to more exotic structures such as a multi-branched device which resembles a Hall-bar or a so-called "fish-bone" structure (refs) and asymmetric EMR devices (ref).

Three approaches have been used to probe the effect of varying the geometry of the device. In the first approach, the geometry effect has been investigated by experimentally varying the device geometry. Here, the considered devices are based on concentric circular geometries[1,6,7,9,24,27,43,45] or bar-shaped geometries.[15,20,46–49] In the second approach, geometries are described by parameters, which are varied theoretically in numerical[3,5,56–62,43,45,50–55] or analytical models[32,43] in order to evaluate the effect on magnetoresistive performance. Here, concentric circular devices[3,5,43,58,59] and bar-shaped devices[50–55,60–62] are also widely addressed. In the third approach, the shape of the shunt is changed into more complex multi-branched geometries in order to enhance the magnetoresistance response by several orders of magnitude.[63,64] The circular, bar, asymmetric and multi-branched device geometries are illustrated in Figure 2.1. In the following subsections we will introduce the various geometries which have been reported in literature and how geometric factors affect their performance.

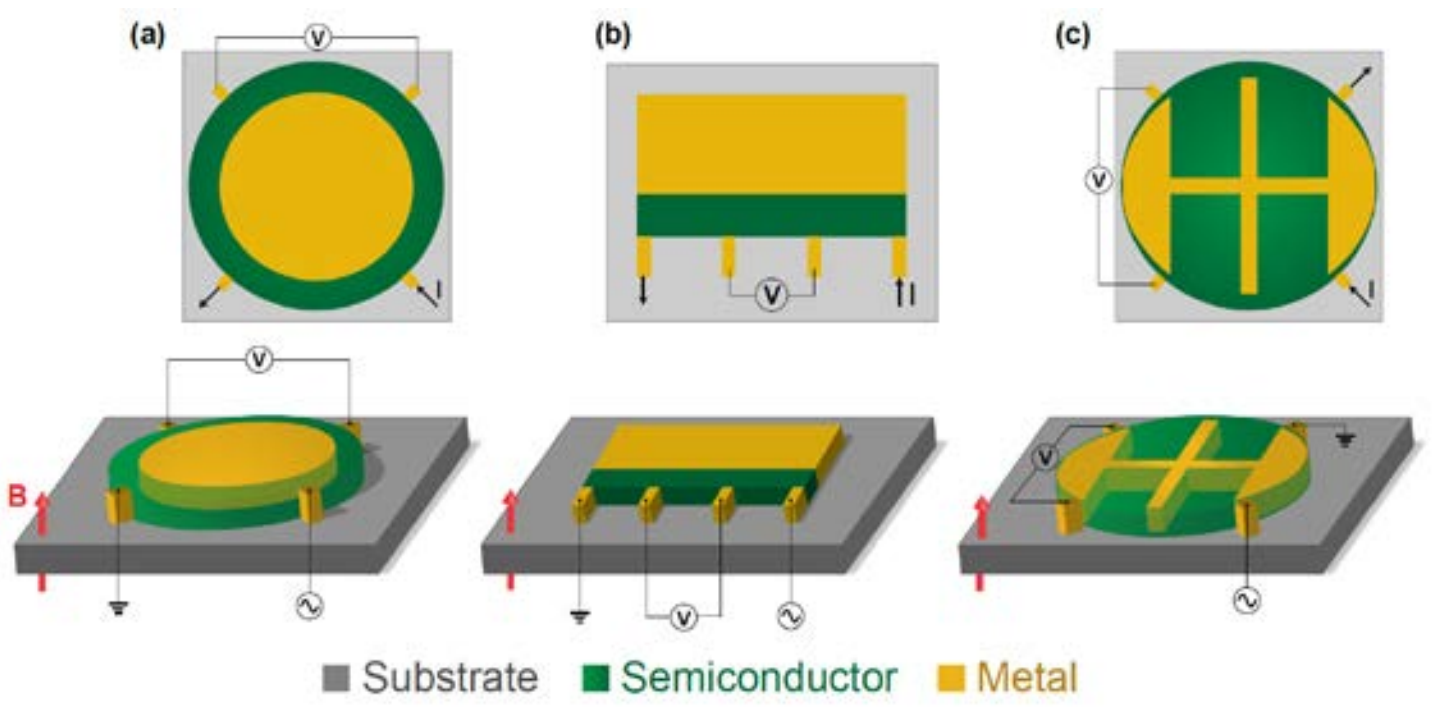

*Figure 2.1: Schematic illustrations of the four most common EMR device geometries: concentric circular device (a), bar-type device (b), asymmetric circular device and a branched device (c).*

## 2.1. Corbino Geometry

While the Corbino disc is generally not considered to be encompassed in the family of extraordinary magnetoresistive devices, it does exhibit a geometric magnetoresistance and merits some brief discussion as it was one of the first examples of geometric magnetoresistance. O. M. Corbino first proposed the Corbino disc geometry in 1911 in a series of papers which explored the nature of charge conduction in metals (refs). The choice of a disc was not novel, as precedence for this design existed in studies by Maggi, Hall, and Boltzmann regarding the case of conduction in metal discs (refs). However, Corbino claimed to have discovered previously unreported magnetoconductance phenomena through the use of this particular geometry (ref). In fact, Corbino's results were later discovered to be not an entirely new effect, but rather a clear and simple demonstration of the Hall effect.[65]

Similar to the EMR devices described by Solin[1], the Corbino disc is a round heterostructure with a metallic center embedded into another material. Unlike the classic EMR device where contacts are located in discrete areas on the outer boundary of the device, in the Corbino disc the entire perimeter forms one metallic contact and the inclusion at the center forms the other. As such it can only be operated in the two-terminal configuration. Current is injected through the inner contact and in the absence of a magnetic field it flows radially towards the perimeter (see Figure 2.2a). When a perpendicular magnetic field is applied, the electrons experience a deflection due to the Lorentz force and the current path spirals, with the degree of spirality determined by the strength of the field and the carrier mobility of the non-metallic material (Figure 2.2b-c). The spiral motion occurs because the geometry of the device creates a condition in which the Lorentz force is not counterbalanced by the Hall field formed by charge buildup at the device boundary and thus the current is allowed to freely deflect. An increase in the resistance of the device can be measured as a direct result of the lengthening of the current path.

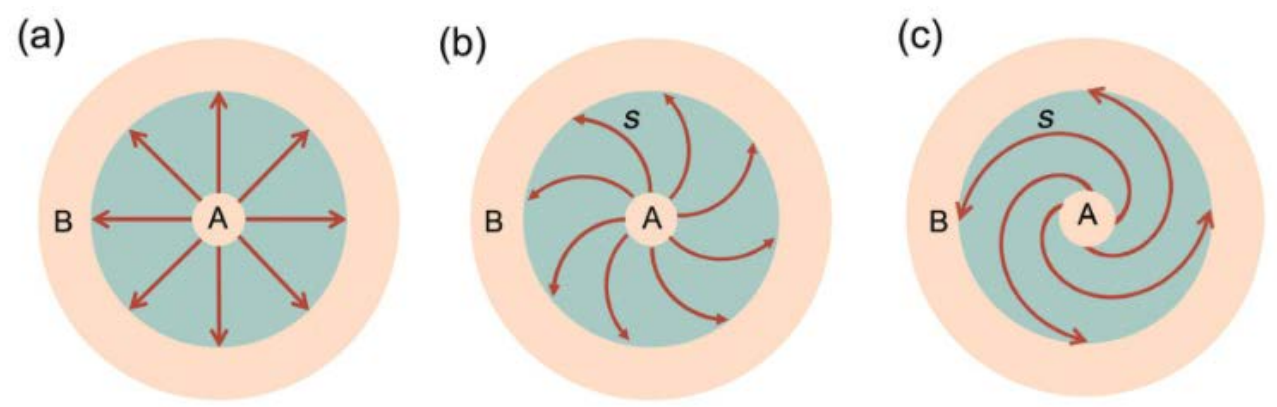

*Figure 2.2: Schematic illustrations of the current flow in a Corbino disc geometry at zero magnetic field (a), moderate magnetic fields (b) and high magnetic fields (c). Figure is adapted from Dong et al.[66].*

Due to its simple geometry, the only geometric parameters that the performance of a Corbino disc depends on are the thickness of the active material and the ratio of the outer and inner radii. Kleinman derived analytical expressions for the voltage drop and sensitivity of a Corbino disc and showed that the radii ratio is the dominant geometric factor, with larger values of the ratio yielding greater magnetoresistances (ref). The thickness of the active layer has some contribution to the value of the zero-field resistance.[67]

The effectiveness of the Corbino disc as a magnetoresistive device can be clearly seen in the work published by Branford et al.[4] To compare the effect of various geometric designs, devices of various shapes were made from InSb. A 1 µm thick film of InSb was grown on a semi-insulating GaAs substrate and then etched to form devices with van der Pauw and Corbino disc geometries.

A high intrinsic magnetoresistance in the InSb was observed in the data from the vdP samples which showed an increase in resistance of 885% at 8 T (see Figure 2.3). While relatively large, this value is much smaller than that of the Corbino disc where the resistance increased by 4,685%, demonstrating the strength of the geometric effect.

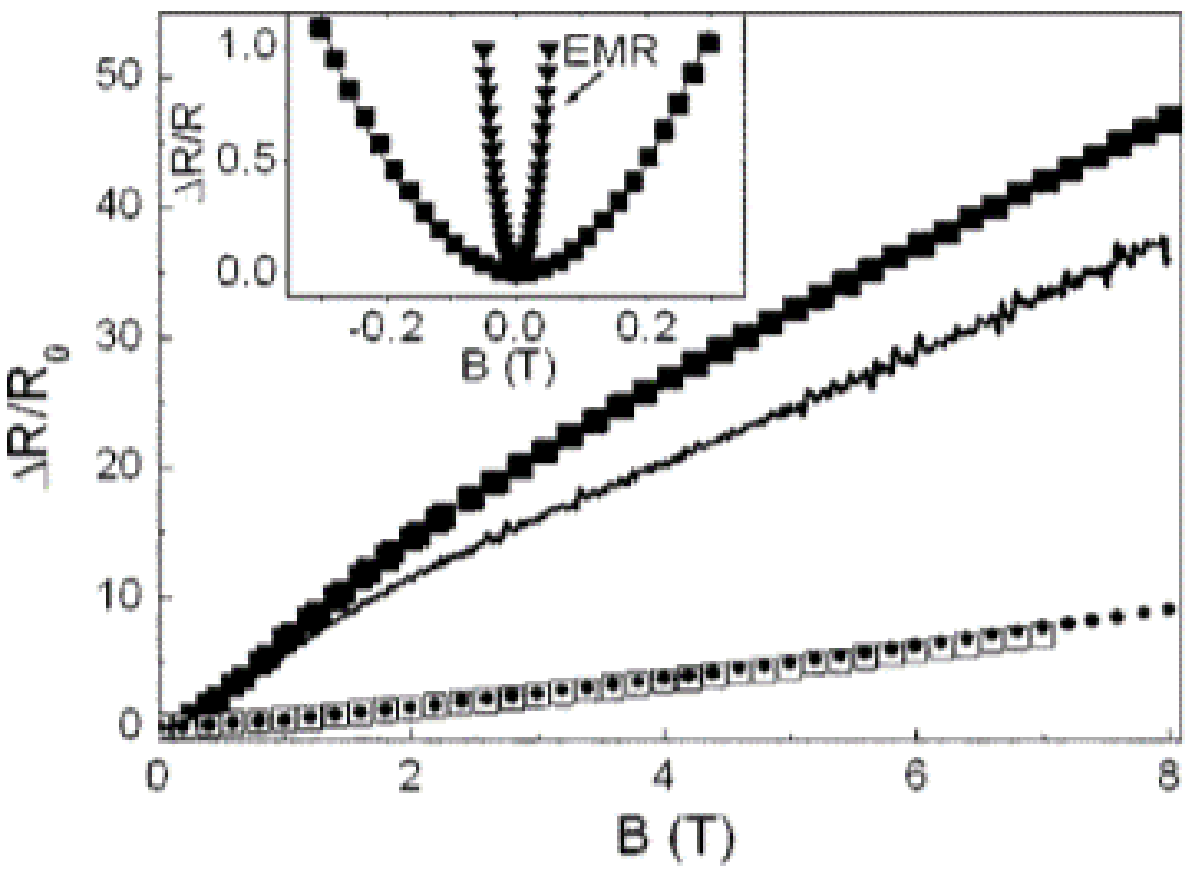

*Figure 2.3: MR(B) for the cloverleaf geometry (solid circles), a 1x1 array in vdP geometry (open squares), and Corbino disc geometry (solid squares). The solid line shows the predicted MR value using measured material parameters from the as-grown samples. Inset: Low-field MR shown for the Corbino disc (squares) and a concentric circular EMR device with a filling factor of 12/16 (triangles). Figure from Branford.*[4]

A sensitivity analysis of the Corbino samples revealed that there are two regimes: at low fields the sensitivity, dR/dB, increases as a function of field until it reaches a maximum value, and at high fields the sensitivity is constant.[4] Despite its ability to demonstrate geometric magnetoresistance, the Corbino disc is not often used in EMR literature due to its relatively low magnetoresistance compared to other designs.

## 2.2. Shunted van der Pauw Disc

The discovery of EMR in 2000[1] was based on the concentric circular geometry with magnetoresistances of more than $10^6$ % at magnetic fields of 5 T. The design consists of a semiconductor disc which is shunted by a highly conductive material (see Figure 1.1). The perimeter of the device features four equally spaced contacts which can be operated in a four terminal configuration in which current is sourced through adjacent contacts and the voltage difference is measured at the opposite pair, or as a two terminal measurement in which only one pair of adjacent contacts is used for both the current and voltage leads.

Devices made from shunted vdP discs are heavily affected by geometric variations in the shape and size of the conductive shunt. The effect of varying the filling factor defined as the ratio of the radii of the shunt and the outer boundary, $\alpha = r_i/r_o$, have been demonstrated in both experiments[1,6,7,9,24,27,43,45] and simulations.[5,32,58,59,63,64] The experimental results and simulations have both shown that increasing α decreases the overall resistance, but improves the magnetoresistance as the R(0) term decreases faster than R(B). This improvement continues up to a threshold value of α, after which the magnetoresistance for a fixed magnetic field drops as the current cannot effectively be deflected around the shunt (see Figure 2.4). The optimum value of α depends on the strength of the applied magnetic fields with larger fields yielding higher values of α. Here, the lack of effective current deflection for large values of α is compensated by a large magnetic field.

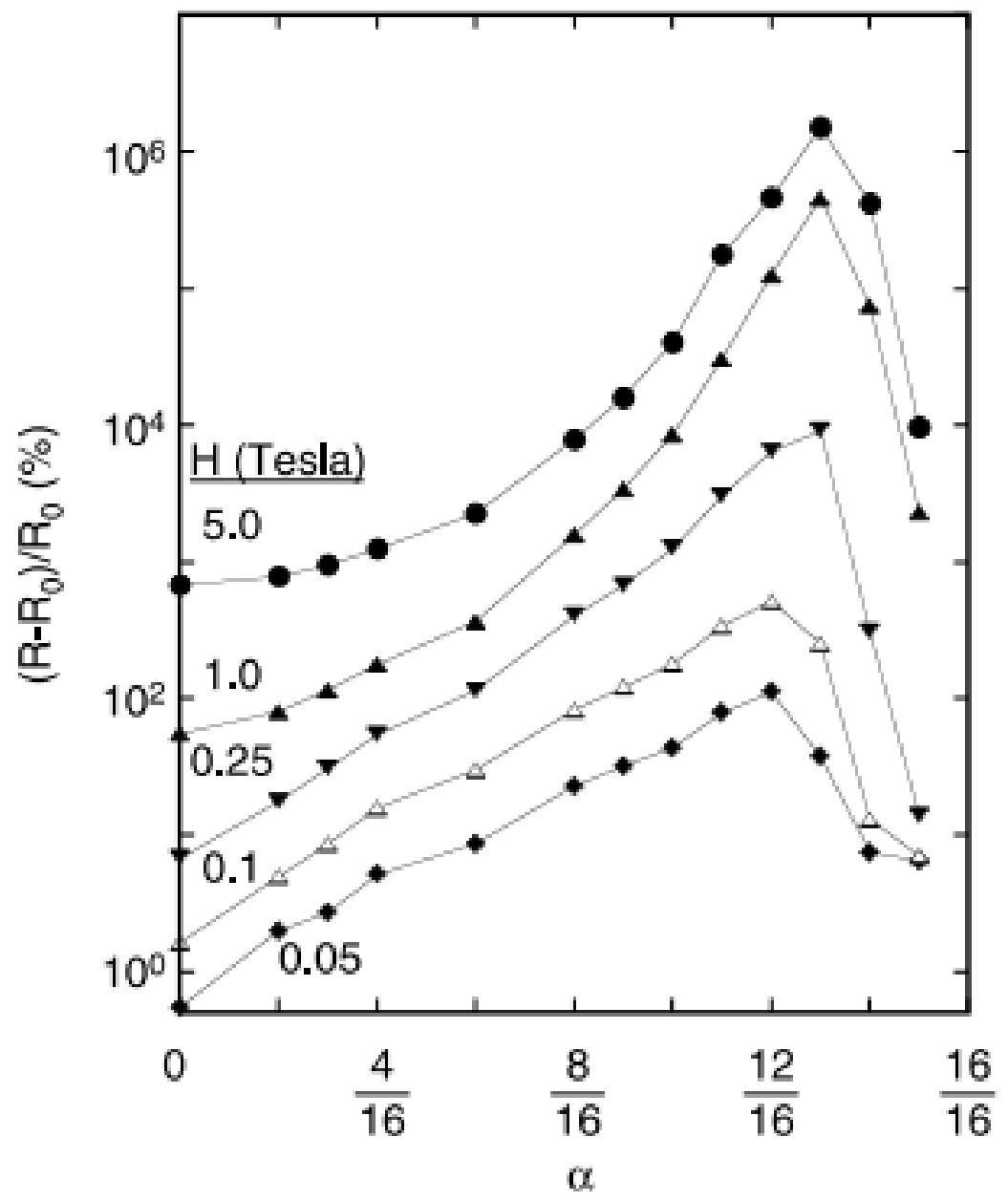

*Figure 2.4: Magnetoresistance as a function of filling factor α for shunted InSb vdP devices. Figure from Solin et al.*[1].

Sun et al. examined the effect of the filling factor on both magnetoresistance and sensitivity[6] and found that the sensitivity decreases as the size of the shunt is increased. The magnetoresistance followed the opposite trend, with high values of α generally producing a dramatically higher magnetoresistance.

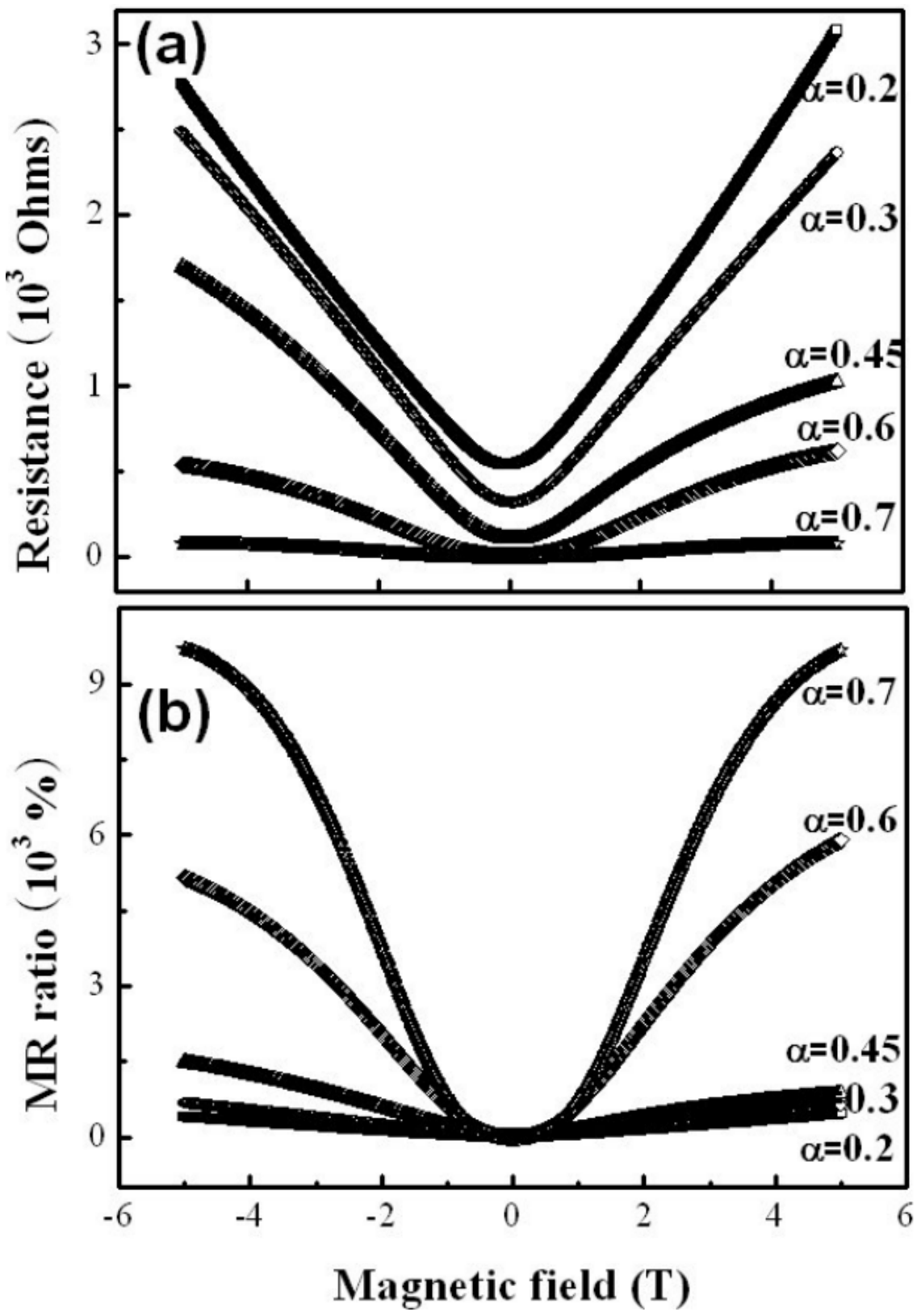

*Figure 2.5: a) Resistance and b) magnetoresistance as a function of magnetic field for shunted InSb vdP discs with differing filling ratios (α).*[6]

Figure 2.5 highlights the competing and often contradictory requirements of different figures of merit. Devices which feature a large shunt have a low zero-field resistance, driving up the magnetoresistance as the term in denominator becomes small. As such, small changes in the device resistance induced by the field can produce large magnetoresistances, but since the magnitude of the change is small the resulting sensitivity is poor. In contrast, if one optimizes the device for high sensitivity, the ideal shunt diameter is much smaller, but produces a larger zero-field resistance and hence a reduced magnetoresistance.

The circular geometry has not received as much interest as some other designs due to difficulties in fabrication. Misalignment of the lithography masks can result in deviations in the concentricity of the two materials which can affect performance.[13] Nonetheless, the geometry shows great flexibility and can yield good magnetoresistive results. In 2013, Sun et al. published an experimental study comparing the circular, bar and Corbino disc geometry in n-doped InAs epilayers.[13] The authors reported that the 4-terminal configurations of the shunted vdP geometry reach higher magnetoresistances, however the 2-terminal bar-shaped device provided a magnetic field resolution of ~12 $nT/\sqrt{Hz}$ which is 15 and 75 times better than the Corbino disc and the 4-terminal bar-shaped counterpart, respectively. Branford also compared these values to the results obtained by Solin et al.[1] on an InSb EMR device with a 12/16 filling factor. It can be seen in Figure 2.3 that in the low field regime the vdP geometry produces a much stronger response than that of the Corbino disc; with the former yielding a 100% increase in resistance compared to a meager 4.5% for the latter under a field of 0.05 T.[4] Overall it can be said that for applications requiring high magnetoresistances, large shunt diameters and a four terminal configuration are preferable whereas if a high sensitivity at finite magnetic fields is sought after, smaller shunt diameters and a two-terminal setup should generally be employed.

## 2.3. Off-centered shunted vdP disc

Variations to the concentric circular vdP disc was considered in a couple of studies (ref).

The most comprehensive study was done using finite elements by Erlandsen et al. who shifted the inner metal disc systematically either vertically or horizontally as displayed in Figure 2.5a and d. A horizontal shift was found to dramatically alter the magnetic field dependence of the resistance, which transitioned from symmetric trace with *R(B)* = *R(-B)* when the metal disc was centered in a device to a highly asymmetric curve with *R(B)* ≠ *R(-B)* upon displacement (Figure 2.5b). In contrast, when the metal inclusion was shifted vertically, the symmetry of R(B) was kept intact and only small variations were found (Figure 2.5e). The authors deduced that breaking the mirror symmetry along the vertical axes was essential for breaking the symmetry of the field-dependent resistance. This symmetry breaking was declared a prerequisite for producing EMR sensors with magnetic field sensitivity at very weak magnetic fields as it entails that $dR/dB \neq 0$. Furthermore, it enabled the detection of both the direction (positive or negative perpendicular fields) and magnitude of the magnetic field. The sensitivity was found to increase as the displacements and the filling factor were enlarged simultaneous with the largest value of approximately 70 Ω/T found when metal disc is as big as the semiconductor disc, but displaced such that it only overlaps with the right-hand part of the semiconductor.

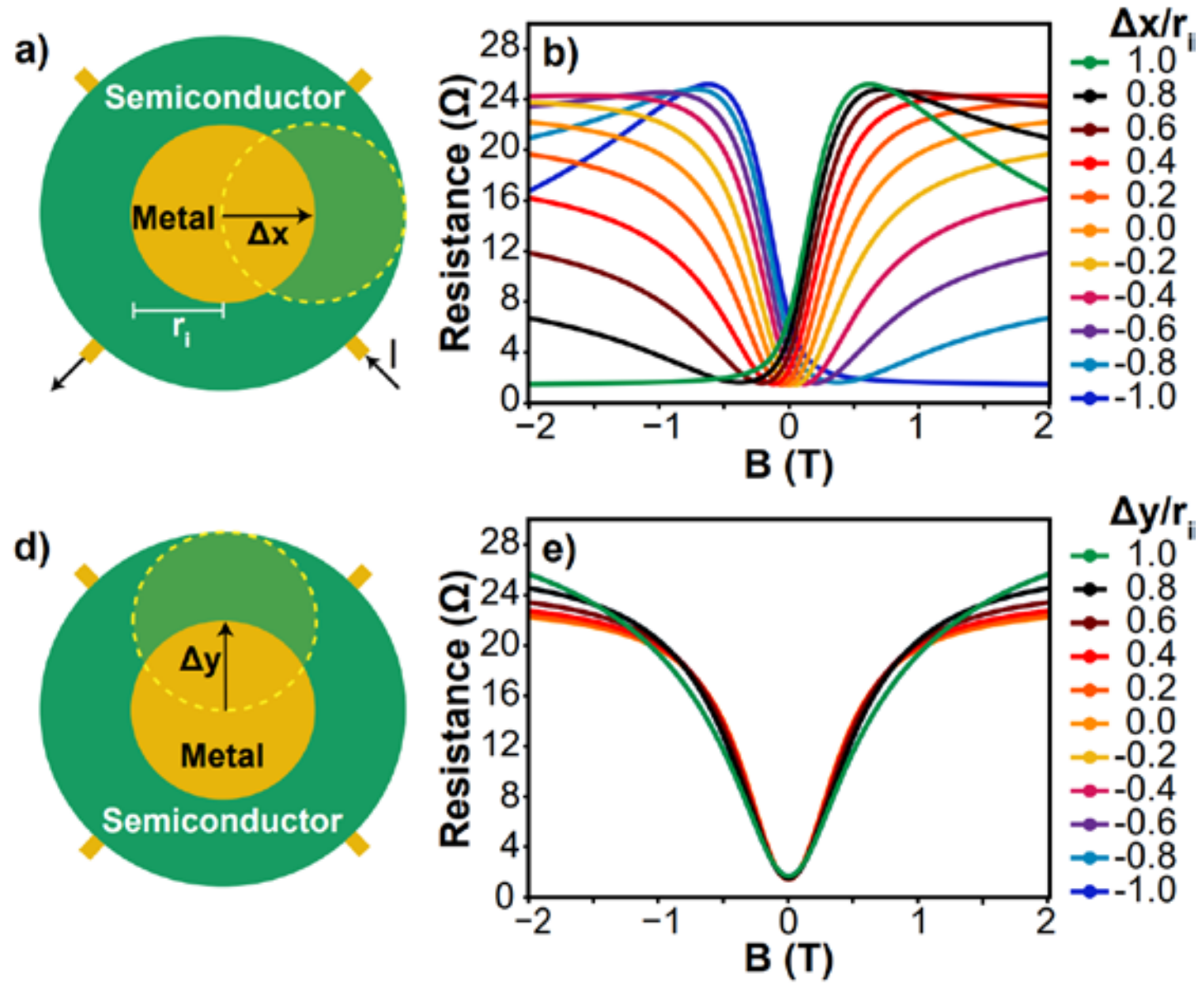

*Figure 2.6: Schematic figures of displacing the gold inclusion horizontally (a) and vertically (c) in a InSb/gold EMR device as well as the effect on the magnetic field-dependent resistance shown in (b) and (d), respectively. (ref)*

Erlandsen et al. further showed that the symmetry could also be broken by dividing the semiconductor into two regions with different electron mobilities. By examining the current distribution in both the symmetric and asymmetric devices, the authors were able to visualize the change in the current distributions when exposed to a magnetic field (Figure 2.5). As expected, a large part of the current is predicted to flow through the metal shunt in absence of a magnetic field. In presence of a perpendicular magnetic field, the current is deflected around the metal inclusion. In the symmetric device, the current is deflected in a symmetric manner when comparing positive and magnetic fields. However, for both the asymmetric devices, the current flow was found to differ significantly. In particular, when applying *B* = -1 T to the device with two semiconductor regions (lower row in Figure 2.5) the current is deflected around the gold inclusion in the semiconductor region with high mobility, but is readily directed into the gold as the current flows into the semiconductor region with low mobility.

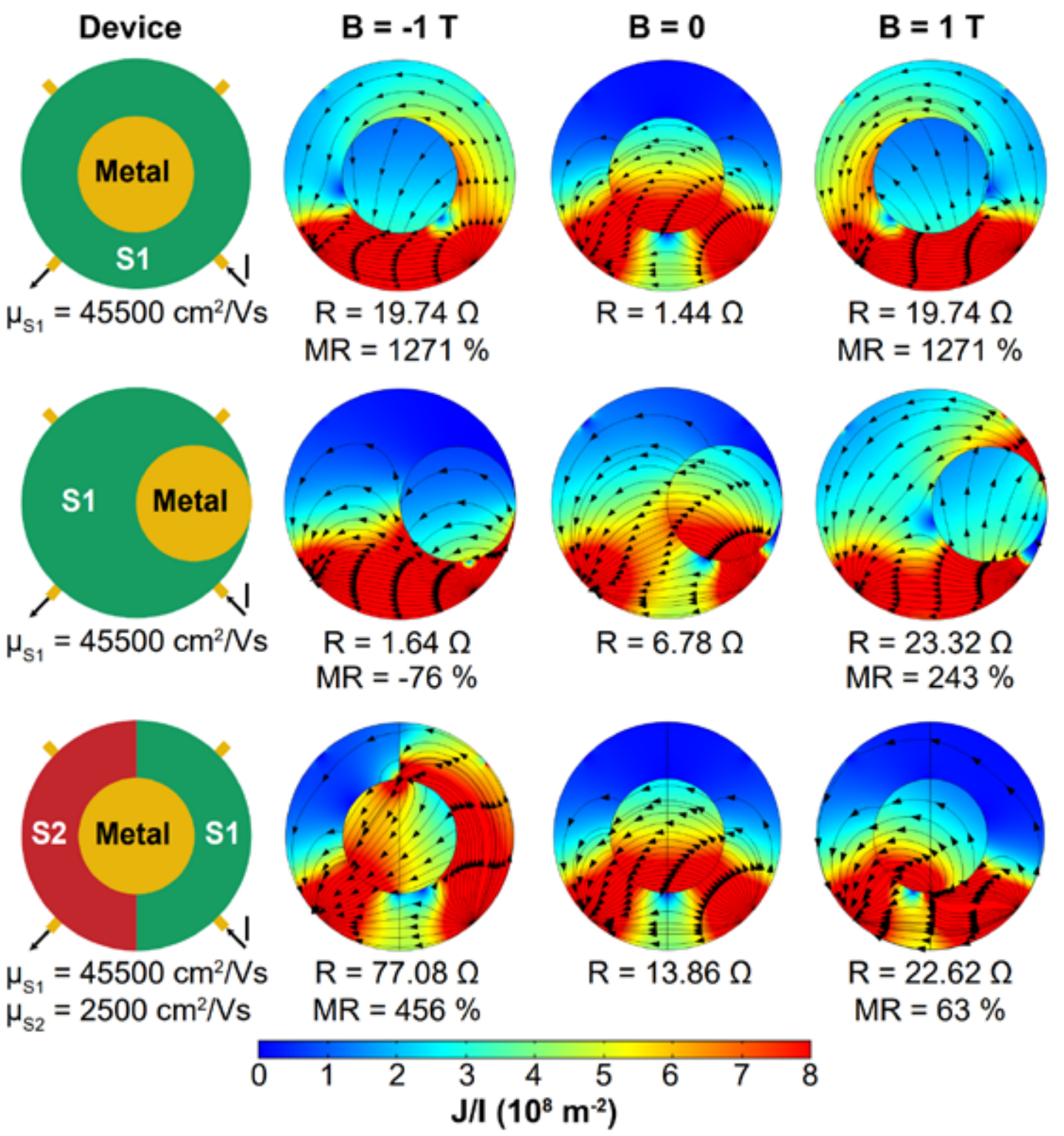

*Figure 2.7: Current distributions of a symmetric EMR device composed of InSb and a concentric gold shunt (top row) as well as two asymmetric device geometries (middle and bottom row). (Ref)*

Solin et al. also produced a device based on the vdP disc but with an off-center shunt and asymmetric lead configuration.[2] The reported magnetoresistance of this design was more than an order of magnitude greater than that of the concentric device with the same sized inhomogeneity.

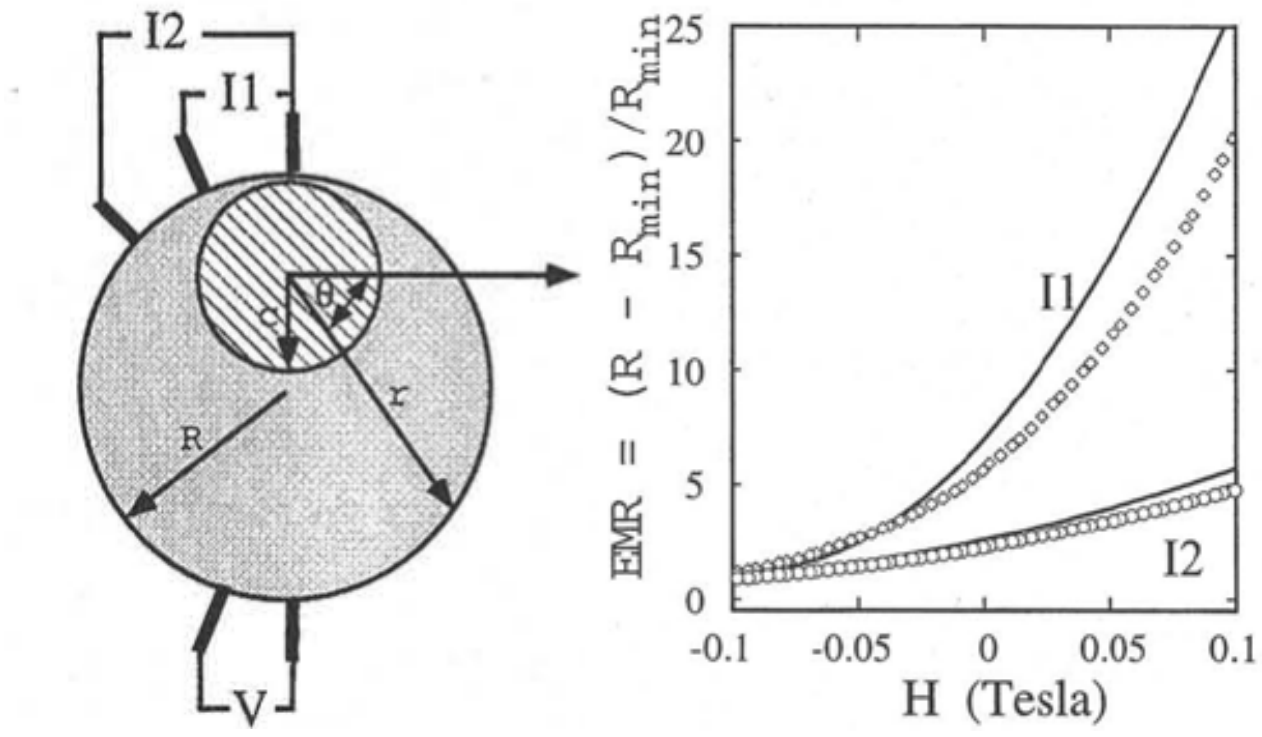

*Figure 2.8: Off-center vdP disc with asymmetric current and voltage leads (left). Calculated (solid lines) and experimental (symbols) results of the magnetoresistance as a function of field for the two different electrode configurations.*[2]

The locations of the voltage and current contacts can greatly affect the performance of the device, as can be seen in Figure 2.15 where a relatively small shift in the location of the current leads results in a 5-fold increase in the magnetoresistance at 0.1 T. The position of the shunt and the leads provides yet more degrees of freedom for the design of EMR devices.

## 2.4. Bar Geometry

The most common EMR device shape is the bar-shaped geometry due to its simple design, simpler fabrication and compatibility with nanoscale sensor fabrication.[8] Solin et al.[44,68] showed how the bar-type geometry can be derived from a circular geometry by using conformal mapping[69] as illustrated in Figure 2.6.

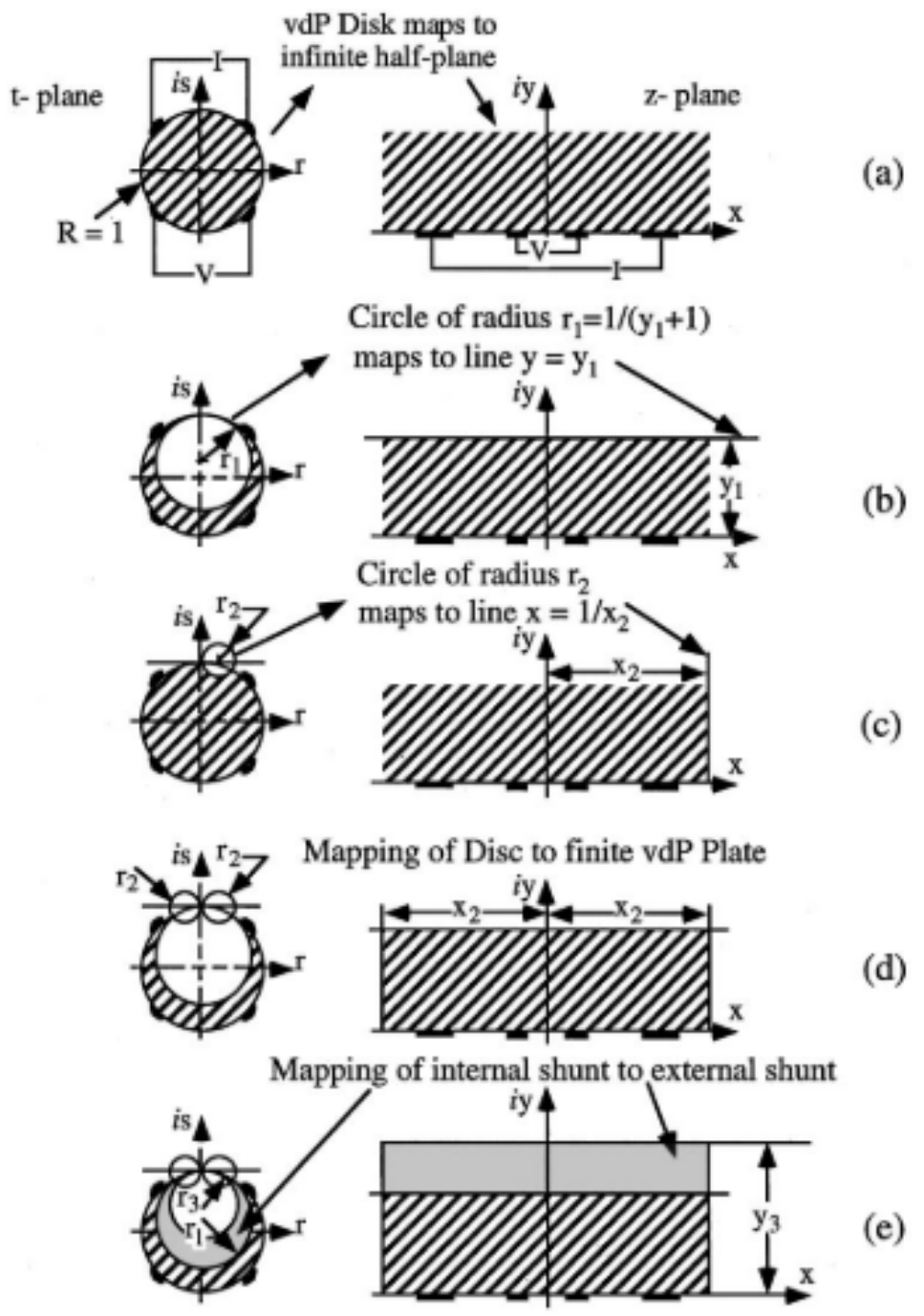

*Figure 2.9: Illustration showing how conformal mapping is used to derive the bar-type geometry from the concentric circular geometry. Figure is adapted from Zhou et al.*[44].

In the case of a bar-shaped device with symmetric leads, an equivalent filling factor can be calculated with the expression:

$$\alpha = \left[\frac{1}{(a \cdot \tan(135°/2)\, y_1 + 1)^2} - \frac{1}{(a \cdot \tan(135°/2)\, y_2 + 1)^2}\right]^{1/2}$$

where $a$ is the horizontal distance from the axis of symmetry to the furthest current lead, $y_1$ is width of the semiconductor, and $y_2$ is the vertical distance from the leads to the top of the shunt.[70] The factors that affect the performance of bar-shaped devices have been investigated in various experimental[15,20,29,47,49] and numerical studies.[12,54,55,60,64,71] Sun et al.[71] showed that the performance of the bar device is highly susceptible to changes in geometry e.g., the location of the contacts, the width of the semiconductor, and the width of both semiconductor and metal. In particular, the effect of length-to-width ratios of the semiconducting region in the bar-shaped device has been widely explored in literature with experiments[15,20,29,47,49] and simulations.[12,54,55,60,71] This parameter is the geometric equivalent of the shunt diameter in the shunted vdP geometry.For instance, Möller et al. demonstrated how magnetoresistance in bar-shaped devices increases exponentially with decreasing semiconductor width.[15] MR values of 93%, 1,900%, and 115,000% were measured at 1 T for devices with length-to-width ratios of 2.8, 10, and 28, respectively. The increase in magnetoresistance is due to the decrease in the zero-field resistance that occurs as the width is reduced.[46] A maximum value in the sensitivity of XX for B = YT was observed for the device with the thinnest semiconductor. Sun et al. further elaborated on the effect of the length-to-width ratio and suggested that the optimal ratio depends on the strength of the field one is trying to measure. They reported that for high magnetic fields, length-to-width ratios of 10 to 20 gave the best results, while a ratio of 5 was optimal for low magnetic fields.[71] Similar results have been reported in the other cited papers.

Huang et al. relaxed the assumption that the cross-section of the semiconductor should be uniform and presented a chevron-shaped design in which the width of the semiconductor decreases as it approaches the axis of symmetry of the device (see Figure 2.7). This creates a constriction in the center which increases the resistance and tripled the MR at 3 T.[64]

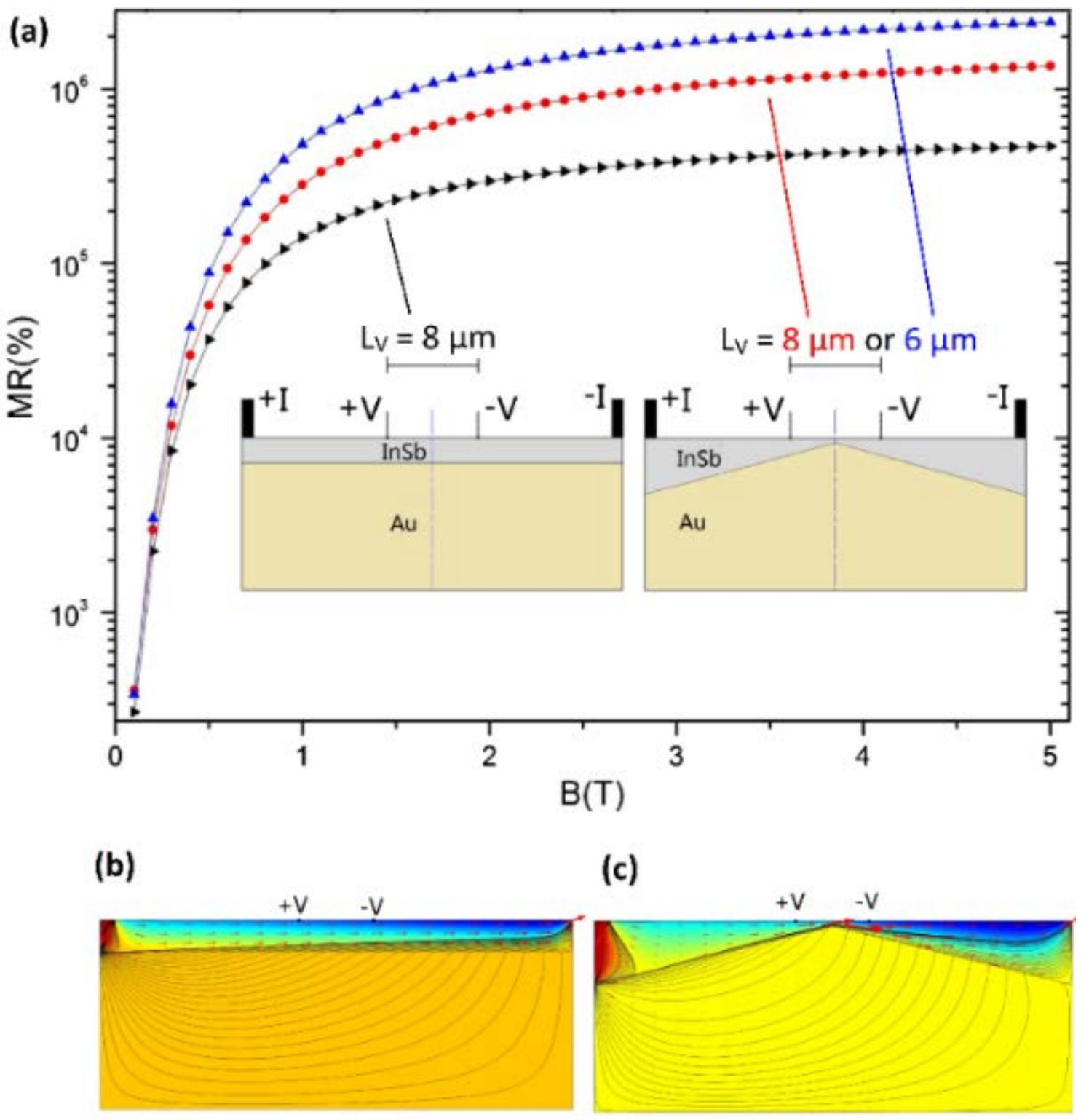

*Figure 2.10: Magnetoresistance response of the bar-type devices with varying probe distance and a modified shunt design. (a), magnetoresistance for conventional bar-type devices with a voltage probe distance of 8 µm (black). The modified bar-type is shown with a probe spacing of 8 µm (red) and 6 µm (blue). (b) and (c) shows the current flow in the conventional and modified bar-type design for B = XT, respectively. The figure is adapted from Huang et al.*[64]*.*

Sun and Kosel[12] investigated bar-shaped devices with metal shunts in various shapes with the use of finite element simulations. Instead of treating the device as if it was strictly 2D, they explored electrodes that were thinner or thicker than the semiconductor (See Figure 2.9) or were lying partly on top of the semiconductor (see Figure 2.10) as is often the case in experimental devices. In the case of devices without an overlap, the magnetoresistance and sensitivity were both found to be constant if the metal thickness was larger than that of the semiconductor ($t_m > t_s$ in Figure 2.9), and decreased only slightly for the case where the thickness of the metal was one tenth that of the semiconductor. In the case where a significant overlap existed between the metal and semiconductor, the output sensitivity is slightly decreased, however, it comes with a simplified fabrication process. Thus, the small decrease in performance could likely be offset by the potentially improved electrical contact and easier fabrication. In addition, these results suggest that bar-type devices are not particularly sensitive to the types of physical deviations from idealized geometries that arise during the fabrication of experimental devices.

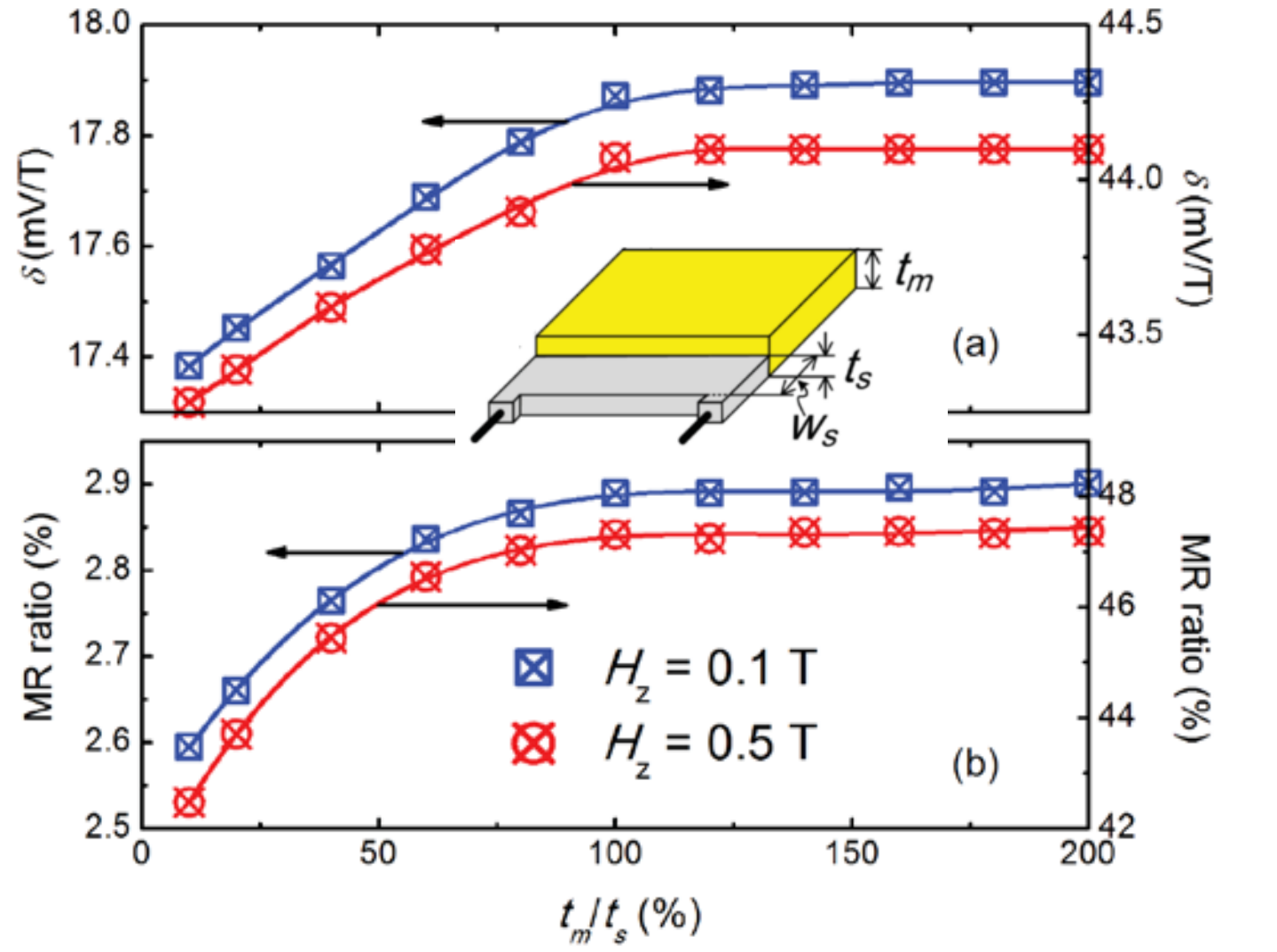

*Figure 2.11: (a) Output sensitivity and (b) magnetoresistance as a function of $t_m/t_s$, which is the ratio between thicknesses of metal and semiconductor, respectively. Data is shown for magnetic fields of 0.1 T (blue) and 0.5 T (red). Here, the material properties correspond to InSb interfaced with gold with a device length of XX, semiconductor width of Y and metal width of Z. The figure is adapted from Sun et al.*[12]

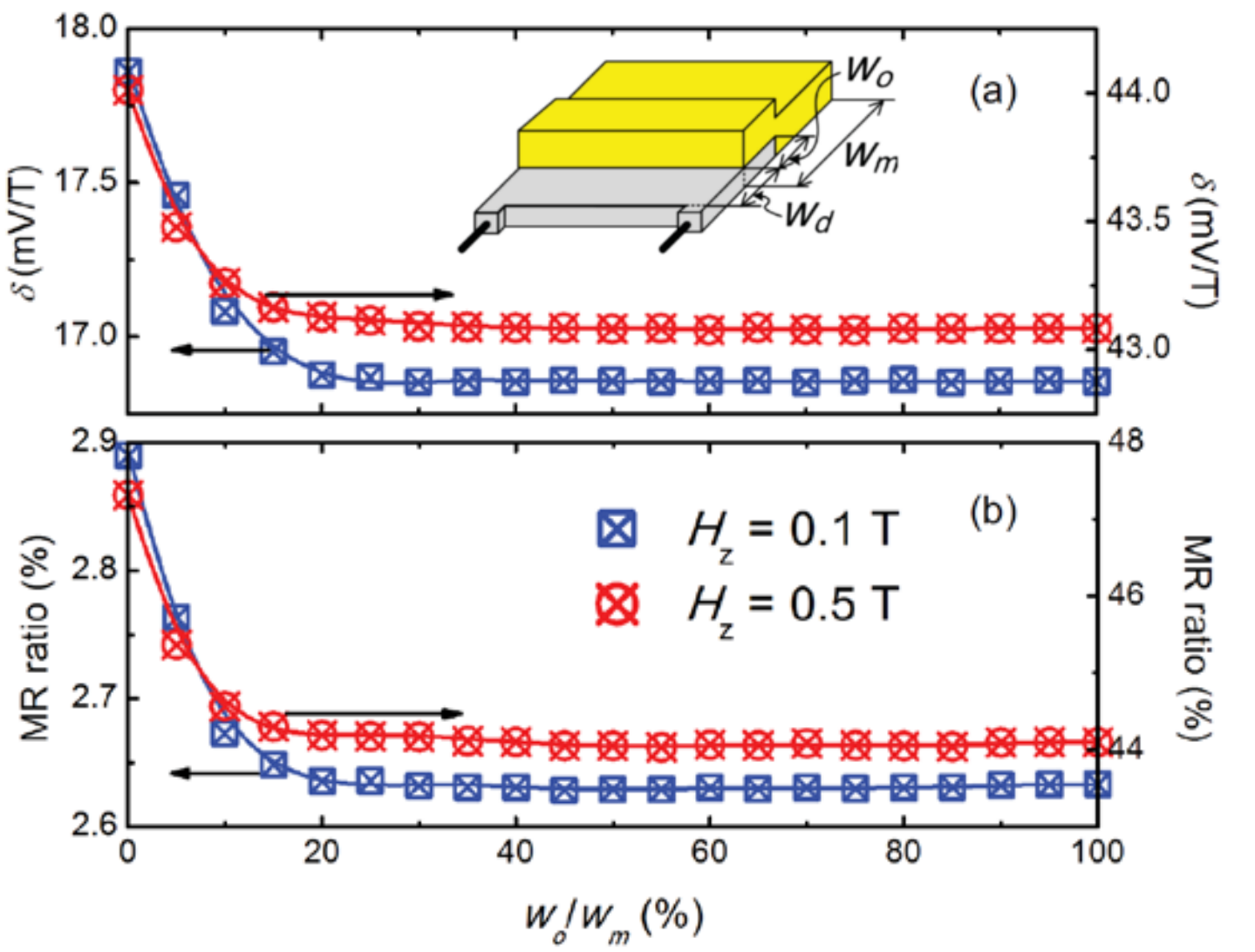

*Figure 2.12: (a) Output sensitivity and (b) magnetoresistance as a function of $W_o/W_m$, which is the ratio of the overlap and metal width, respectively. Data is shown for perpendicular magnetic fields of 0.1 T (blue) and 0.5 T (red). Here, the material properties correspond to InSb interfaced with gold with a device length of XX and a semiconductor width of Y. Figure is adapted from Sun et al.*[12]*.*

Oszwaldawski et al.[8] also published an experimental procedure to produce bar-shaped devices which a simplified fabrication process, however, using metal contacts and shunt deposited on top of the semiconductor (see Figure 2.11). This configuration was reported to yield an even better magnetoresistance compared with its conventional counterpart due to the very low resistance at zero magnetic field, which is mostly attributed to the low metal-semiconductor contact resistance arising from the large contact area. They claim that the new configuration can be utilized to upscale the production of EMR sensors that have historically been difficult to fabricate. The new configuration also provides an accessible platform for investigating EMR in promising thin film and 2D materials such as graphene.

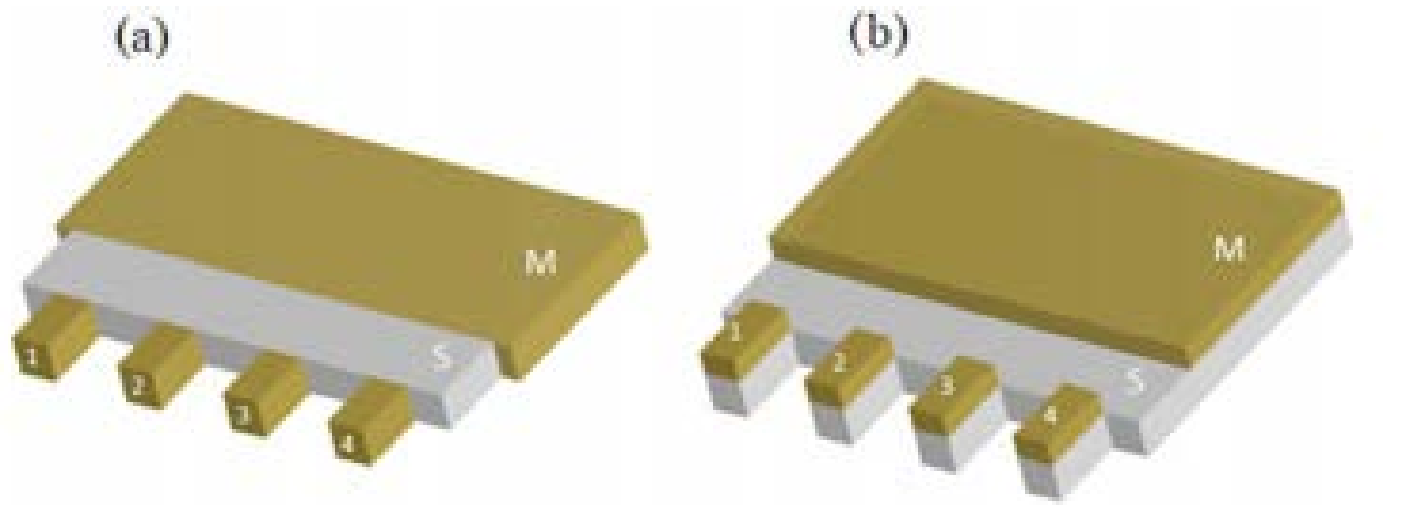

*Figure 2.13: Illustration of how the conventional bar-type (a) can be produced using a new configuration where metal is deposited on top of the semiconductor (b), which simplifies the fabrication process. Figure from Oszwaldawski et al.*[8]

In order to estimate how the devices would perform at higher fields the devices were simulated using FEA software.[26] Experimental results matched the output from the simulations for the range of magnetic fields tested, allowing the authors to simulate the performance at higher fields. The simulations predicted a magnetoresistance of 55,000% at 9 T, similar to what was reported by Lu et al. with a shunted vdP geometry[24], which lends credence to the proposed planar geometry.

## 2.5. Multi-branched Structures

As Huang showed in the case of the chevron shaped shunt in the bar-shape device (see Figure 2.7), there are no requirements around the uniformity of the shunt geometry. Several groups have tried to optimize EMR devices by changing the shape of the metallic shunt (refs). One common type of alternative shunt geometry is what we refer to as multi-branched structures. The multi-branched structure is based off of the shunted vdP disc, but the shape of the shunt resembles a Hall bar with multiple arms.

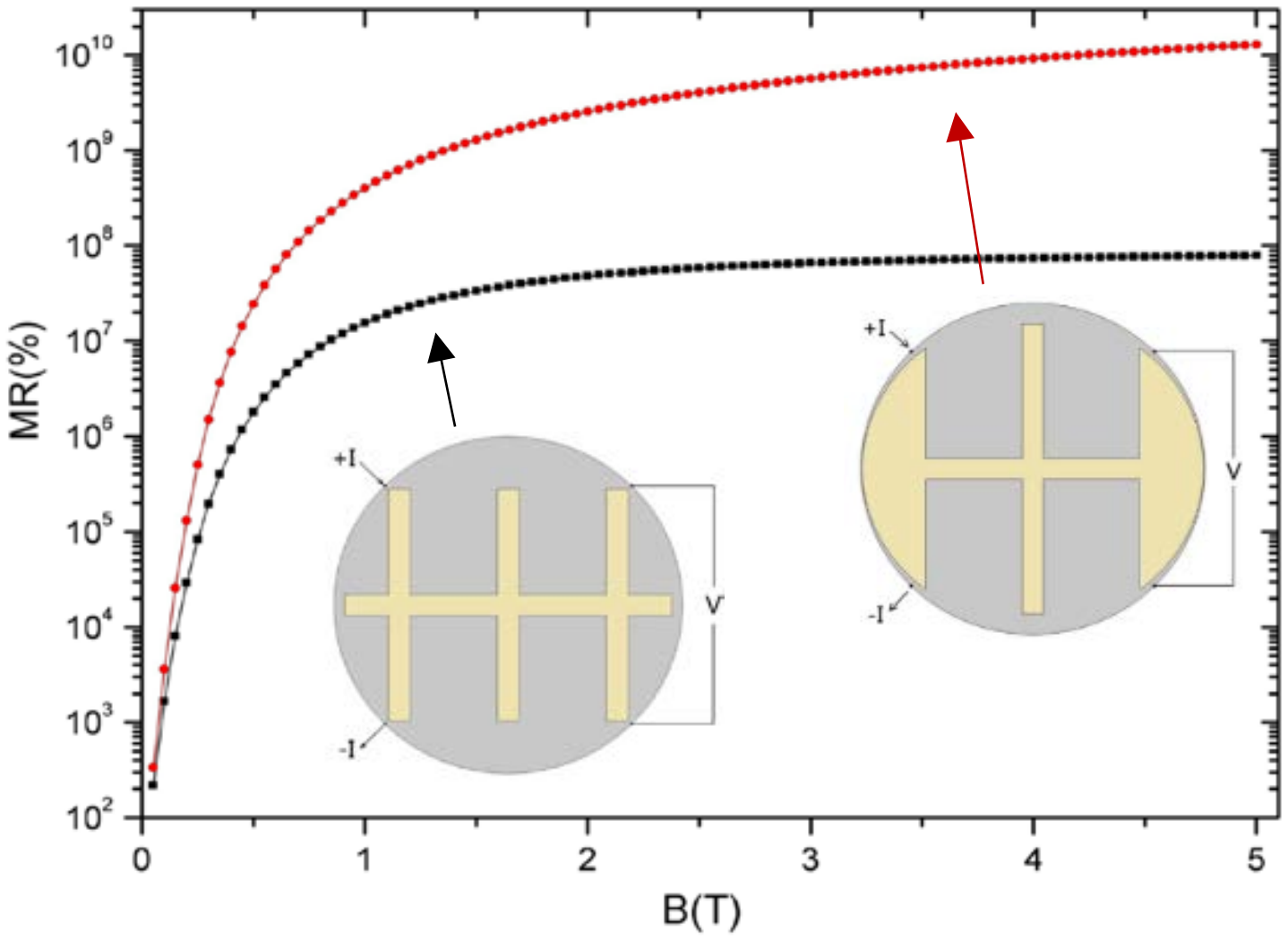

*Figure 2.14: The magnetoresistance as a function of magnetic field for the multi-branched device geometry (black) and the combined branched and elliptic device (red). Here, the material properties correspond to InSb interfaced with gold. The figure is adapted from Huang et al.*[64]

Branched structures were originally proposed by Hewett et al. who showed using FEM simulations that the multi-branched structure similar to device shown in black in Figure 2.12 performed significantly better than the conventional shunted vdP disc.[63] This study was later expanded by Huang et al.[64] that numerically showed that by combining the branched structure with the outer parts of an ellipse, an incredible enhancement of the MR by several orders of magnitude could be achieved (see Figure 2.12). The rounded outer parts from the elliptic geometry are crucial to this structure, since they keep the resistance at zero-field extremely low given that almost the entire current pathway at zero-field is covered in metal. At the same time, the voltage drop at finite magnetic fields is large since the current deflects away from the metal and into tight constrictions between the voltage probes in the semiconductor, leading to even larger resistance values compared to the other multi-branched structure. Hewett et al.[63] additionally demonstrated that structures with smaller filling factors experience the largest gains when converted into multi-branched geometries.

Moktadir and Mizuta[59] proposed a multi-branched device with additional arms which they called a fish-bone structure. The performance of the device composed of a graphene/metal hybrid was studied numerically and yielded large magnetoresistances exceeding $10^8$ % (see Figure 2.13). This devices geometry was further used to study the effect of an inhomogeneous graphene conductor composed of p-type and n-type puddles with equal mobility and charge carrier density, but with different sign of the charge. By varying the area fraction of n-type puddles ($f_n$), the authors concluded that the largest magnetoresistance was obtained when the system approach a homogeneous conductor composed either solely of electrons or holes (see Figure 2.13).

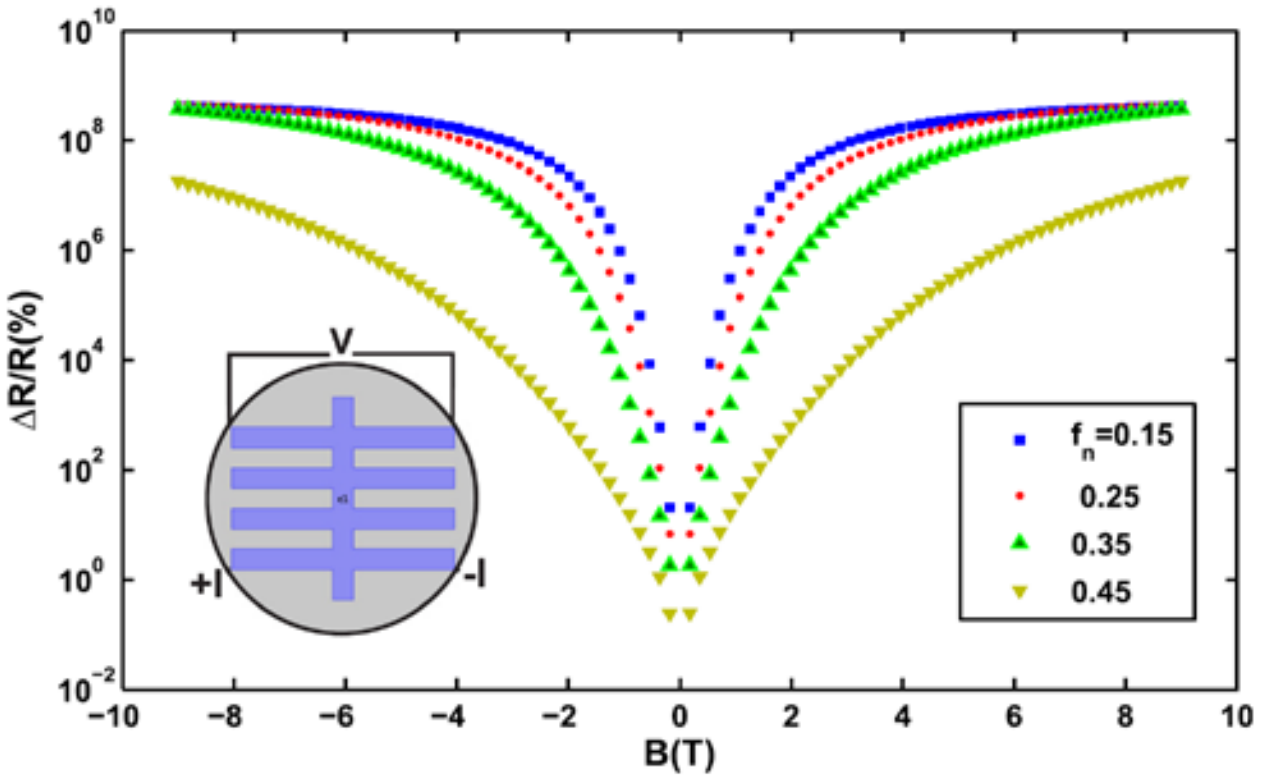

*Figure 2.15: Magnetoresistance as a function of magnetic field for varying fractions of n-type puddles ($f_n$) in the multi-branched device composed of graphene. Figure is adapted from Moktadir et al.*[59]

While experimental results of multi-branched devices have not been reported in literature, their incredible performance in numerical studies suggests that the geometries which have been explored to date are far from ideal and that massive improvements may be achieved through geometric optimization.

## 2.6. Other Geometries

A variation of the bar-shaped device geometry was investigated by Pugsley et al.[72] who considered a square version of the device; i.e. a square metal inclusion inside a square modeled with finite element simulations. The authors concluded that at 1 T the optimal ratio between the side lengths of the inclusion to that of the semiconductor is 8/10, a value close to that of the optimal filling factor of 13/16 for the shunted vdP geometry. The authors also showed that by separating the square inclusion into two rectangular regions, the magnetoresistance at low fields (0.04 T) can be increased by two orders of magnitude, without affecting the magnetoresistance at high fields (see Figure 2.14).[72]The 3D version of the system in which a cubic metal inclusion is embedded inside of a cubic semiconductor was also calculated, with the results showing magnetoresistances on the same order of magnitude as that of the corresponding 2D system.

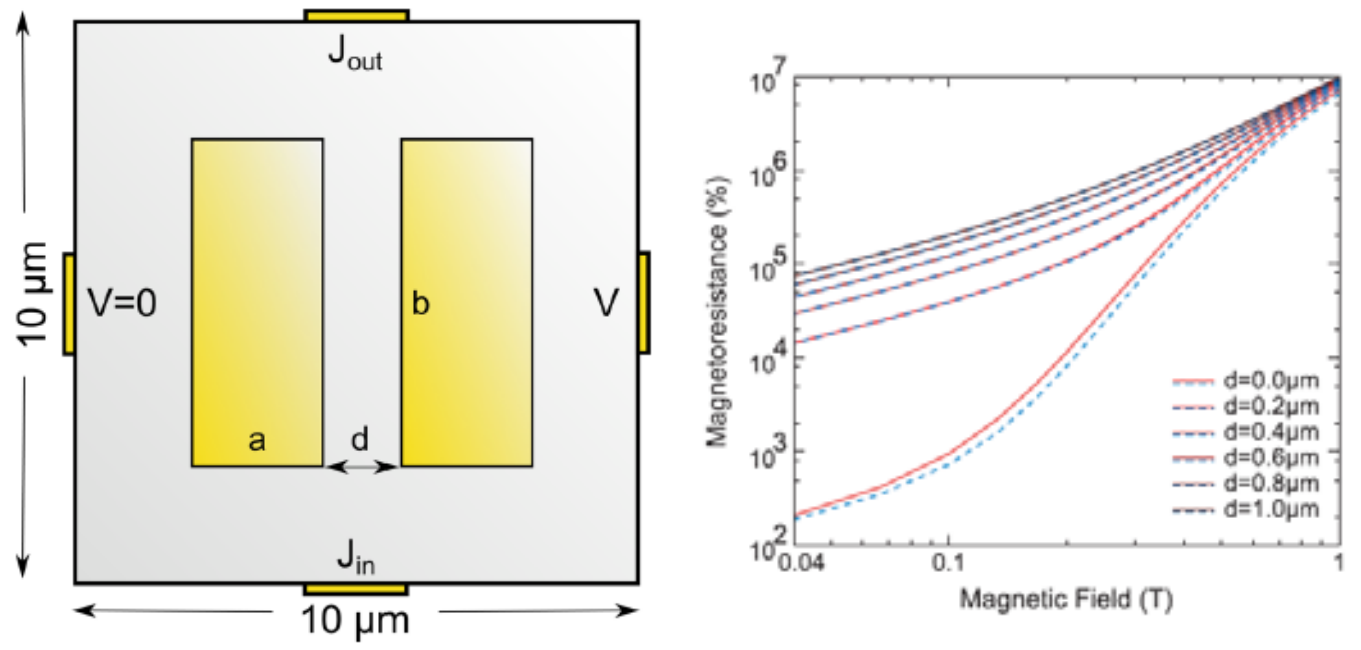

*Figure 2.16: Schematic illustration of the square device (left) with corresponding plot showing MR(B) for varying separation, d, between the two metallic inclusions (right), where the dashed lines represent the MR values for negative magnetic fields. Figure is adapted from Pugsley et al.*[72]

Hong et al. took a different approach to generate an EMR response than what is typically done with the metal/semiconductor hybrid devices that have been discussed previously.[14] Rather than creating a device with inhomogenous material properties in order to control the current flow in a magnetic field, an inhomogenous magnetic field was used to control the current path in a homogenous material. The device was produced by creating a two dimensional electron gas in an InAs quantum well with a GaSb cap layer. The deposited layers were then etched to form a vdP square with contacts on each of the four corners. Ferromagnetic gates were then made by depositing either Fe or Co layers in a strip over the center of the devices (see Figure 2.16a).

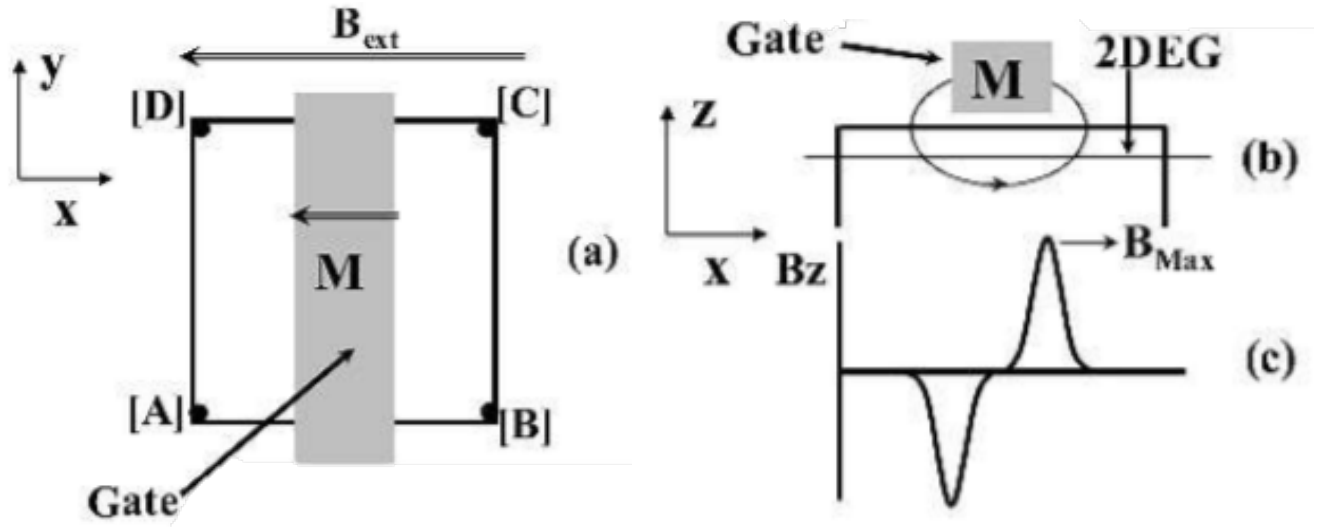

*Figure 2.17: Schematic of the device and top gate (a). Distribution of fringe fields(c) when the device is viewed from the side (b).*[14]

The operating principle of the sensor is that the ferromagnetic gate is capable of magnetizing in the presence of an external field, and in doing so, it generates fringe fields at the edges. The fringe fields at opposite ends of the gate are oriented in opposite directions and act as magnetic barriers (see Figure 2.16b and c). This change in the local magnetic landscape causes the electrons to move preferentially along the channels formed by the peaks in the local field since the Hall angle in the region near the edges is close to 90°, thereby lengthening the conduction path in the presence of an external magnetic field and increasing the measured resistance.

Two different configurations were used during the testing of the device. In the diagonal configuration, the A and C terminals functioned as the current source and drain and the potential was measured at the B and D terminals. In the side configuration, A and D were used as the source and drain and the potential was measured at B and C. Electron mobility and carrier density were determined to be 193,800 cm$^2$V$^{-1}$s$^{-1}$ and 9.46×10$^{11}$ cm$^{-2}$ at 4.2 K and 31,000 cm$^2$V$^{-1}$s$^{-1}$ and 2.28×10$^{12}$ cm$^{-2}$ at room temperature.

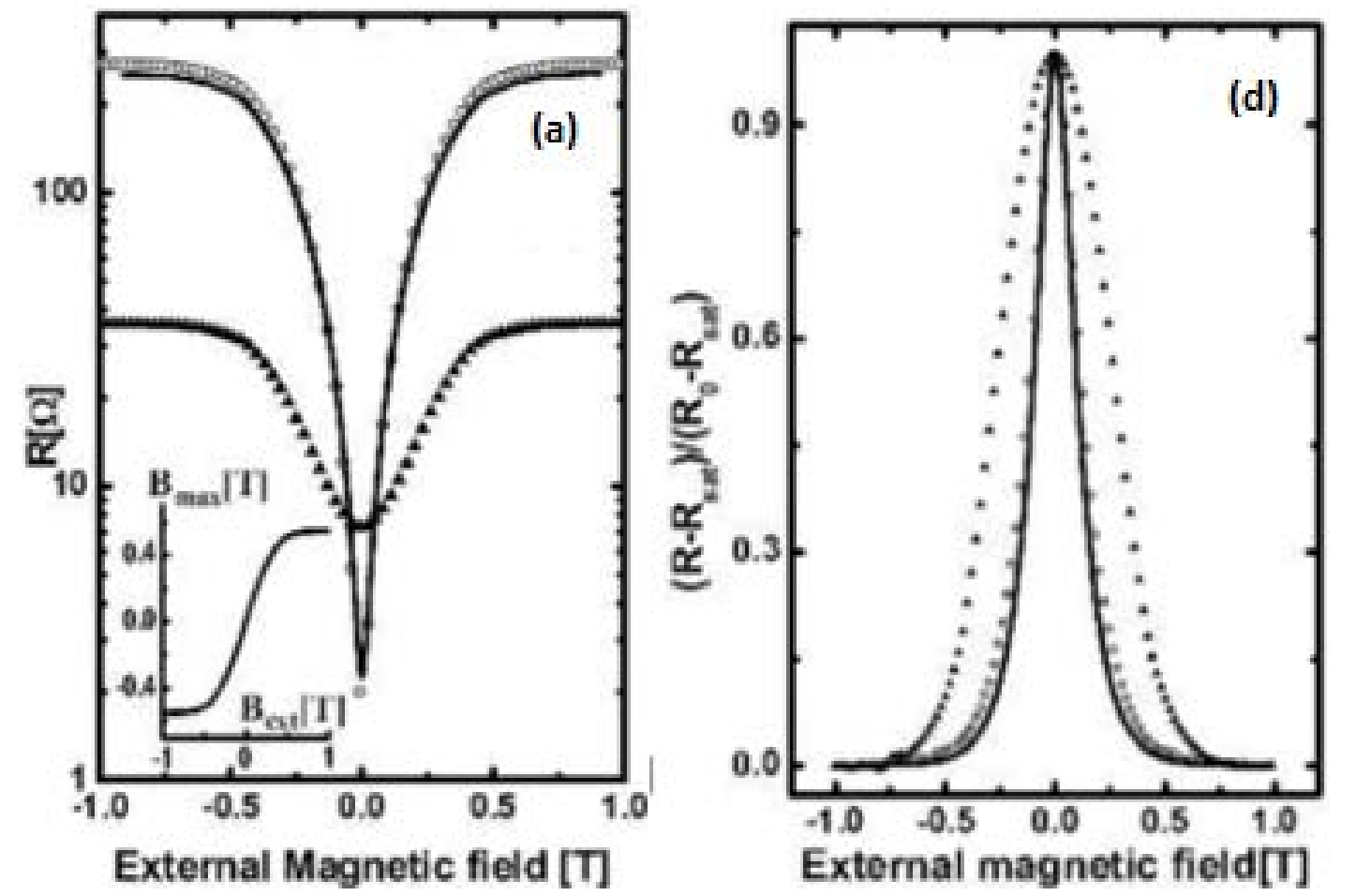

*Figure 2.18: a) Resistance as a function of magnetic field for the diagonal configuration at 4.2 K (○) and room temperature (▲). d) Magnetoresistance of the device operated in the two probe side configuration at 4.2 K (○) and room temperature (▲).*[14]

Interestingly, it was observed that the device was capable of producing a positive magnetoresistance when operated in the diagonal configuration but a negative magnetoresistance when the side configuration was used (see Figure 2.17). When operated in the diagonal configuration, the magnetoresistance at 1 T was 800% at room temperature and 12,000% at 4.2 K. A maximum sensitivity of 78 Ω/T and 681 Ω/T at 0.3 T were recorded for the room and low temperature measurements respectively. When tested in the side configuration, magnetoresistances of -27% and -1450% at room and low temperatures were observed.

Both the diagonal and side configurations show a saturation in the magnetoresistance at an external field strength somewhere between 0.5 and 1 T. This result from the magnetization of Fe typically saturating around 0.6 T, which represents a ceiling to the strength of the magnetic barrier which the gate can generate. This means that there is a limit to the maximum field strength that can be measured with this device. However there are some unique advantages to this design, namely that it is easy to fabricate and that its performance does not depend on contact resistance as there are no internal interfaces. Similar strategies for controlling the electron trajectory using inhomogeneous magnetic fields have also been addressed in several other studies as reviewed by Nogaret (ref).

## 2.7. Geometry of Contacts

### 2.6.1. Contact Positions:

Similar to the geometry of the metal inclusion, the position, number and size of the contacts are of great importance for the response of EMR devices. Most studies regarding the order or location of the contacts focus on the bar geometry. Conformal mapping of a vdP disc where the current probes are adjacent to one another into a bar-shaped device results in a *VIIV* order of the voltage (*V*) and current (*I*) probes (ref Zhou 2001). Sun et al. reported that changing the contact configuration between the two symmetric cases of *IVVI* and *VIIV* yielded almost identical device performance.[55]
Huang et al.[64] presented simulations which predicted that the bar-shaped device could be optimized by varying the distance between voltage probes. The spacing between the innermost leads in the *IVVI* configuration was changed from 18% of the total device length to 13% which led to a doubling in the MR at 3 T (see Figure 2.7), demonstrating that the location of the probes is an important geometric parameter.
The effect of contact spacing was also examined by El Ahmar et al. using a ten terminal device (see Figure 2.18) in the *VIIV* configuration.[24] It was determined that a terminal arrangement of $I_{3,t}V_{1,4}$ produced an EMR response almost 3 times higher than $I_{4,t}V_{3,5}$ regardless of the value of *t*, signifying that the location of the second current terminal does not have a large effect, whereas the location of the first terminal is a significant factor.

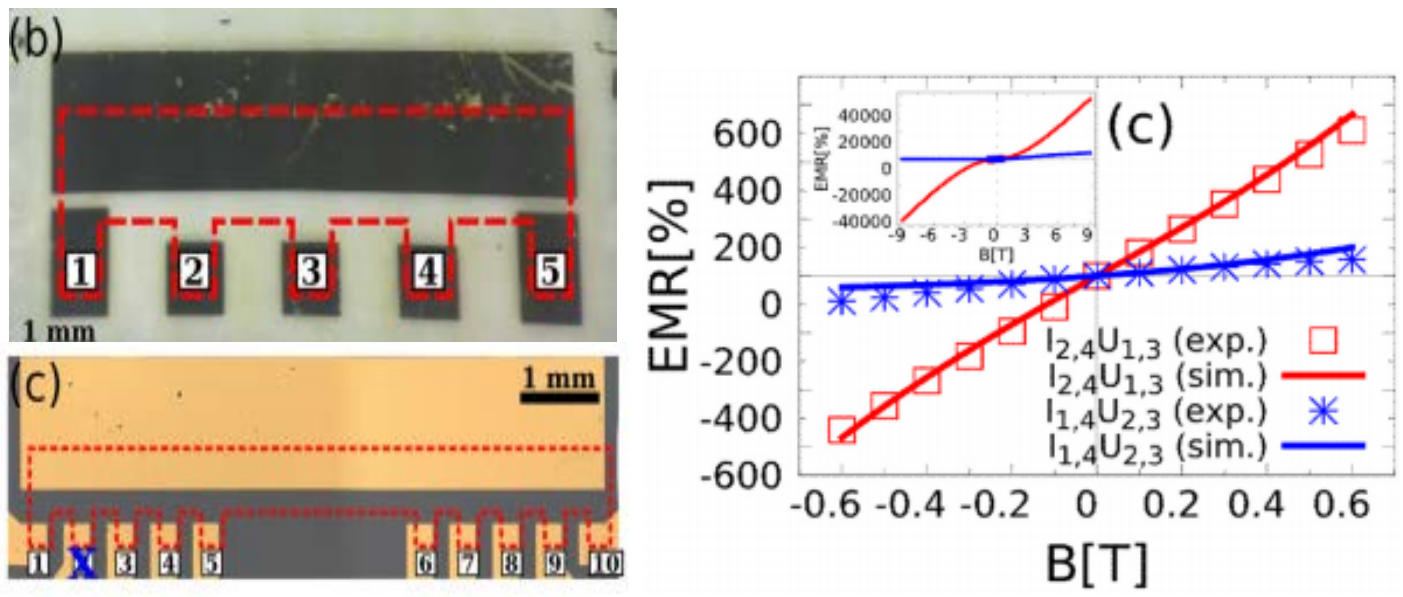

*Figure 2.19: EMR devices using monolayer epitaxial graphene with 5 terminals (top left) and 10 terminals (bottom left). FEM simulations and experimental data (right) showing EMR as a function of magnetic field for different terminal configurations Figure is adapted from El-Ahmar et al.*[26]

In addition to changing the probe spacing, the probe order can also be varied. Several works have explored the effect of staggering the probes both experimentally[8,21,26] and with the use of simulations.[56,57] Troup et al. showed a nearly four-fold increase in the magnetoresistance and three-fold increase at XX T in the sensitivity of a device when switching from an *IVVI* to an *IVIV* configuration.[21] The magnitude of the improvement in the performance of the sensor was similar to what was observed by El-Ahmar et al. (see Figure 2.18).[8,26] It should be noted that when bar-shaped devices are operated in the *IVIV* configuration, the EMR signal is not symmetric around 0 T. The authors claim that due to the asymmetry of the device, the effectiveness of the metal shunt differs depending on the direction of the applied field. Ultimately El Ahmar et al. were able to verify results from previously established literature and recommend using an asymmetric *VIVI* terminal configuration, placing the first current terminal at a distance of 25-35% of the length of the device from one edge and the second terminal at a distance of 10-20% from the opposite edge.[26]

The effect of asymmetric contact placement was also studied by Holz et al. for the case of a bar shaped device in the *IVVI* configuration.[51] When the voltage contacts were placed in a mirror symmetric manner (V2-V4 in Figure 2.19a), a symmetric resistance with $R(B) = R(-B)$ was obtained. If the mirror symmetry was broken (V2-V3), slight asymmetries were observed. In both cases, the resistance appeared to approach the value obtained without a metal shunt for large magnetic fields, signifying a complete expulsion of the current from the metal. If the voltage contacts were positioned in a *VIVI* configuration (V1-V2 and V1-V4 in Figure 2.19b), the overall measured resistance approached the linear Hall coefficient at large magnetic fields. For small values of μB, non-linearities were observed that resembled the EMR and enhanced the sensitivity, $(dR/dB)$, above that of the Hall effect while simultaneously benefitting from a zero-field sensitivity which is greater than zero, which is characteristic of Hall sensors. Thus, the interplay between the Hall effect and conventional EMR may be combined to yield superior device performance. That hypothesis was also reinforced by Sun et al.[11] who studied a 3-terminal bar-type device which exhibited enhanced low-field sensitivity while retaining the strong EMR effect at high fields. They found that using three terminals would produce a larger sensitivity as the Hall effect also contributes to the total output sensitivity.

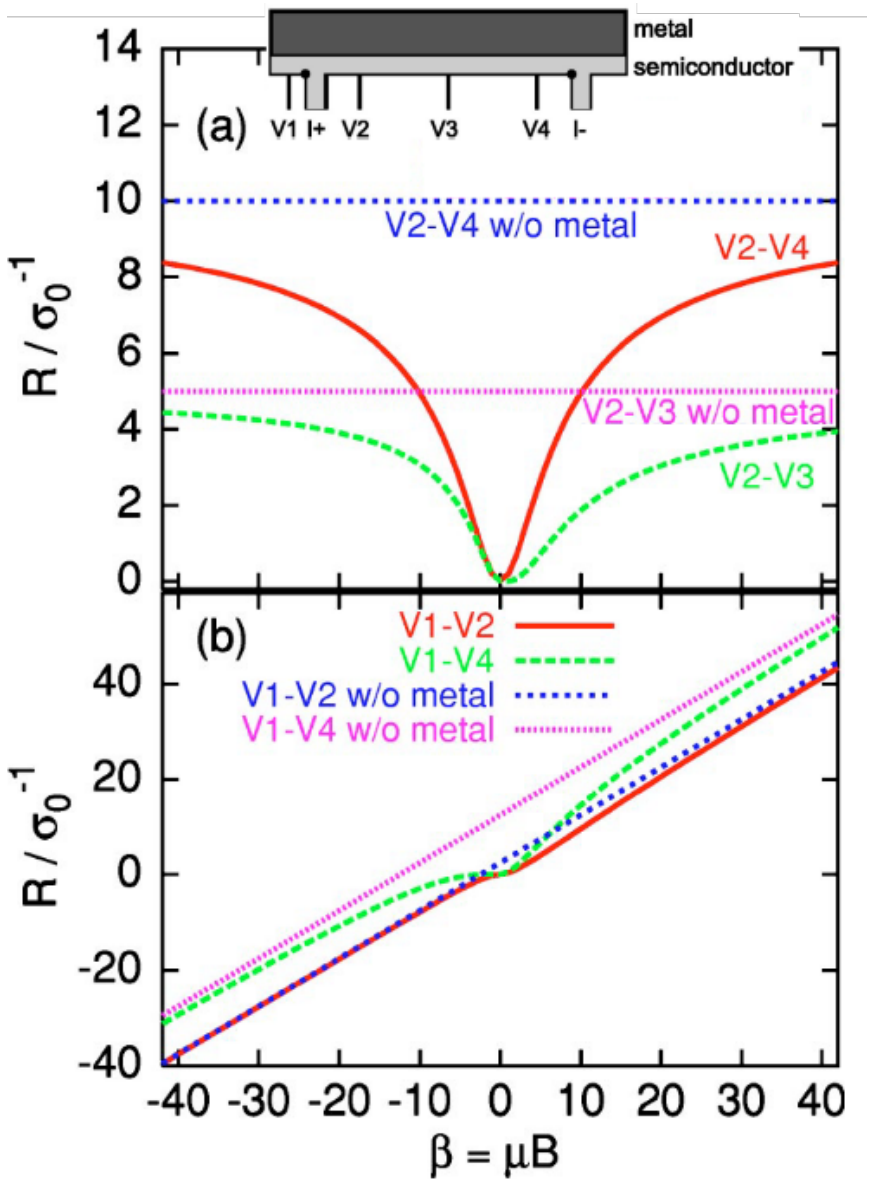

*Figure 2.20: The numerically obtained resistance and its dependence on the product of the mobility and magnetic field for a bar-shaped device with properties similar to an InAs/InGaAs quantum well shunted by a metal with infinite conductivity. The resistance is normalized by the zero field conductivity* ($\sigma_0 = ne\mu$) *and calculated using various combinations of voltage probes. Figure is adapted from Holz et al.*[51].

Holz[29] and Sun[71] suggested that for the weaker magnetic fields around 50 mT, an asymmetric voltage probe placement boosts the magnetoresistance and sensitivity by an order of magnitude compared to the symmetric voltage probes. However, Solin[45] showed that asymmetric contact configurations in a bar-type device could also lead to significant enhancements in higher magnetic fields (see Figure 2.20).

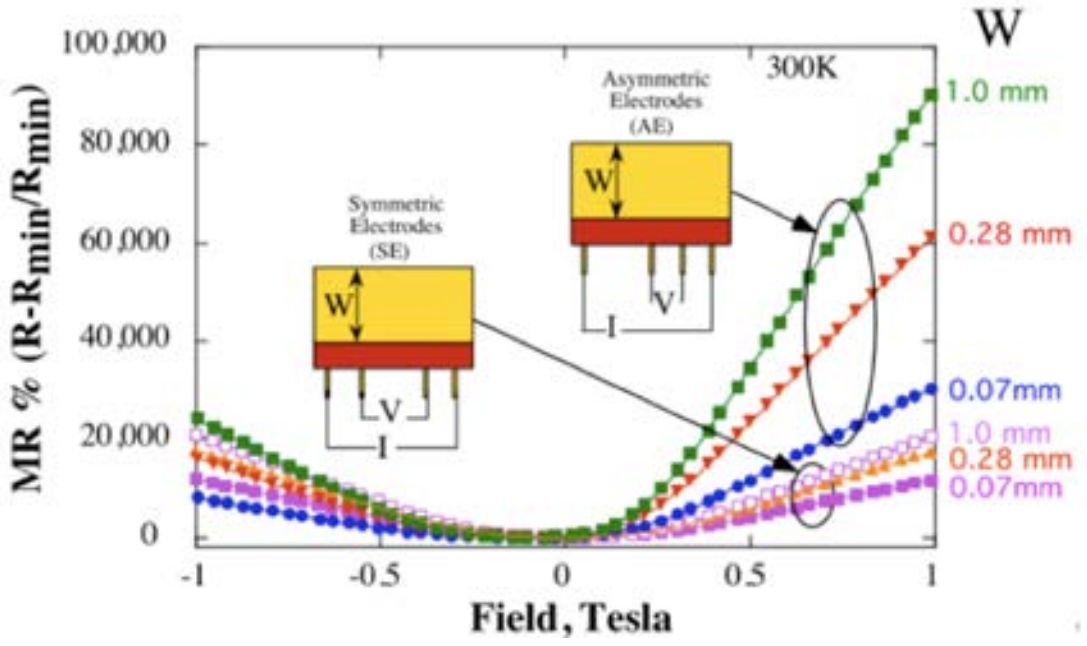

*Figure 2.21: MR(B) for bar devices with symmetric and asymmetric probe configurations with varying metal thickness, W. Figure from Solin*[45].

For the concentric circular device, the influence of the contact positions was numerically investigated by Huang et al.[73]. Using filling factors ranging from 11/16 to 13/16, the magnetoresistance in a field of 0.1 T was simulated for various symmetric voltage probe configurations. It was found that the magnetoresistance could be increased by about a factor of two by narrowing the angular span between the two voltage contacts. An asymmetrical voltage contact placement was also briefly investigated with only a single simulation indicating that the magnetoresistance might be increased slightly by an asymmetrical voltage contact placement in the case of the shunted vdP geometry.

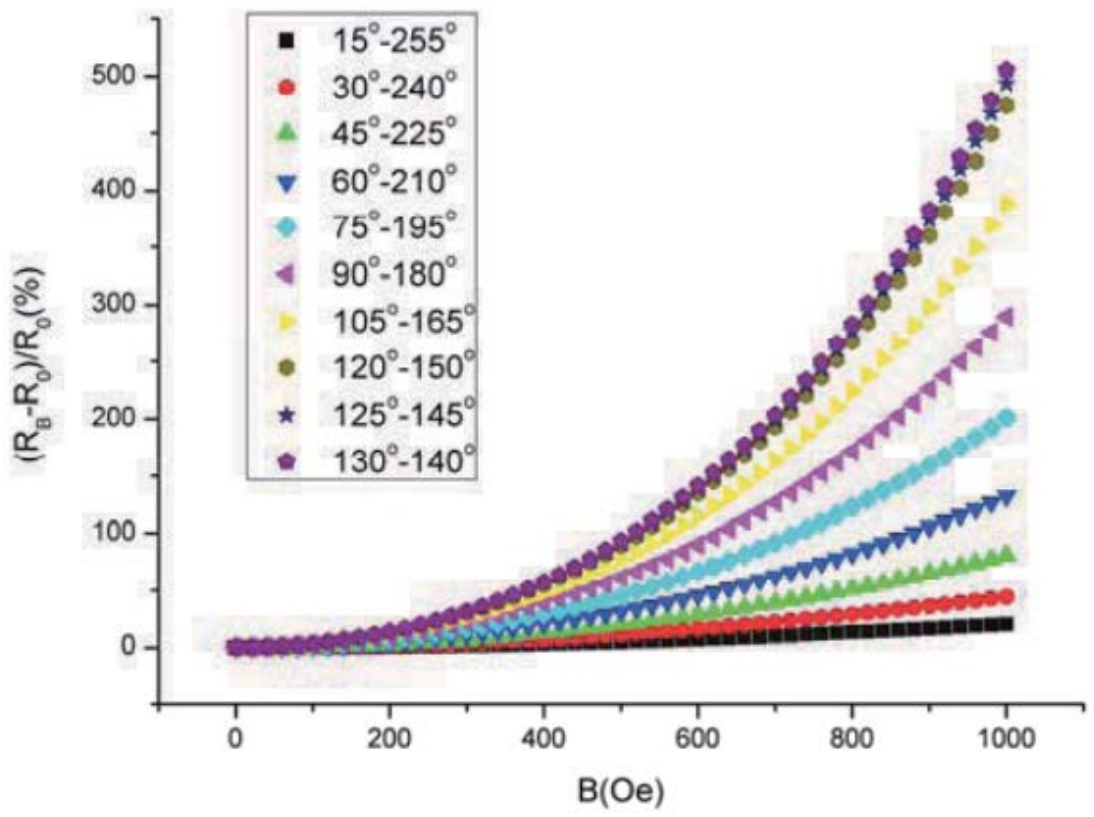

*Figure 2.22: Plot showing MR as a function of magnetic field for various angular distances between voltage probes. Figure from Huang et al.*[73]

### 2.6.2. Top vs. Side Contacts:

Electrical contacts to the active materials can be made by either depositing metal on the top surface or along the side-wall of the structure. Top contacts are easier to produce than side contacts and require less precise alignment of the etch masks but in some cases can result in lower contact quality (ref). Sun et al.[74] investigated the difference between top and side contacted EMR devices using two bar device geometries (see Figure 2.22) to determine if the manufacturing process could be simplified by using top contacts. The first device was a conventional bar-shaped device where Si-doped InAs was patterned into a strip followed by an aligned metal deposition to define the metal shunt and electrodes. The metal contacts the semiconducting bar from the side with a slight overlap at the top. The second device is composed of an unpatterned semiconductor with metal deposition on top. To avoid current leakage between the two electrodes, an insulating layer of silicon nitride was added below the electrodes. Both devices showed an EMR effect when exposed to an out-of-plane magnetic fields with the top contacted device yielding a lower resistance at zero field (see Figure 2.22).

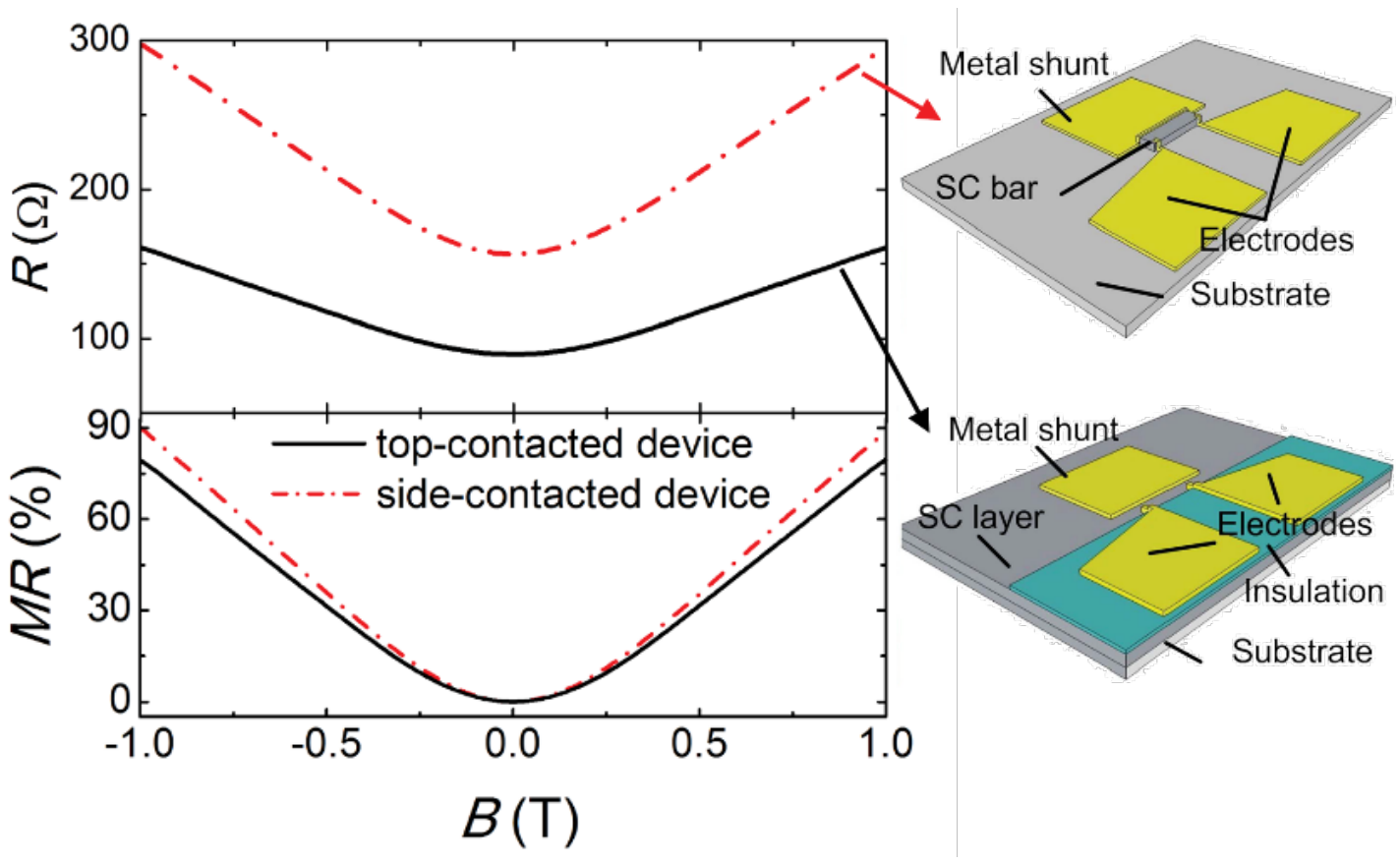

*Figure 2.23: Resistance and magnetoresistance as a function of the magnetic field for EMR devices with the metal shunt and electrodes contacting the side (red) or top (black) of the semiconducting bar. Figure is adapted from Sun et al.*[74]

Both devices showed a similar magnetoresistance, but the sensitivity (dR/dB) was approximately a factor of 2 lower for the top-contacted device. The authors further investigated the magnetic field sensor properties and found that the top-contacted device resulted in a thermal voltage noise of 1.7 nV/√Hz, which was reduced slightly compared to the side-contacted device (2.3 nV/√Hz) due to its lower resistance. The larger sensitivity of the side-contacted device resulted in an approx. 30% better magnetic field detection limit, yielding a value of 19 nT/$\sqrt{Hz}$ between 0.4 and 1 T and 4.3 μT/$\sqrt{Hz}$ at 0 T.

### 2.6.3. Size of Contacts:

Poplavskyy examined the effect of contact size by studying the magnetoresistance of the concentric circular geometry by using the analytical expression.[43] In this study, the contact size was found not to be a critical parameter, as the magnetoresistances for point contacts, 8 degree wide contacts and 16 degree wide contacts were very similar, and only a small reduction was found when considering contact widths up to 32 degree (see Figure 2.23). The case of point contacts further led to a significant simplification of the analytical expression.

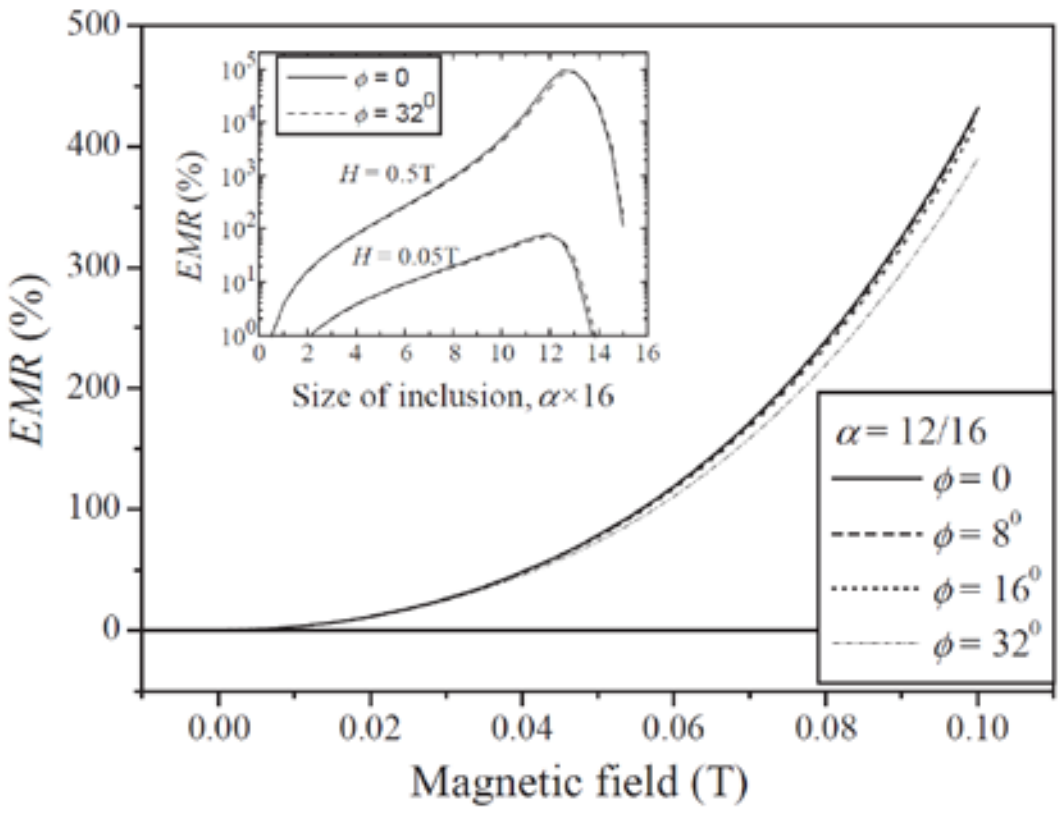

*Figure 2.24: EMR as a function of magnetic field for point contacts and contacts of increasing width with a filling factor α=12/16. Inset shows EMR response as a function of α for 0.5 and 0.05 T using point contacts (solid line) and 32° wide contacts. Figure adapted from D. Poplavskyy.*[75]

# 3. Material Parameters

In this section we review how the choice of materials and material properties affect the performance of EMR sensors. Firstly, we examine how individual material properties affect device performance, after which experimental procedures and results are detailed for various material systems.

## 3.1. Electronic Transport Parameters

The electronic properties of the constituent materials in EMR devices have a strong impact on the EMR device performance. The electrical conductivity, $\sigma$, is given by:

$$\sigma = en\mu$$

where $e$ is the elementary charge, $\mu$ is the charge carrier mobility, and $n$ is the density of charge carriers which can be positive or negative depending on the charge state of the carrier species. Both the mobility and carrier density of the constituent materials as well as the contact resistance to the metal shunt can effectively turn the magnetoresistance in EMR devices from a high value to being nonexisting as outlined in the following sections. The materials that have been used for making EMR sensors are primarily III/V semiconductors and graphene as outlined in Table 1 in the introduction.

Another vital parameter which determines the performance of EMR sensors is the interfacial contact resistance between the shunt and the semiconductor and whether interface forms an ohmic contact or Schottky barrier. If the energy gap between the Fermi level in the metal and the conduction band in the semiconductor is low, the resulting contact is ohmic. For many semiconductors however, energy gap can be quite sizable and the wave function of electrons in the metal decays into the semiconductor band gap, resulting in pinned electronic states within the band gap of the semiconductor which can result in a relatively high resistance.[76] This state is known as a Schottky barrier and results in a non-linear current-voltage response which is directionally dependent.

### 3.1.1. Carrier Mobility

The carrier mobility is a measure of how quickly charge carriers move through a material in response to an external electric field, $E$. For electrons and holes the mobility is given by:

$$\mu = \frac{v_d}{E} = \frac{-e\tau_s}{m^*}$$

Where $v_d$ is the drift velocity, $\tau_s$ is the relaxation time between scattering events and $m^*$ is the effective mass of the carrier. The most significant factor affecting the mobility is the scattering processes present within the system such as scattering off impurities, phonons and crystal defects. The carrier mobility is particularly relevant for EMR devices as it both lowers the zero-field resistance and directly affects the Hall angle ($\varphi_H = \tan^{-1}\mu B$) by entering the off-diagonal elements of the conductivity tensor.

Finite element models were used to investigate the effect of the carrier mobility of the semiconductor in numerous studies.[29,54,60,77,78] In the low-field regime, higher mobilities translate to higher magnetoresistance values due to the increase in the Hall angle via the $\beta$ terms in the magnetoconductivity tensor allowing for a more effective redistribution of the current. High field strengths $\beta$ and consequently the Hall angle will always be large and thus the improvement in magnetoresistance begins to saturate at high field and mobility values since all of the current is forced from the shunt.[54,60]

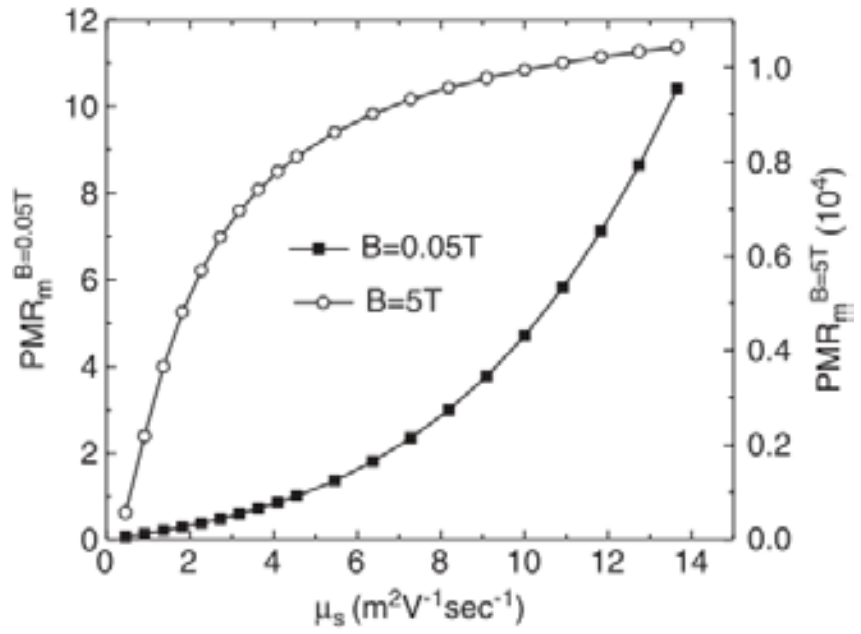

*Figure 3.1: Magnetoresistance using point contacts (PMR) for a three terminal device as a function of mobility at the low and high field regimes.*[54]

The higher the mobility of the semiconductor, the faster the magnetoresistance saturates due to the larger Hall angle at a given magnetic field. For mobilities exceeding 20,000 $cm^2$/Vs saturation can be observed within 5 T (Figure 3.5).[77] In the case of very large mobilities, the semiconductor conductivity can approach that of the shunt and reduce the effectiveness of the device.[54]

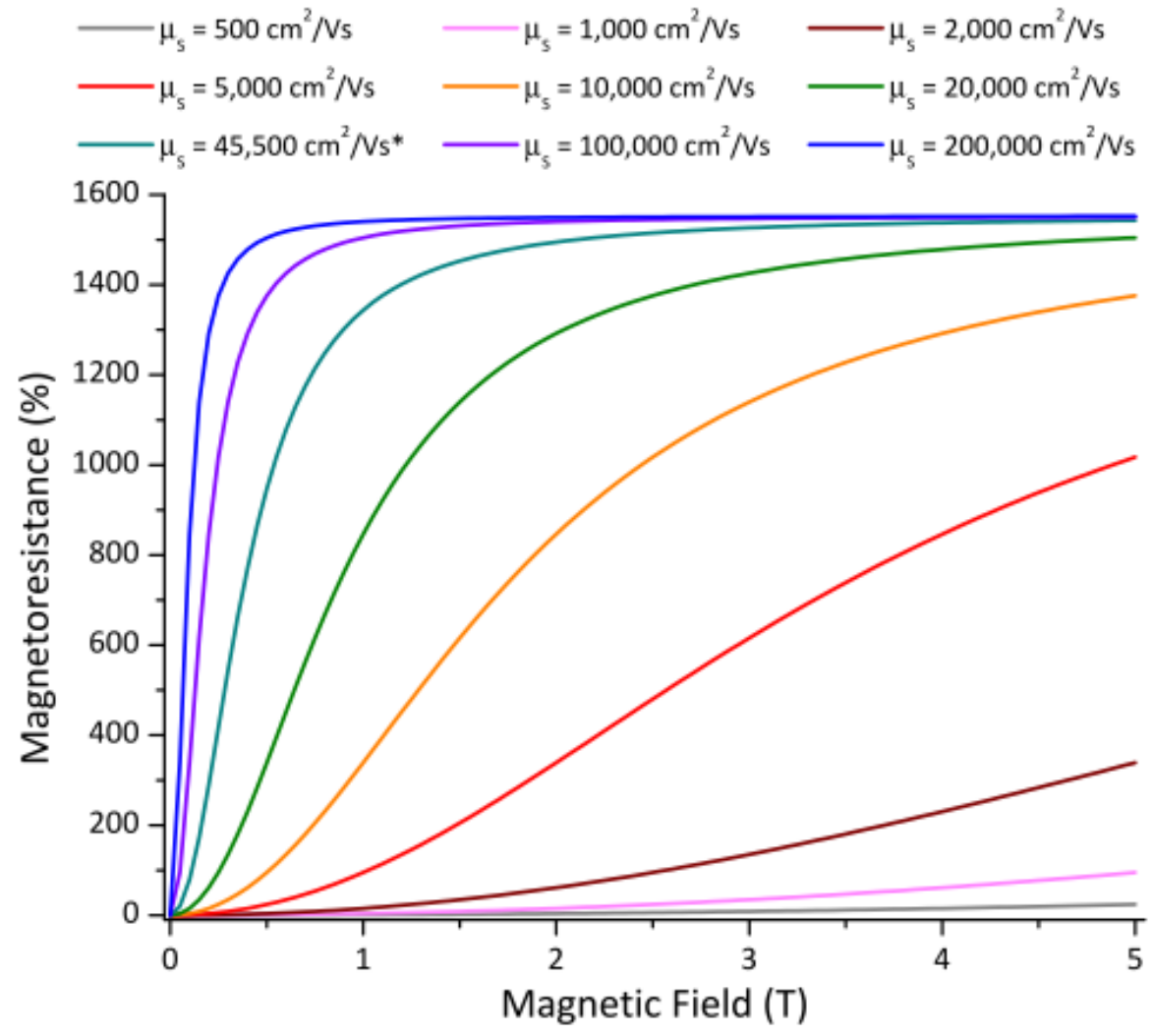

*Figure 3.2: Magnetoresistance as a function of the applied magnetic field for different values of the carrier mobility in the semiconductor.* [77]

The effect of mobility on the sensitivity of EMR devices was also considered. It can be seen in Figure 3.6 and Figure 3.7 that the device resistance increases as the carrier mobility is decreased. However, this decrease is not linear under a finite magnetic field and as a result both the magnetoresistance[29,60] and sensitivity[78] curves of the sensors show a clear maximum as a function of mobility. The mobility which maximizes these two figures of merit becomes lower as the magnetic field strength is increased. Holz determined that the maximum sensitivity, dR/dB, scales with μB and under their conditions the sensitivity reaches a maximum when μB is equal to 0.8.[78]

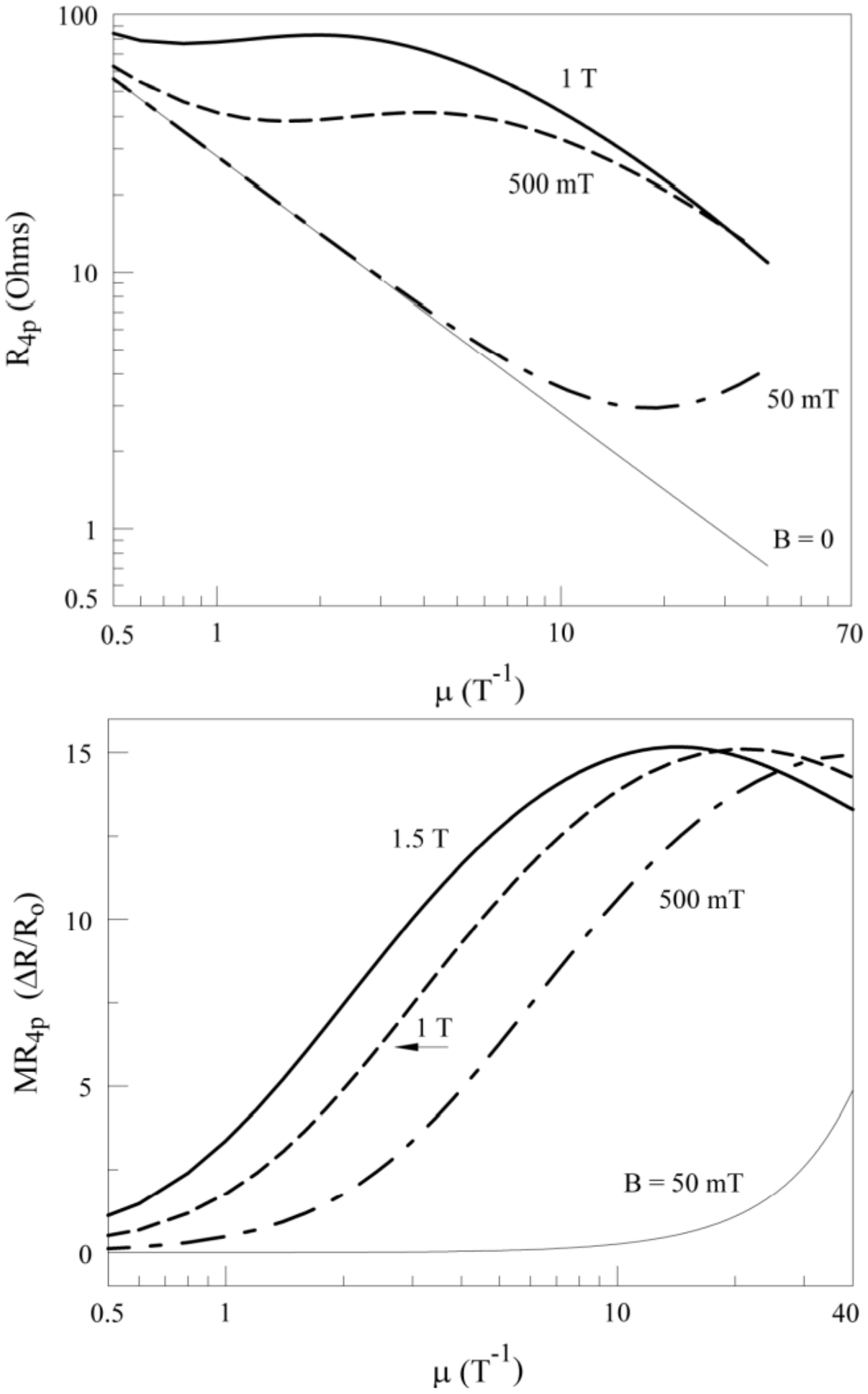

*Figure 3.3: Resistance (top) and magnetoresistance (bottom) of a four terminal bar device for 3 magnetic fields as a function of carrier mobility.*[60]

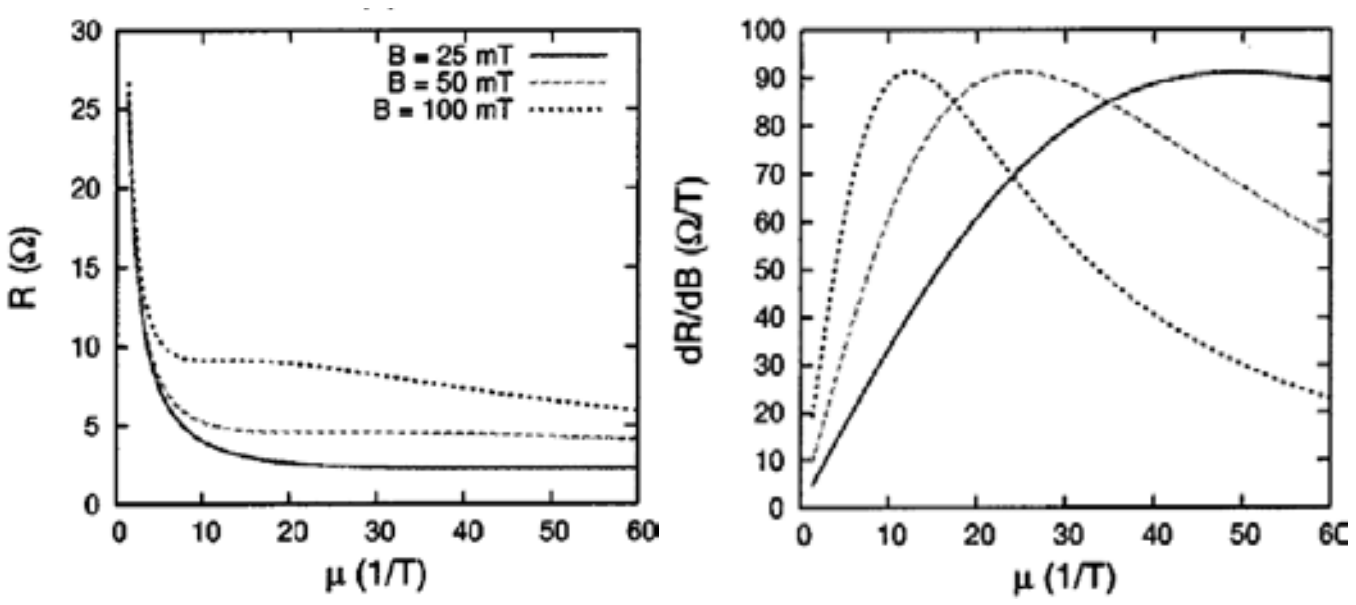

*Figure 3.4: Resistance (left) and sensitivity (right) of a bar shaped EMR sensor for 3 magnetic fields as a function of carrier mobility.*[63]

In a two-terminal device, no such saturation in the resistance as a function of mobility was observed (see Figure 3.8). As a result, the magnetoresistance of the devices only increased with higher mobilities. Whereas four-terminal devices have an optimum mobility depending on the field range that one would like to measure, for two-teriminal devices a higher mobility yields better performance at all field ranges tested.[60]

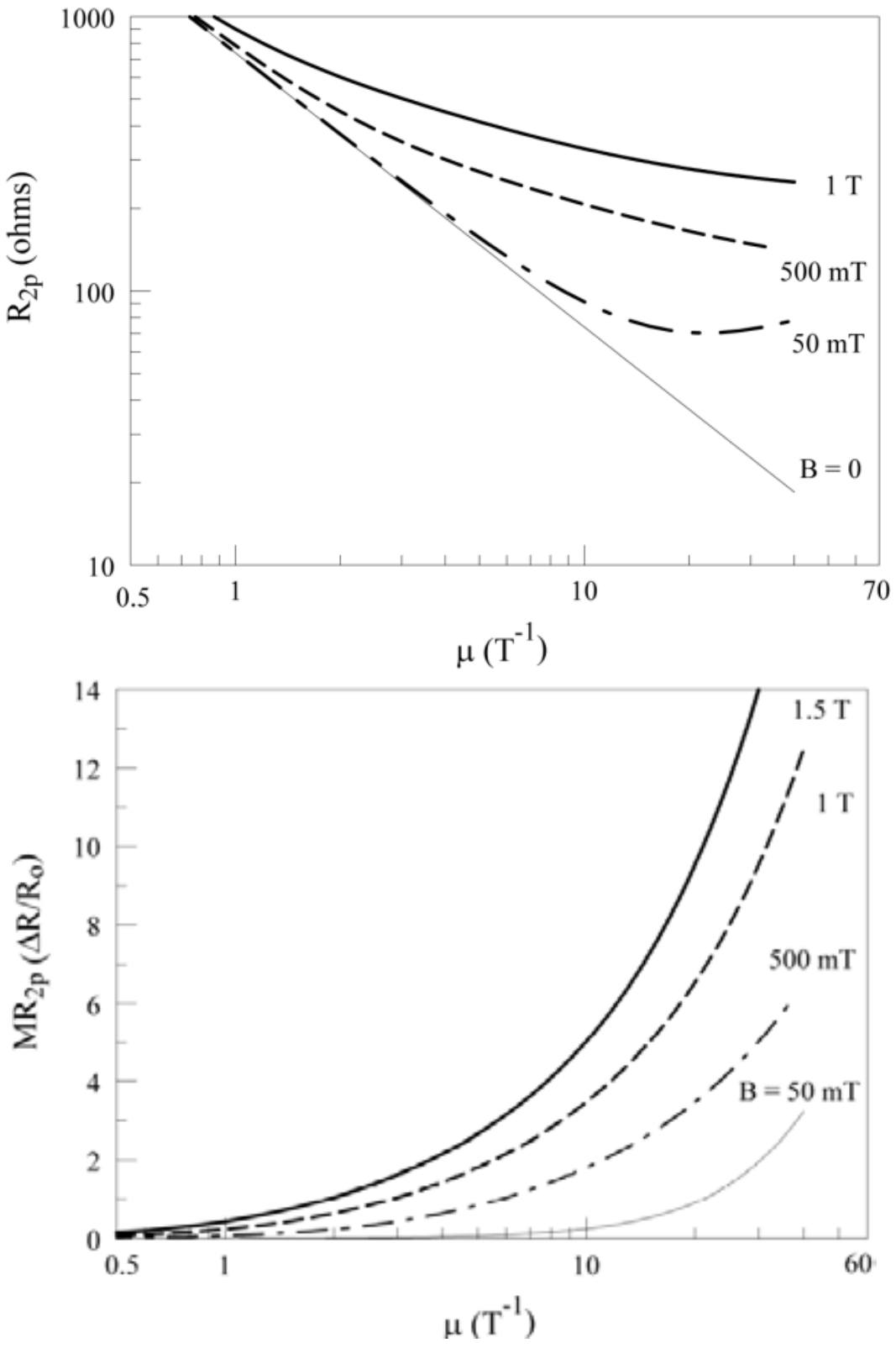

*Figure 3.5: Resistance (top) and magnetoresistance (bottom) of a two terminal bar device for 3 magnetic fields as a function of carrier mobility.*[60]

### 3.1.2. Carrier Density

The effect of varying the carrier density of the semiconductor was investigated in various papers which modeled bar shaped devices with two[60], three[54], and four contacts.[60,78] In all three contact configurations, the magnetoresistance was found to be invariant as a function of the semiconductor carrier density provided that the resulting conductivity of the semiconductor remained below a threshold value (see Figure 3.6). Below this threshold, varying the carrier density of the semiconductor produces a inversely proportional relationship between the device resistance at a given magnetic field and the carrier density, which is in line with what is expected theoretically.[60,78] Since the low and high field resistance curves are parallel, the magnetoresistance remains unchanged.

However, when the carrier density is increased past a threshold value, the resulting conductivity of the semiconductor becomes comparable to or higher than that of the metal. The metal therefore no longer acts as a shunt which reduces the magnetoresistance of the device until it eventually vanishes.[60]

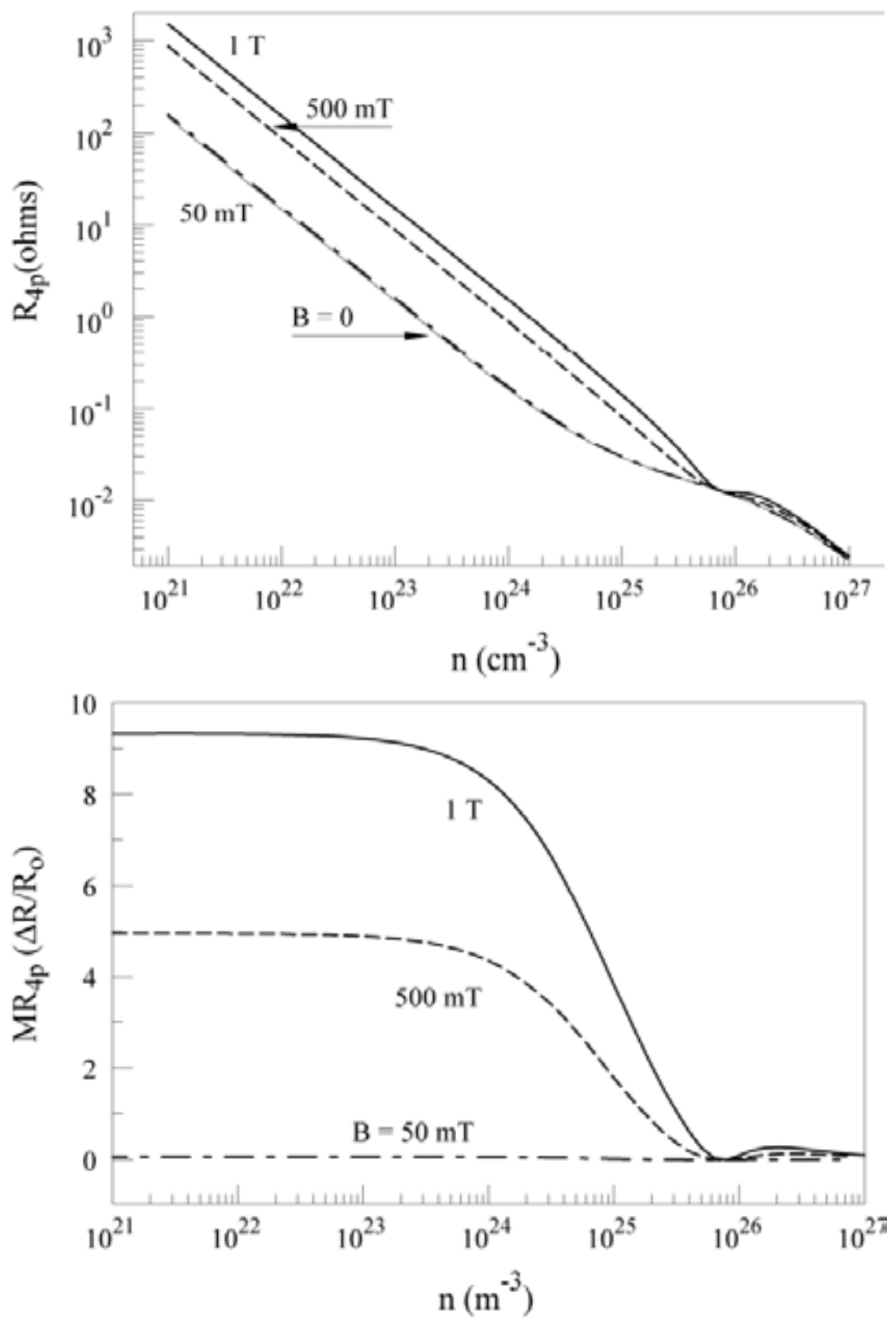

*Figure 3.6: Resistance (top) and magnetoresistance (bottom) of a four terminal bar device for 3 magnetic fields as a function of carrier density.*[60]

A critical carrier density was also observed when the device was operated in the two-terminal configuration, however the magneto-resistance did not vanish at high carrier densities up to $10^{-27}$ $m^{-3}$.[60]

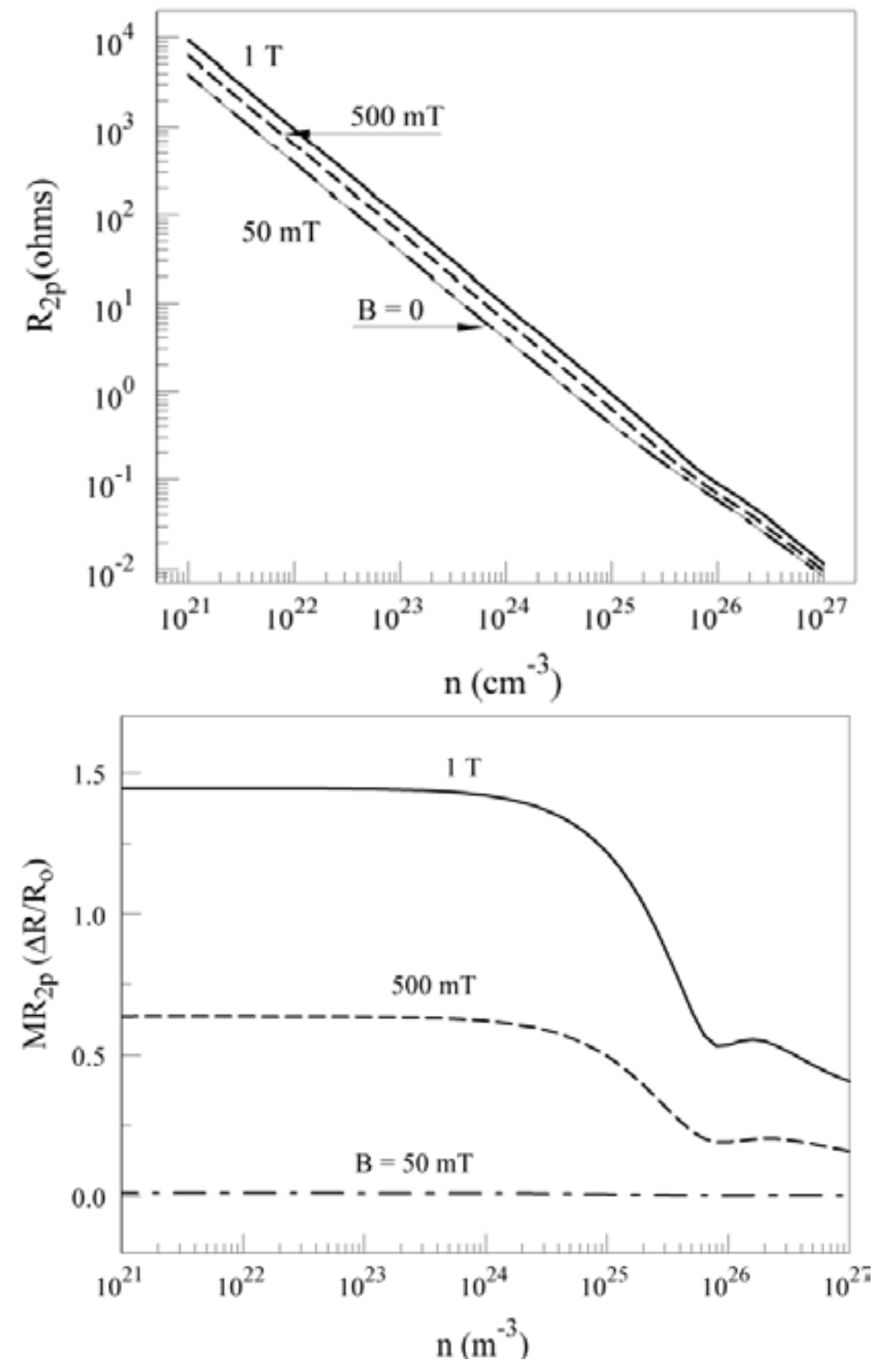

*Figure 3.7: Resistance (top) and magnetoresistance (bottom) of a two terminal bar device for 3 magnetic fields as a function of carrier density.*[60]

Lower carrier densities also result in a higher sensitivity with an inverse linear relationship between the two up to a threshold carrier density (see Figure 3.3). The lower carrier densities result in a higher semiconductor resistance which increase the overall device resistance in the presence of a field. However, because the device resistance as a function of carrier density has the same slope as the current sensitivity, the voltage sensitivity defined as $(1/R)\ dR/dB$ shows the same behavior as the magnetoresistance and saturates below the threshold carrier density.[78]

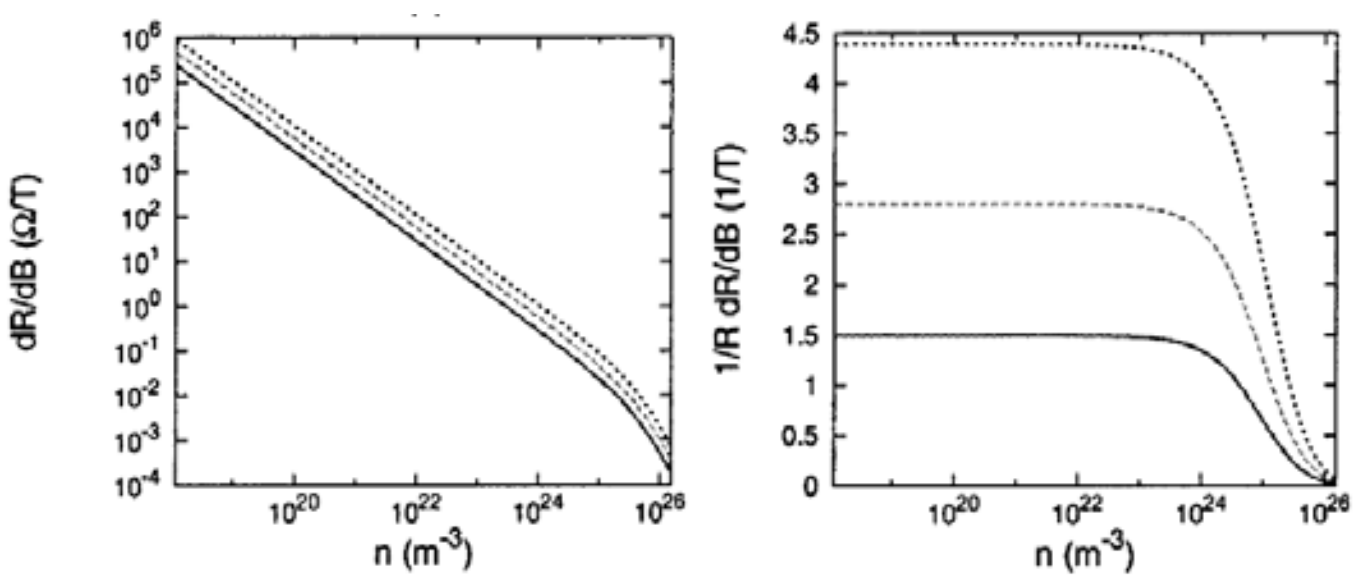

*Figure 3.8: Current and voltage sensitivity for a four terminal device as a function of the carrier density of the semiconductor.*[78]

### 3.1.3. Shunt Conductivity

FEA models were used in various studies in order to examine the role of the conductivity of the shunt material.[54,60,77,78] Here, the conductivity contrast between the semiconductor and shunt regions was varied by changing the carrier density in the shunted region. Holz[78] and Rong[54] found that the conductivity of the shunt itself does not play a decisive role in determining the magnetoresistance, but rather the ratio of the shunt and semiconductor conductivities, $\sigma_m/\sigma_s$. When the ratio of the shunt to semiconductor conductivity is increased above $10^7$, the resistance of the device changes little and the magnetoresistance saturates. As such, it is not critical to find extremely high-conductivity metals or superconductors in order to observe a strong effect. However, if the conductivity of the two materials is similar, the resistance of the device increases rapidly and the magnetoresistance and sensitivity of the device approaches zero as no current flows through the shunt. Both papers claim a threshold in the conductivity ratio of $10^4$, above which the the device produce a significant EMR effects.

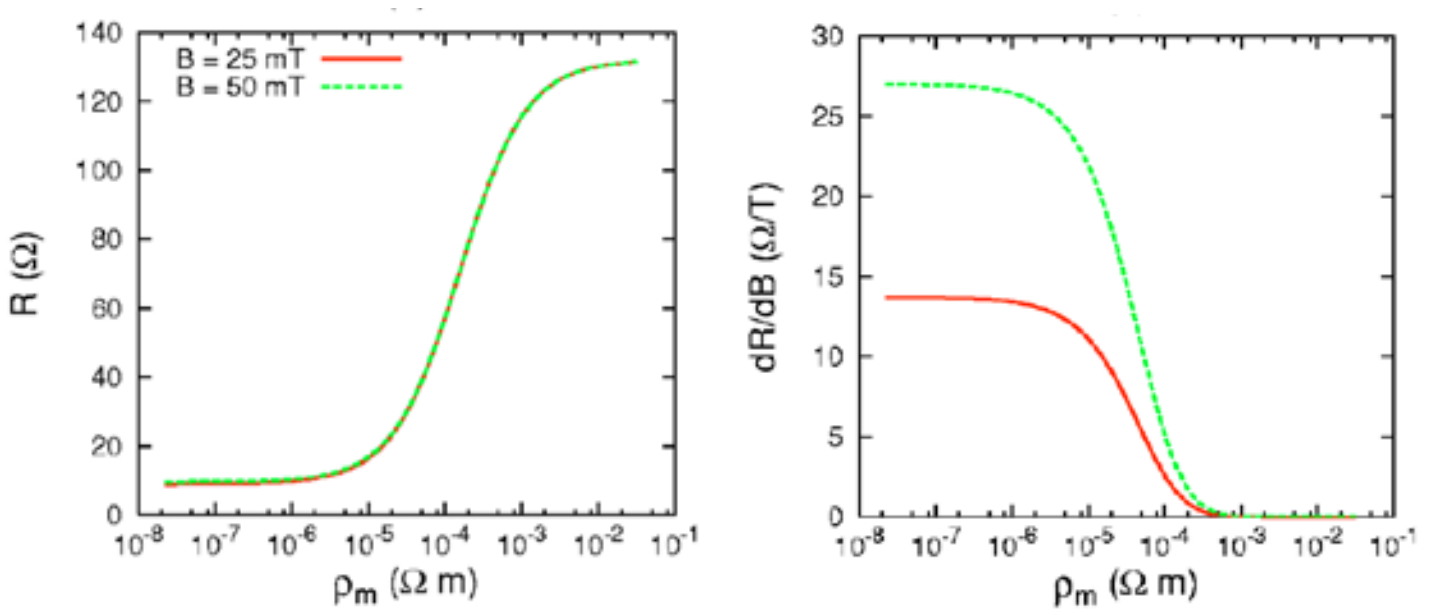

*Figure 3.9: Resistance (left) and sensitivity (right) of a bar shaped EMR sensor as a function of shunt resistivity.*[78]

Hewett[77] and Nunnally[60] performed the same study by modeling concentric vdP disc and bar shaped devices, respectively. In these papers it was reported that the magnetoresistance response began to saturate when the shunt was only approximately 100 times more conductive than the semiconductor. Considering that the original paper by Solin[1] used a device with a conductivity ratio $\sigma_M/\sigma_S$ of 2430, these results seem to be more realistic.

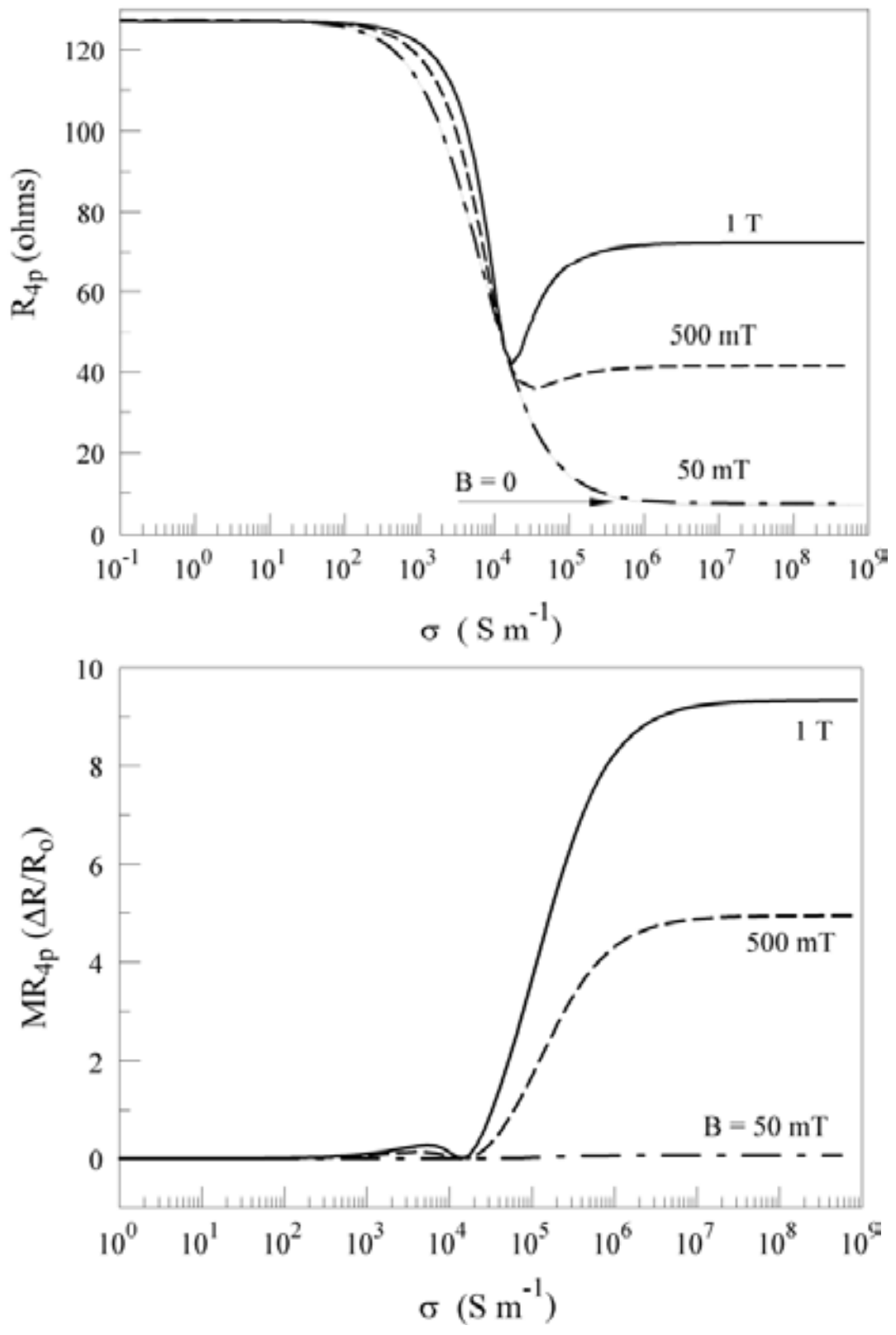

*Figure 3.10: Four point resistance (top) and magnetoresistance (bottom) as a function of shunt conductivity for three different magnetic fields.*[60]

Nunnally conducted the aforementioned simulation for a two-terminal bar device and observed similar behavior in the magnetoresistance response. However, the saturation of the magnetoresistance begins when the shunt is only on the order of 10 times more conductive.[60] Interestingly, there is a finite magnetoresistance in the device regardless of the conductivity of the shunt. The two terminal configuration may, therefore, be more appropriate for material systems with a low conductivity contrast.

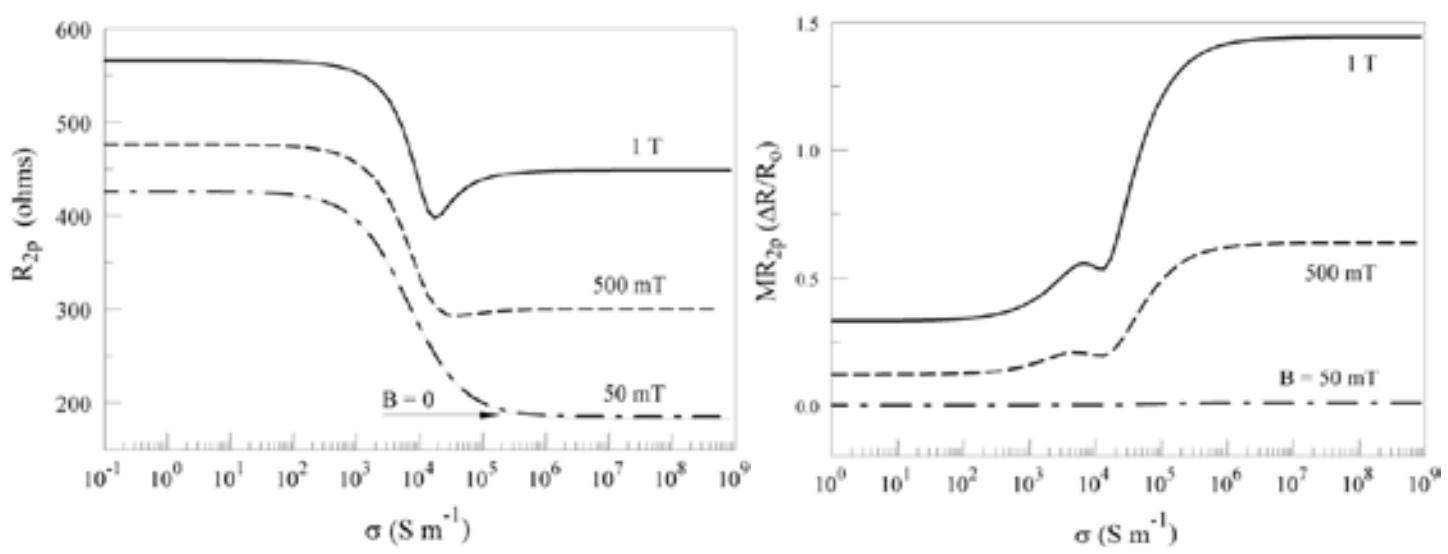

*Figure 3.11: Two point resistance (left) and magnetoresistance (right) as a function of shunt conductivity for three different magnetic fields.*[60]

### 3.1.4. Interface Contact Resistance

The effect of a contact resistance between materials was considered numerically in several papers for cases featuring ohmic contacts[61] and Schottky barriers.[59,64,77,78] In addition, the shunt has also been considered to pin the Fermi level of the graphene and forming a p/n-junction in graphene in vicinity of the graphene/metal interface (ref Bowen).

In the modeling of ohmic contacts, a contact resistance was applied as a constant value to the interface between the two materials.[61] For the simulations involving a Schottky barrier, a thin material region was placed between the two materials, at either 0.3%[59] or 1%[64,77] of the semiconductor thickness. To this material region a resistivity tensor was applied with the following form:

$$\hat{\rho} = \begin{bmatrix} \rho_c & 0 \\ 0 & \rho_c \end{bmatrix}$$

where $\rho_c$ is the specific interface resistance. Unlike the magnetoresistivity tensor, the interface resistance is treated as independent of the magnetic field.

For the case of ohmic contacts it was found that as long as the contact resistance was lower than $10^{-8}$ Ω/cm$^2$ there was no influence on the magnetoresistance, but above this value the magnetoresistance decreases in an S-shaped curve, dropping by a factor of 100 at $10^{-5}$ Ω/cm$^2$ (see Figure 3.12). However, it is interesting to note that the sensitivity, while displaying the same behavior as a function of contact resistance, remains high until a contact resistance of $10^{-6}$ Ω/cm$^2$ before decreasing.[61]

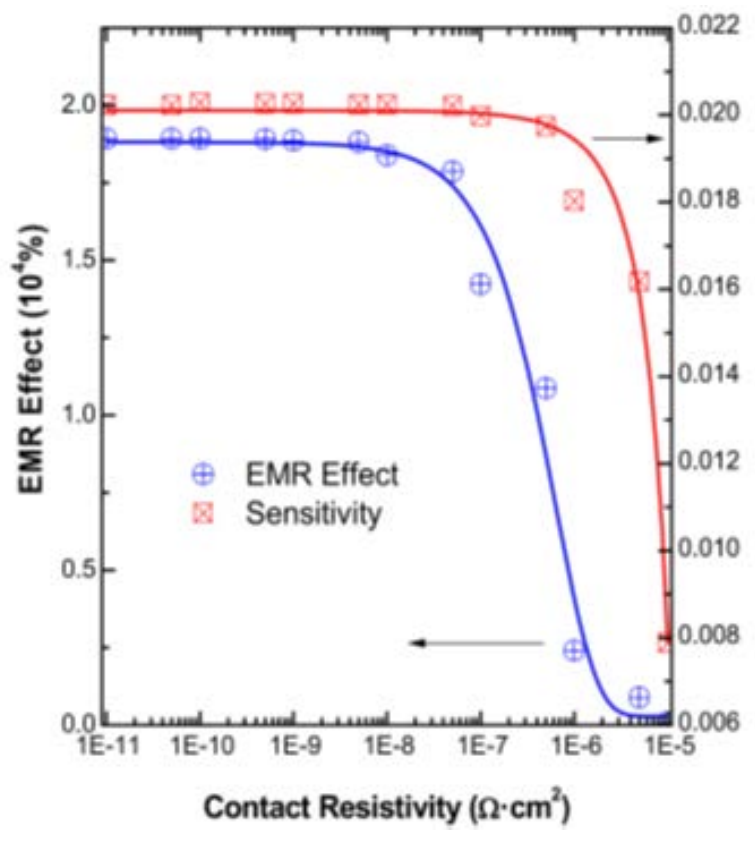

*Figure 3.12: Magnetoresistance and sensitivity at 1 T as a function of the interface resistivity.*[61]

Models which considered the contact resistance with a Schottky barrier found that very low values of the contact resistance do not affect the resistance of the device, but once the contact resistance approaches $10^{-6}$ Ω/cm$^2$ the magnetoresistance depends critically on the value of the contact resistance.[77,78] At intermediate or high values of the contact resistance the current is effectively blocked from entering the metal, causing to flow through the semiconductor which increases the resistance of the device.

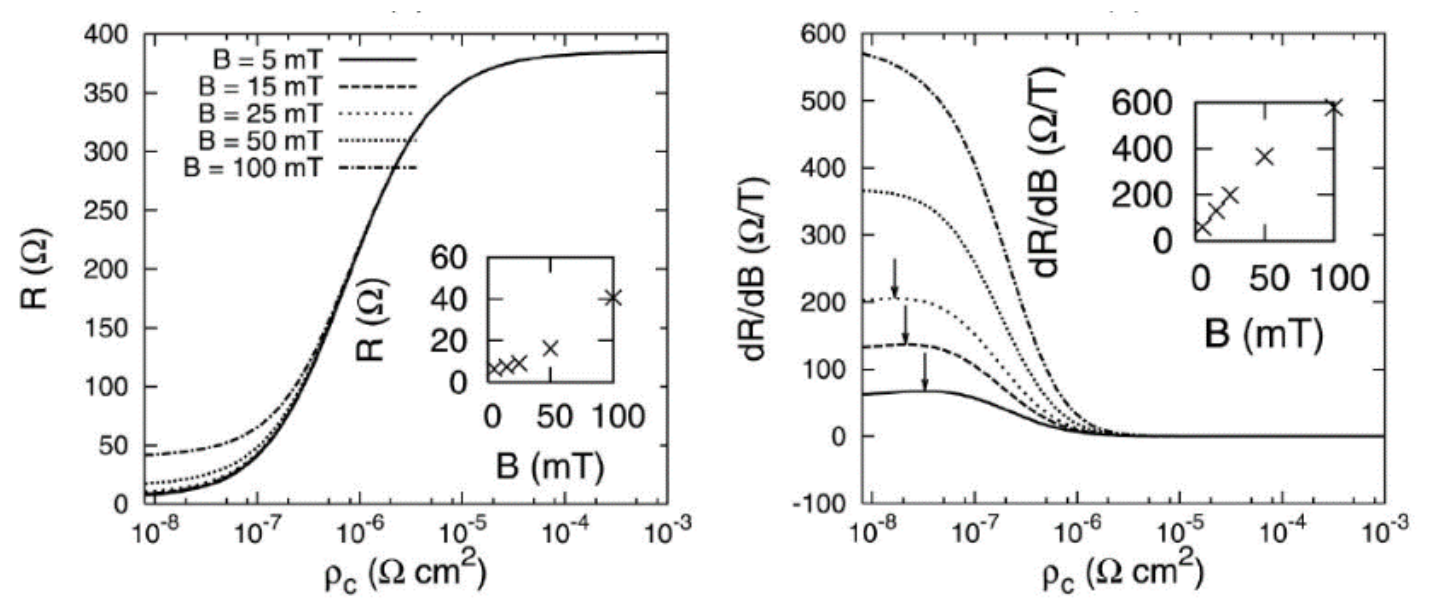

*Figure 3.13: Resistance (left) and sensitivity (right) as a function of the interface resistivity.*[78]

It should be noted that the critical value of the contact resistance depends on the geometry of the device, specifically the width of the semiconductor layer, with the trend that the critical value of the contact resistance increases as the width of the semiconductor decreases. Additionally, the authors also observe a shift in the peak sensitivity of the device as function of contact resistance for different widths of the semiconductor layer, i.e. the sensitivity is not highest at zero contact resistance but at some small but finite value of the contact resistance which varies with the width of the semiconductor layer.[78] These findings were consistent with previous experimental reports of EMR devices.[47,50,78]

Holz et al.[78] also estimated the contact resistance for the best case scenario of a high mobility 2DEG with the possibility of ballistic transmission of electrons via the Sharvin resistance:

$$R_{Sh}^{2D} = \frac{h}{2e^2} \cdot \frac{\pi}{k_F a_c}$$

where $k_F = \sqrt{2\pi n}$ is the Fermi wave number for a 2DEG system, $n_s$ is the sheet carrier density, and $a_c$ is the width of the semiconductor-metal interface. The specific contact resistance is thus given by:

$$\rho_c^{Sh} = R_{Sh}^{2D} \times a_c \times t_c$$

where $t_c$ is the thickness of the quantum well. For the values $a_c =$ 200 µm and $t_c = 4$ nm they estimate a realistic lower bound for the contact resistance of 8.5×10$^{-9}$ Ωcm$^2$.

The role of contact resistance in EMR devices was also studied experimentally by Möller et al on InGaAs/InAs/InGaAs quantum wells.[47] Several bar-shaped devices of varying feature sizes and contact quality were produced for the experiment. Figure 3.16 shows the performance of two devices with nearly identical zero-field resistances but different feature sizes (200×20 vs. 20×1.9 µm) and contact resistivities (1.6×10$^{-8}$ vs. 7×10$^{-8}$ Ω/cm$^2$). Despite being an order of magnitude smaller in physical size, the higher contact resistance in the second device resulted in an overall zero-field resistance that was similar to the larger device with a higher quality contact. At 1 T, the devices showed similar magnetoresistances of 1950% and 1290%, demonstrating that at high fields the properties of the semiconductor dominate the electronic flow and the measured magnetoresistance is thus mainly determined by the zero-field resistance. However, the slope of the response is significantly higher at small fields for the low contact resistance device suggesting that in the low field regime the interface resistance plays a crucial role in determining the effectiveness of the metal shunt.

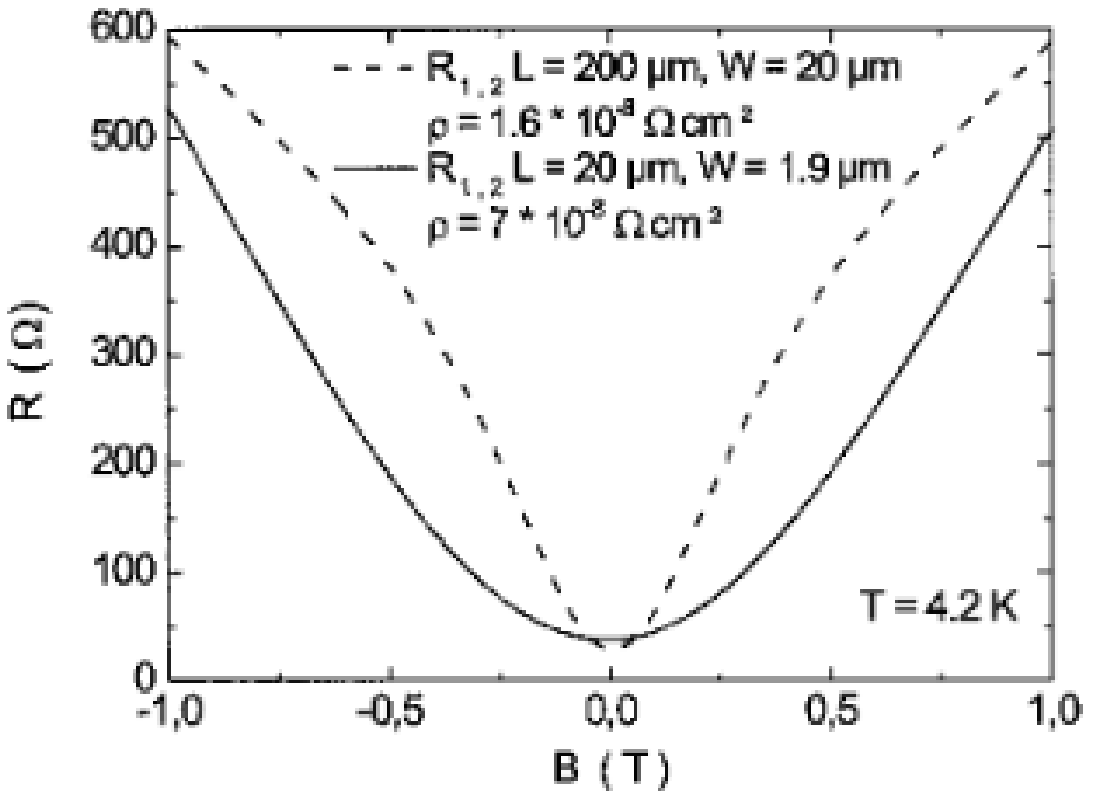

*Figure 3.14: Resistance as a function of magnetic flux density for two devices with different length scales and contact resistances.*[47]

The same study also examined the interplay between the semiconductor width and specific contact resistance in determining the zero-field resistance and sensitivity (see Figure 3.17). In these experiments all of the devices were 200 µm long but differed in terms of the semiconductor width and the quality of the ohmic contacts. For the devices with high quality contacts, the zero-field resistance decreases sharply as the semiconductor is narrowed since the current has to travel through a shorter length in the higher resistance material.

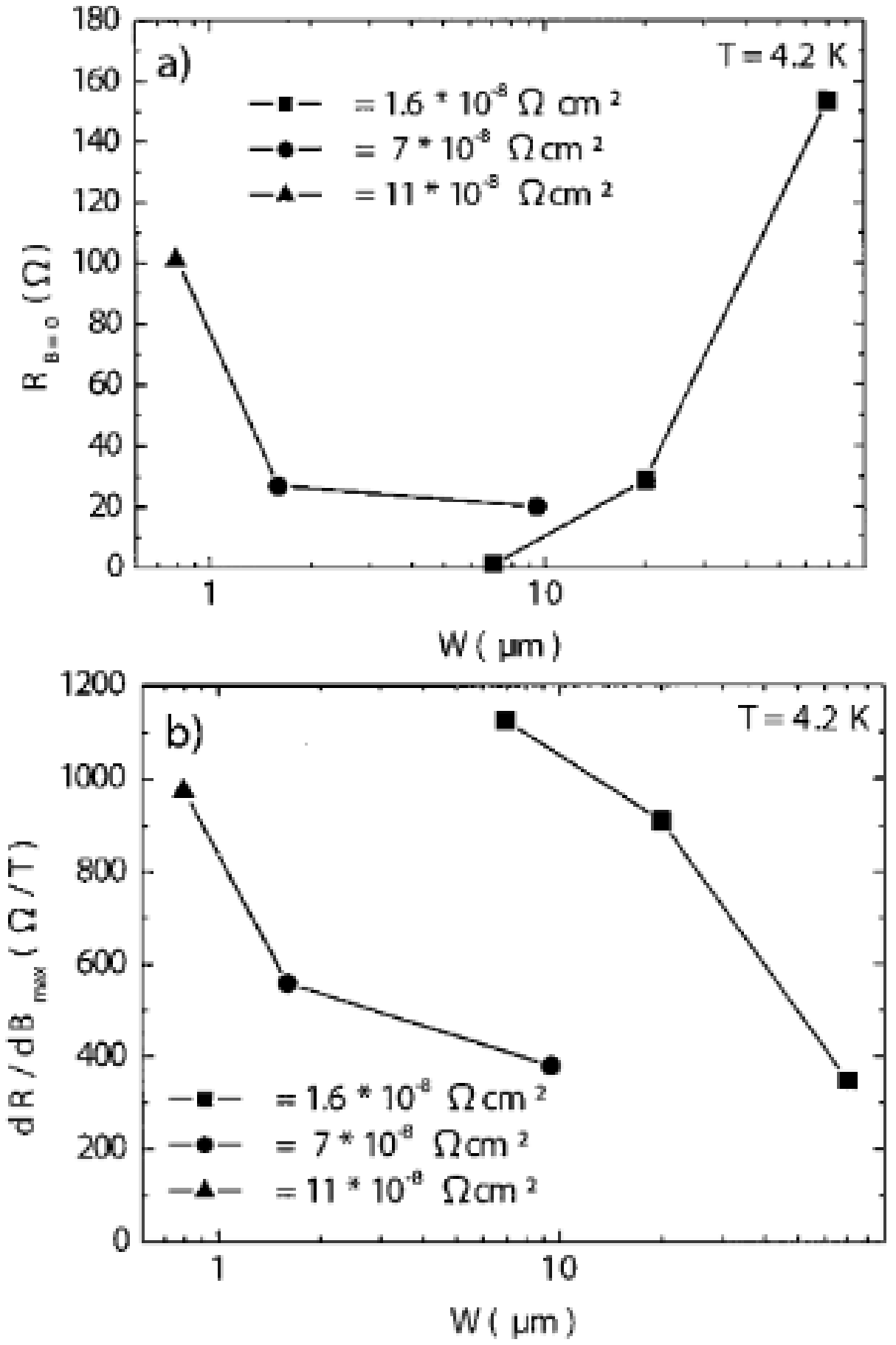

*Figure 3.15: Zero-field resistance (top) and maximum sensitivity (bottom) for devices with varying semiconductor widths (W) and specific contact resistivities.*[47]

The effect of the contact resistance can be clearly seen when comparing devices of similar widths but different contact quality. Despite the relatively small difference in size between the 7 and 9.6 µm-wide devices, the contact resistivity is 4.5 times higher in the latter. The resulting zero-field four-terminal resistance however is one order of magnitude greater for the device with lower quality contacts. Another important observation is that for the devices with an intermediate contact resistivity of $7 \cdot 10^{-8}\ \Omega\text{cm}^2$, the zero-field resistance remains roughly invariant when changing the width from 10 μm to below 2 μm. This suggest that the current passes efficiently into the gold shunt for all small semiconductor width so that the four-terminal voltage drop in the device center does not vary significantly. The resistance is, however, increased when the contact resistance is increased to $11 \cdot 10^{-8}\ \Omega\text{cm}^2$ .The authors also observed that the sensitivity is improved by decreasing both the semiconductor width and the contact resistivity.

It can be seen from the results detailed above that the contact resistance is a key parameter in determining the performance of EMR devices. Solin directly addresses the influence of contact resistance in his discussion of nanoscale scanning EMR magnetometers, stating that for applications where there is a need for a high spatial resolution this requirement may dictate the minimal size of the active area and choice of materials, thus rendering the specific contact resistance even more crucial.[5] In this case, the nanoscopic sizes may reduce the extraordinary magnetoresistance by lowering the efficiency of the metal shunt as well as increasing the noise by the increased resistance of the device. Despite the high mobilities which can be achieved with the InSb-based devices used for the nanoscale EMR sensors, Solin argued for using InAs for an improved contact resistance since the material naturally forms ohmic contacts.

## 3.2. Materials

In the previous section we examined the critical material parameters which influence EMR device performance. It was found that a low carrier density, high mobility, high contrast in shunt/semiconductor conductivities, and low contact resistance are all important for producing EMR devices with a high magnetoresistance. In this section we will detail how different materials have been used within literature to experimentally fabricate EMR devices. Table 1 in the introduction provide an overview of the active materials used in experimental EMR devices. The focus in the section below will be on the material properties, the fabrication methods employed, and how the choice of materials contributed to the results. While we focus only on thin films, heterostructures and 2D materials, care should also be taken when selecting a substrate material to minimize undesired leakage currents through the substrate.[3,5]

### 3.2.1. Thin Films

Thin films are a natural choice of material platform for EMR sensors as typically only the perpendicular component of the magnetic field affects their performance since electronic motion within these structures is more or less confined to a plane. In particular, thin films of the narrow bandgap III-V semiconductors InSb[1,2,4,6–9] and InAs[9–13] have attracted the attention of researchers in the field as they possess some of the highest known room temperature electron mobilities in bulk materials and well as low contact resistances to metal shunts.

InSb thin films can be prepared by various methods. They are typically grown on semi-insulating GaAs substrates due to its low conductivity and low cost, but the large lattice mismatch (14%) between the two materials produces a defect-rich region in the InSb close to the interface. The presence of defects causes electron scattering to occur and significantly lowers the mobility (ref). To counteract this, epitaxial InSb films with thicknesses exceeding 1 μm are used in order to display high carrier mobilities. All of the experimental works cited in this review reported InSb film thicknesses between 1 and 1.4 μm. As an alternative described in the next section, more complex thin film stacks can be used where buffer layers mitigate the mobility degradation form by the dislocation defects without the need for an active layer that is very thick.

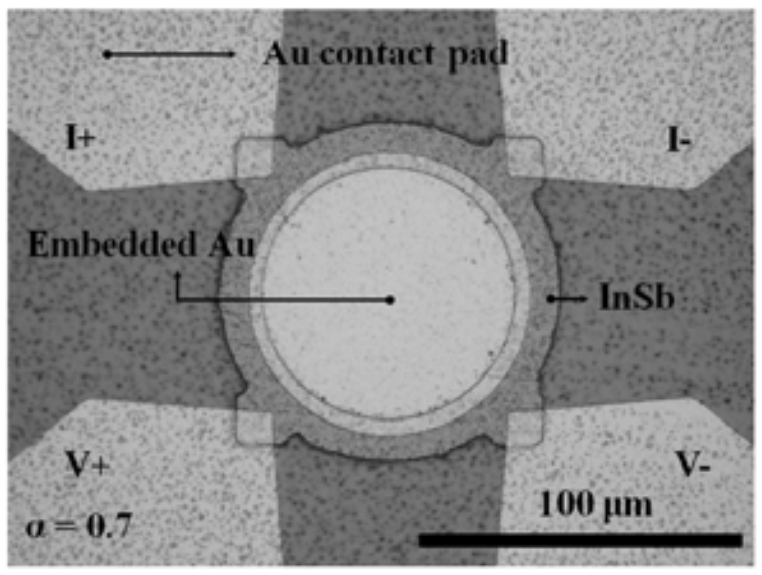

*Figure 3.16: SEM micrograph of an InSb vdP EMR device.*[6]

Single-crystalline films are preferred to further avoid scattering at grain boundaries. These epitaxial layers are grown via metal organic vapor phase epitaxy[1,2] or molecular beam epitaxy[4], which are used to produce EMR devices with high semiconductor mobility values of 45,500 and 38,000 $\text{cm}^2\text{V}^{-1}\text{s}^{-1}$, respectively. While epitaxial thin films can possess excellent quality, the necessary equipment and fabrication are, however, expensive. Cheaper, albeit polycrystalline, thin films have been grown using thermal[6] and flash[7–9] evaporation methods. Both techniques involve vaporizing pure sources of the constituent elements and allowing nucleation to occur on a target substrate. EMR devices with relatively high semiconductor mobilities of 20,000 $\text{cm}^2\text{V}^{-1}\text{s}^{-1}$ and 12,200 $\text{cm}^2\text{V}^{-1}\text{s}^{-1}$ can still be achieved with films prepared via flash evaporation,[8] and for thermal evaporation, respectively.[6]

Thin film mobility is an important parameter but the fabrication methods used to shape the films into devices can also be a key determinant of device performance. The effect of device properties on performance can be seen by comparing the results of several published experiments with similar values of α, around 0.7, at 1 T. Solin et al. fabricated a device made from an epitaxial thin film of InSb which showed a zero-field resistance of 0.08 Ω and a subsequent magnetoresistance of 25,000% at X T.[1] The conductivity of the thermally evaporated film prepared by Suh et al. was only around half that of the epitaxial film but the zero-field resistance of 1.5 Ω was approximately 30 times higher, resulting in a magnetoresistance of only 700% at X T.[6] To explain this discrepancy we can compare these two results to those from flash evaporated films prepared by El-Ahmar et al. which were shunted with two different methods: using either edge contacts or top contacts (see Figure XX).[8] Though the InSb film used for the edge-contacted device had a conductivity that was 40% higher than the film used in the planar device, the edge-contacted device showed a zero-field resistance of 0.9 Ω and a magnetoresistance of 800% at X T compared to values of 0.05 Ω and 22,500% at X T for the top-contacted device.

This comparison directly shows the importance of contact resistance and device construction on performance. The thermally evaporated and edge-contacted devices showed similar transport parameters when compared to the top-contacted device, yet the latter vastly outperformed the former and approximated the behavior of the epitaxial device. While the metal deposition methods used by Solin et al. may have been able to produce lower contact resistances, El-Ahmar et al. were able to reproduce similar results by increasing the contact area with a top-contacted configuration. Flash evaporated films were also produced by Mansour et al.[9] and though no characterization data is provided, a high zero-field resistance of 74 Ω and low magnetoresistance of 50% suggests the presence of a high contact resistance.

While the high mobility of InSb makes it an attractive material for EMR applications, it should be noted that a Schottky barrier may form when it is contacted with metals. As the thermal energy to overcome the barrier is reduced at low temperatures, problems may arise especially for cryogenic applications. InAs is often used for EMR devices because, despite its lower mobility, it tends to form ohmic contacts with metals.[5] Preparation of InAs thin films is quite similar to that of InSb. Thicknesses of over 1 μm are also required for accommodating the lattice mismatch with GaAs; and growth techniques such as molecular beam epitaxy[10–13] and flash evaporation[23] have been used to produce films. Contact resistances as low as $1\times10^{-7}$ Ωcm$^2$ have been demonstrated in MBE grown films.

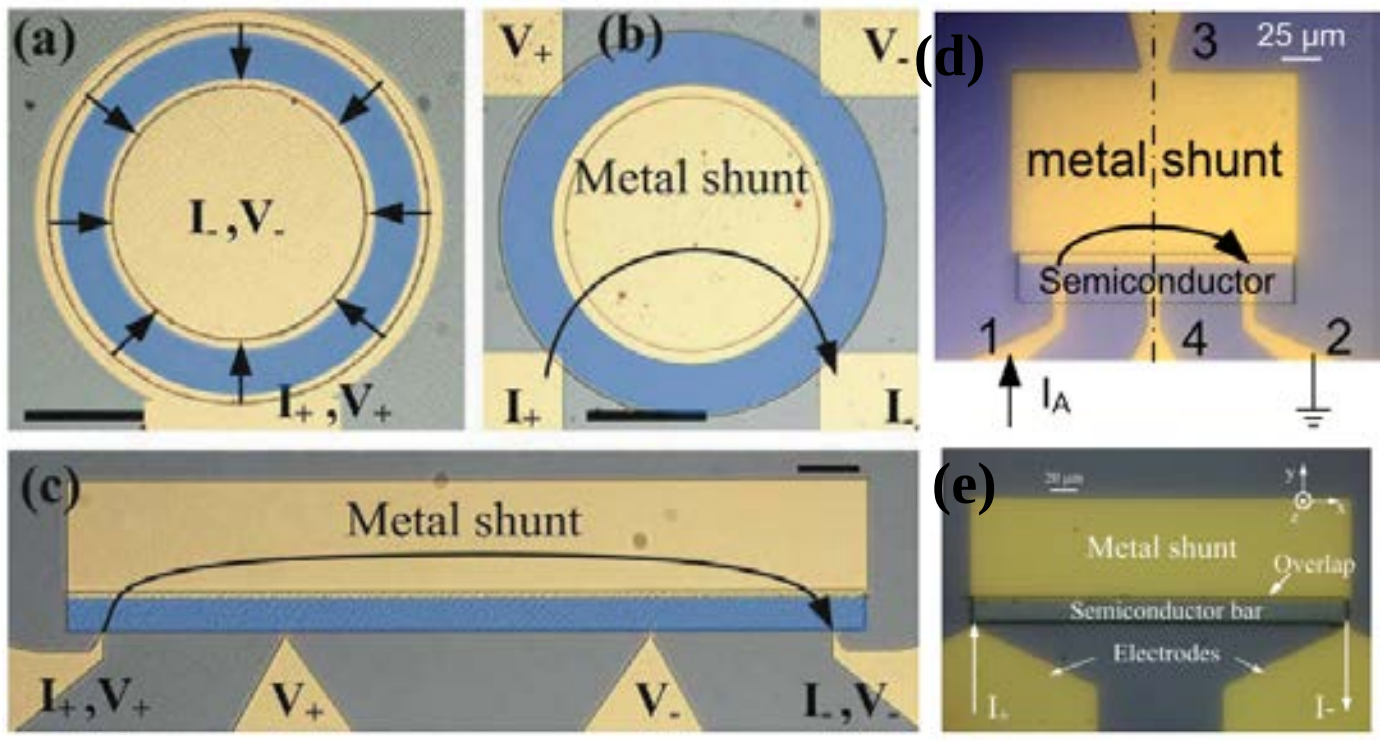

*Figure 3.17: EMR devices made from InAs thin films with different geometries and contact configurations. a) Corbino disc[13] b) shunted vdP disc[13] c) four-terminal bar[13] d) three-terminal bar[11] e) two-terminal bar[17]*

As with most semiconductors, the electron mobility of InAs has a strong temperature dependence. Sun et al. studied the effect of temperature on epitaxial InAs films.[13] The mobility was found to increase from around 8,000 cm$^2$V$^{-1}$s$^{-1}$ at room temperature up to a maximum value of 25,000 cm$^2$V$^{-1}$s$^{-1}$ at 75 K, below which the mobility decreased (see Figure 3.18). Above 75 K, scattering is dominated by lattice vibrations but below this temperature, scattering with charged impurities and lattice defects become the most significant factors. The performance of their two-terminal bar-shaped EMR devices followed the same trend, with magnetoresistance values increasing with decreasing temperatures until around 100 K. The highest magnetoresistances were reported for the shunted vdP geometry, with values at 1 T around 9,000% at room temperature and 30,000% at 100 K.

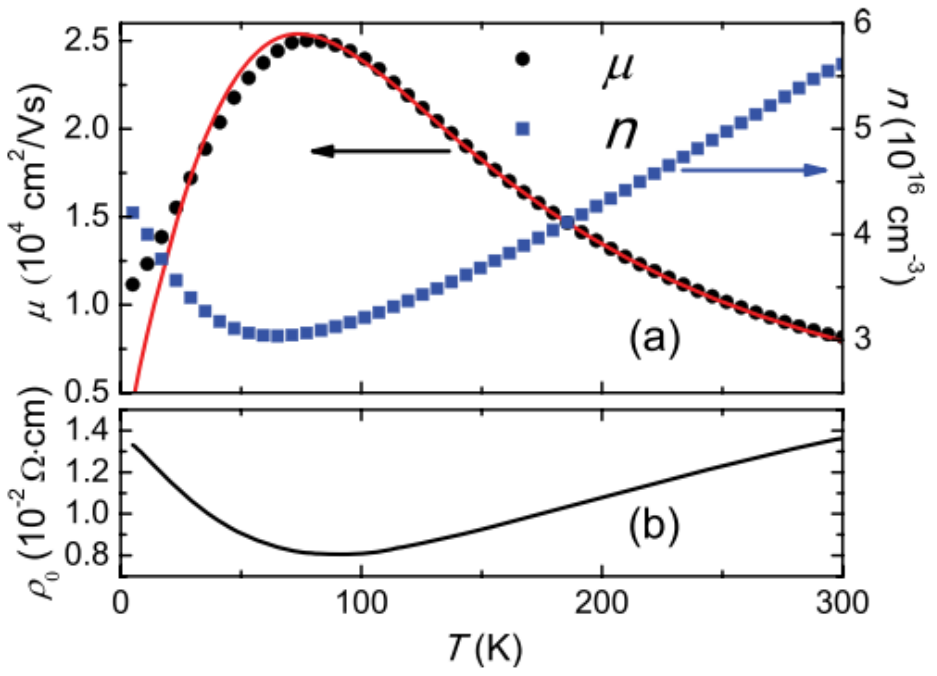

*Figure 3.18: a) Mobility and carrier density and b) resistivity of InAs epitaxial films as a function of temperature.*[13]

Interestingly, Sun et al. reported the observation of intrinsic magnetoresistance in samples which were shaped into unshunted vdP discs (see Figure X).[13] The authors suggest that these results can best be explained by considering a two-band conduction model, as the presence of two carrier species has been demonstrated to produce intrinsic magnetoresistance.[79] Most of the materials discussed in this review are considered to possess only a single conduction band with only one carrier species . However, when measuring the Hall coefficient for their InAs samples the authors noted that the value was found to decrease with increasing magnetic fields and eventually saturated, instead of maintaining a constant value as would be expected for a material with a single conducting band.[13]

The presence of a second conduction band can result from the formation of a charge accumulation layer at the surface of thin films as a consequence of native surface defects. This surface layer acts as a parallel conduction channel, with a carrier density and mobility which is different from that of the bulk. The thickness of the conducting layer was estimated by using the Debye length to be around 33 nm at room temperature. A nominal mobility and carrier density for the entirety of the thin film was measured to be 8,160 cm$^{-2}$V$^{-1}$s$^{-1}$ and $5.6\times10^{16}$ cm$^{-3}$. In a two-band model, which accounts for the contributions from both bands, the averaged zero-field mobility and carrier density for the device is given by:

$$\mu_{avg}(0) = \frac{\mu_b^2 n_b d_b + \mu_s^2 n_s d_s}{\mu_b n_b d_b + \mu_s n_s d_s}$$

$$n_{avg}(0) = \frac{(\mu_b n_b d_b + \mu_s n_s d_s)^2}{d(\mu_b^2 n_b d_b + \mu_s^2 n_s d_s)}$$

where $\mu_s$ and $\mu_b$ are the carrier mobilities; $n_s$and $n_b$ are the carrier densities; $d_s$ and $d_b$ are the thicknesses of the surface and bulk layers respectively.

The mobility and carrier density were deduced to be 3,360 cm$^{-2}$V$^{-1}$s$^{-1}$ and $3.93\times10^{18}$ cm$^{-3}$ for the surface layer and 16,000 cm$^{-2}$V$^{-1}$s$^{-1}$ and $2\times10^{16}$ cm$^{-3}$ for the bulk.
The authors observed that at low fields the magnetoresistance of unshunted vdP discs followed a quadratic dependence, but at higher fields the response became linear. The crossover point between the quadratic and linear regimes occured when $\mu B = 1$ . The explanation for this behavior can be seen in the analytical expression for the intrinsic magnetoresistance of a material with two bands:

$$\frac{\Delta\rho}{\rho_0} = \frac{(\mu_b - \mu_s)^2 G_s G_b B^2}{(G_b - G_s)^2 + B^2(\mu_s G_b + \mu_b G_s)^2}$$

In the low field regime $B^2$ is small and the field independent term dominates the denominator, resulting in a quadratic dependence on the field due to the $B^2$ term appearing in the numerator. At high fields, the field independent term in the denominator becomes negligible and the magnetoresistance as a function of field strength saturates.[13] Regardless of the choice of material, one should take into consideration the possibility of multiple conducting channels when using thin films as they can influence the overall transport behavior within the device. Multiple conducting channels or species can produce an intrinsic magnetoresistance effect in materials that otherwise would not display this behavior.

A major impediment to the widespread adoption of EMR sensors based on the material systems described in most of the literature is the lack of commercially available systems capable of mass producing them. EMR devices were produced using materials and processes that are standard to the silicon-based semiconductor industry by Troup et al.[21] The devices were fabricated by etching a doped Si wafer into mesas with a bar geometry and then sputtering Ti onto the sample (see Figure 3.19). The shunt region was created by rapid annealing in a Ni atmosphere at high temperature which caused any unmasked areas to form a stack where the bottom 70 nm were a metastable state of C49 $TiSi_2$, the next 10 nm were $Ti_5Si_3$, and the top 20 nm were Ti/TiN. The Ti/TiN layers were selectively etched, leaving behind only the titanium silicide phases and the Si. The remaining C49 phase is a material approximately 24 times more conductive than the *n*-doped Si.

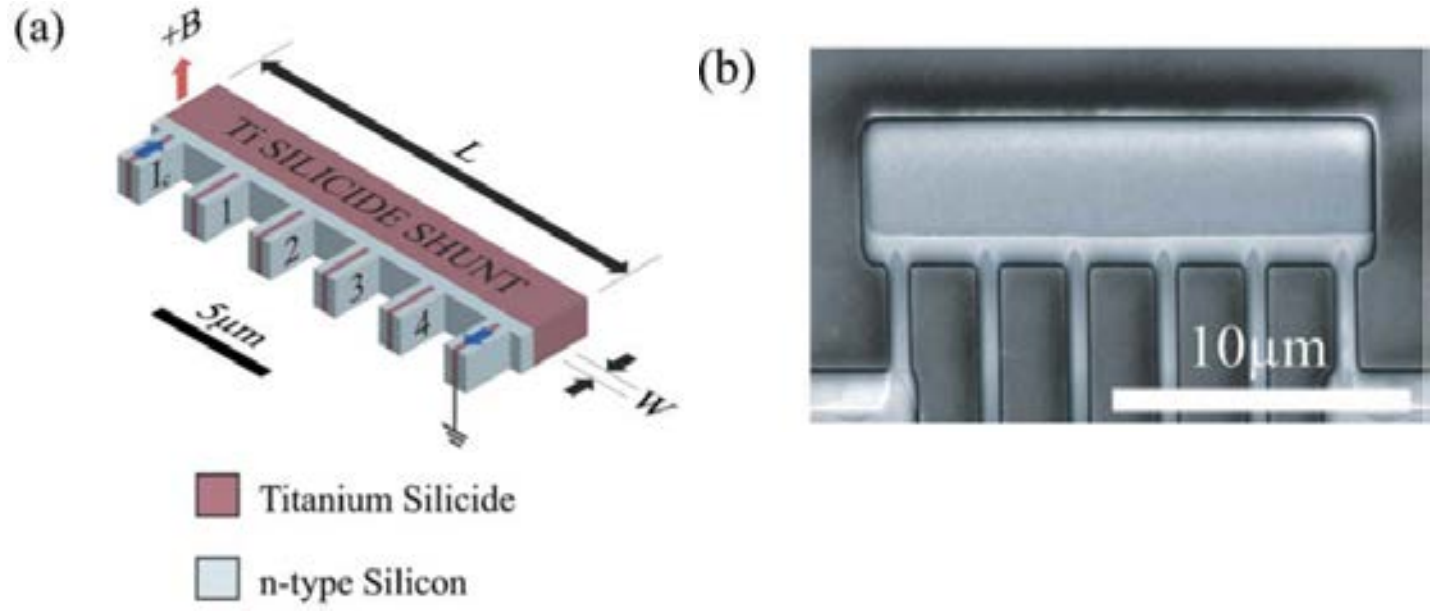

*Figure 3.19: a) Diagram and b) SEM micrograph of a Si/TiSi EMR device.*[21]

The highest magnetoresistance for this device was only 15.3% at X T, a value much lower than what was observed in other material systems. However, it is still interesting to note that a measurable effect can still be achieved in systems with relatively low values of the mobility and conductivity contrast between the shunt and semiconductor.

### 3.2.2. Two-Dimensional Electron Gasses (2DEGs)

As shown in previous sections, materials with high electron mobilities are essential for realizing devices with a strong EMR effect. The highest recorded carrier mobility values have been observed in two-dimensional electron gasses (2DEGs). Due to their exceptional mobilities these material systems are natural candidates for EMR applications.

In a 2DEG the motion of electrons is tightly confined to a 2D plane as the energy for motion in the third dimension is quantized. Typically, this confinement potential is is formed at a surface or introduced through bandgap engineering of semiconductor heterostructures. For the latter, materials are chosen such that they possess similar lattice spacing to minimize defects but different bandgaps. The bandgap mismatch can form discontinuities in the conduction and valence bands which can confine carriers in quantum wells in the vicinity of the interface (see Figure 3.20).[80] It further enables the donors to be spatially separated from the charges in the quantum well, which further reduces the scattering and enhances the mobility.

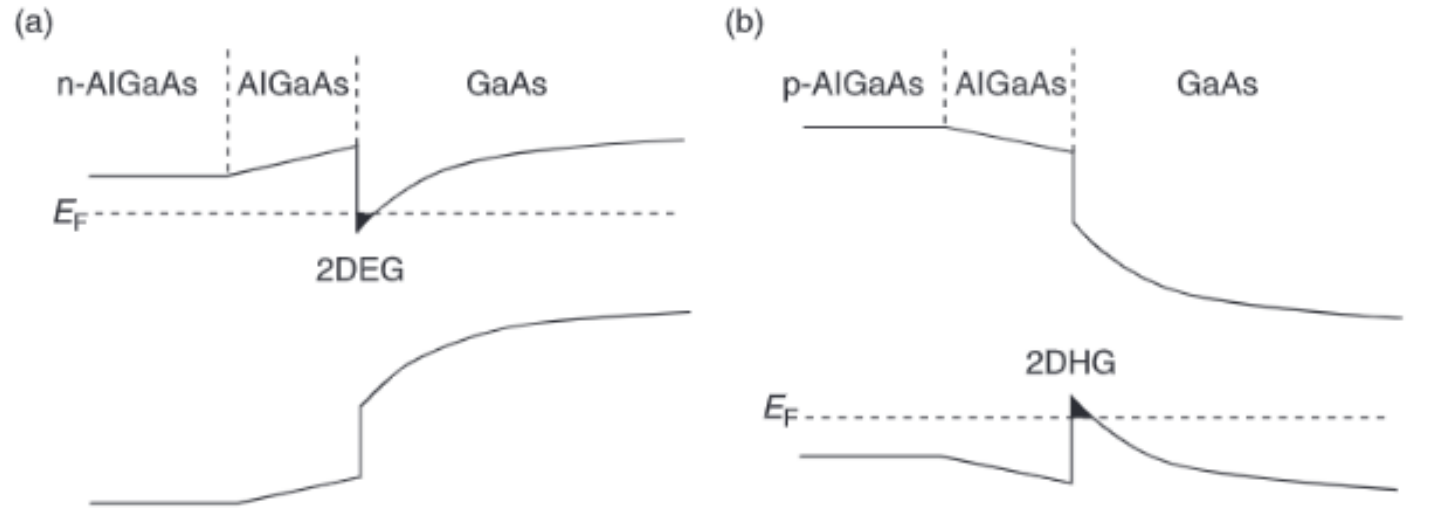

*Figure 3.20: Band diagrams for a GaAs/AlGaAs heterojunctions with quantum wells hosting a) a 2 dimensional electron gas in an n-doped system and b) a 2 dimensional hole gas in p-doped system.*[80]

Experimental devices based on 2DEGs started being produced shortly after Solin first discovered the EMR effect.[3,5,14–20] Devices fabricated from 2DEGs have been reported with material combinations including InSb embedded in InAlSb[3,5], InAs embedded in either InGaAs[15–17], AlSb[18,19], or GaSb[14], in addition to the interface between InGaAs and AlGaAs.[20] In order to ensure epitaxy and high mobility, the heterostructure layers were all grown using MBE and then turned into EMR devices using etching and metal deposition to form side contacts.

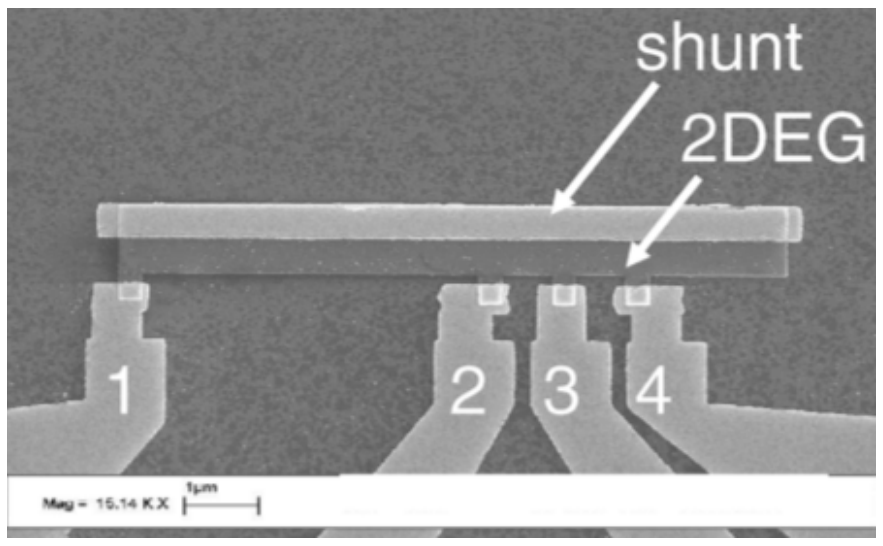

*Figure 3.21: A four-terminal bar-shaped device fabricated from an InAs/InGaAs quantum well.*[18]

The electron mobility in the InSb/InAlSb heterostructures at room temperature (XX,XXX $cm^2V^{-1}s^{-1}$) was found to be lower than in bulk films of InSb (XX,XXX $cm^2V^{-1}s^{-1}$), despite both being prepared with MBE.[3] However, for InAs-based 2DEGs, several groups report electron mobility values exceeding that of thin films, with room temperature values as high as 21,000 $cm^{-2}V^{-1}s^{-1}$ in InAs/InGaAs[16] and 31,000 $cm^{-2}V^{-1}s^{-1}$ in InAs/GaSb.[14] At 4.2 K, the electron mobility reached values as high as 150,000 $cm^{-2}V^{-1}s^{-1}$ and 194,000 $cm^{-2}V^{-1}s^{-1}$ for InAs/InGaAs and InAs/GaSb, respectively. Due to the formation of ohmic contacts, contact resistances in InAs/InGaAs devices as low as $10^{-8}$ $\Omega cm^2$ have been reported.[15,17]

Yet despite the high mobility values and low contact resistances observed in 2DEG EMR devices, there are few notable results with regards to device performance. Solin et al. showed an appreciable magnetoresistance effect at low fields in an InSb/InAlSb device with a 35% increase in device resistance at 0.05 T.[3] Möller et al. reported magnetoresistance values as high as $10^5$% in InAs/InGaAs devices at 1 T and X K. When the devices were tested at room temperature the magnetoresistance was found to be only 1,900% at 1 T.[15] Though many of the devices possessed relatively high sensitivities between 500-1000 Ω/T,[5,14,15,18,20,81] large zero-field resistances led to low magnetoresistances. For example, the zero-field resistances at room temperature were as high as 30 Ω in the device described by Möller et al.[15], or 430 Ω in the case of the InAs/AlSb device described by Boone et al.[18] These values are orders of magnitude greater than what was seen in the best performing thin film devices.

Although producing high-performing EMR devices from 2DEG systems has proven to be quite challenging, their low dimensionality and high carrier mobility allows for quantum mechanical phenomena to be observable at cryogenic temperatures. The clearest example can be seen in the results from Kronenworth et al., who compared the behavior of a bar-shaped device and an unshunted vdP disc fabricated from InAs/InGaAs heterostructures at 4 K.[46] Above 1 T strong Shubnikov-de Haas (SdH) oscillations could be seen in both devices (see Figure 3.22). The SdH effect is characteristic of high-mobility conductors at low temperatures and occurs due to the quantization of the energy levels of electrons in the presence of a magnetic field. The resistance curves for the hybrid and pure semiconductor devices are comparable at high fields, lending credence to the position argued by Solin that at high fields the shunt is no longer effective and the behavior of the device is solely determined by the properties of the semiconductor. The EMR effect is clearly visible in the bar-shaped device where the resistance steadily increases in the low field regime whereas the resistance of the unshunted vdP device is nearly constant in this region. The onset of quantum mechanical behavior at high fields should therefore be considered when designing devices with high electron mobility as the resistance of the device may be dominated by non-EMR phenomena.[46]

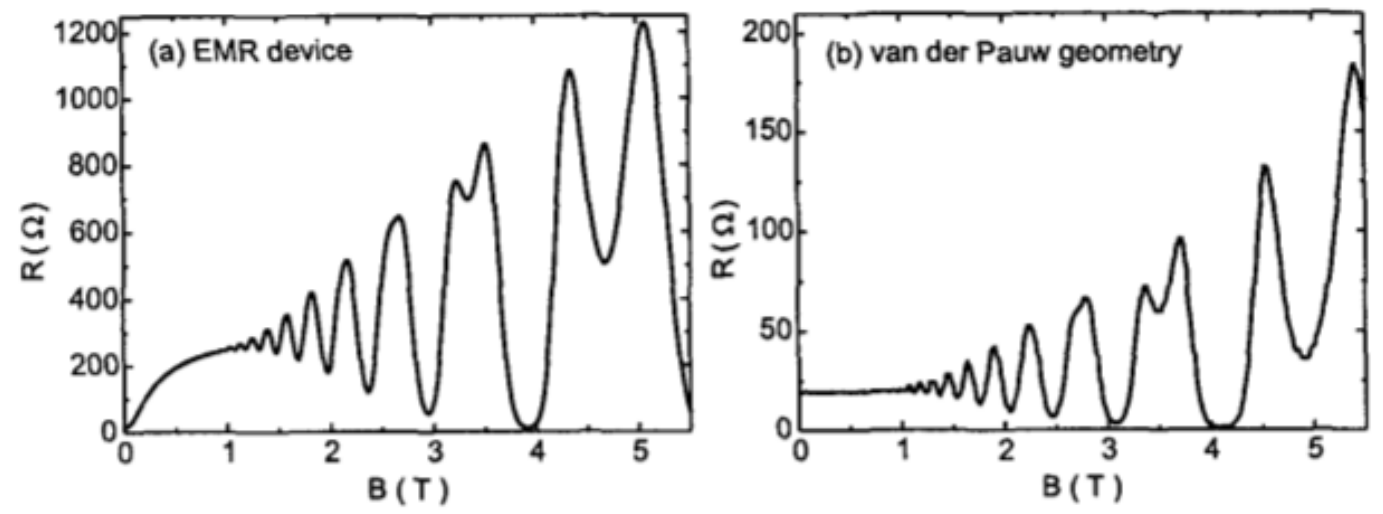

*Figure 3.22: Low-temperature behavior of an EMR device (left) and unshunted vdP disc (right) made from InAs quantum wells as a function of magnetic field.*[46]

Boone et al. calculated the mean free path of electrons in InAs/AlSb heterostructures as a function of temperature.[18] It is interesting to note that although the mean free path of electrons was calculated to range from 80 to 325 nm as the temperature was lowered from 300 to 5 K, the EMR response increased monotonically with decreasing temperature. At low temperatures, the mean free path is larger than the smallest features of the device and thus a significant portion of electronic transport may occur in the ballistic or quasi-ballistic regime. Despite this, it is promising to observe that the magnetoresistive effect continues to persist into the ballistic regime.

### 3.2.3. van der Waals Materials

Like 2DEGs, van der Waals materials are natural candidates for EMR sensors because of the exceptional electronic properties found in some of these materials. Among vdW materials, graphene in particular has long been seen as an attractive candidate for EMR devices due to its intrinsically high carrier mobility and its ability to reach extremely low carrier densities.

The first documented use of graphene in EMR sensors was reported by Pisana et al. in 2010.[22,23] Since then several works have been published describing EMR devices made from graphene.[24–28] Graphene has a thickness of only one atomic layer, and its transport properties are easily influenced by local conditions as the entire material is essentially a surface. Due to the lack of electronic screening, graphene is highly susceptible to charged impurity scattering and electronic doping. Even the choice of substrate can have a large influence, as phonons and impurities in the substrate material can scatter electrons in graphene.[82]

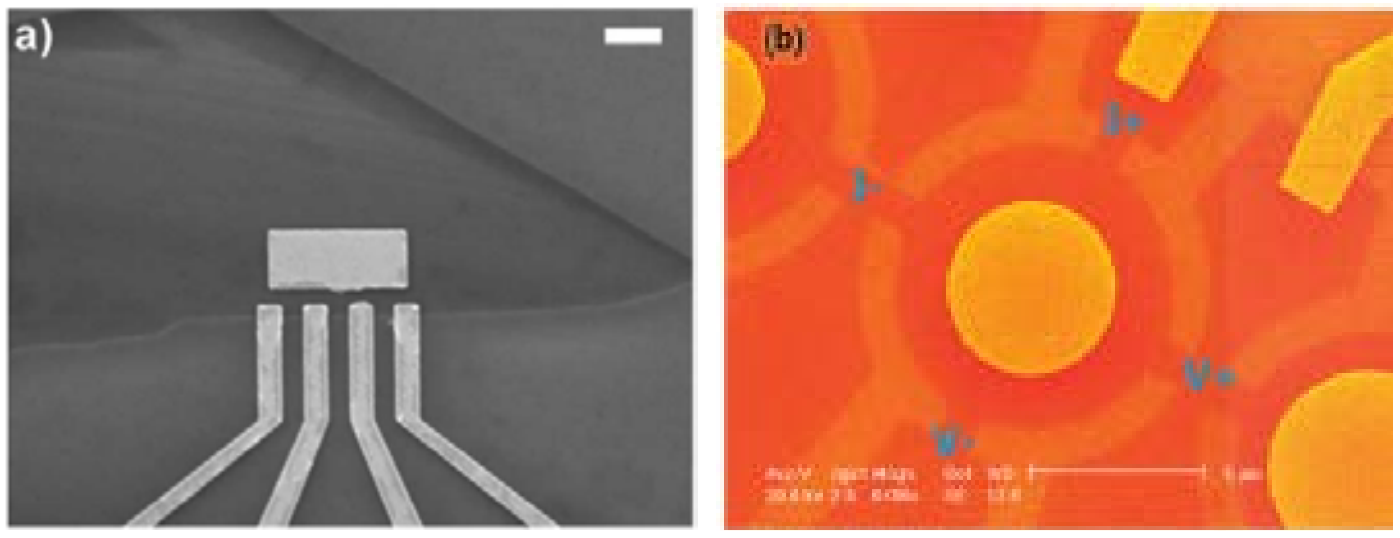

*Figure 3.23: SEM micrographs of a planar bar EMR device on a graphene flake (left)*[22] *and of a vdP disc graphene device (right).*[24]

The first graphene EMR devices were made by exfoliating graphene layers from graphite and depositing them onto $Si/SiO_2$ substrates.[22–24] Graphene-on-oxide systems have a relatively low electron mobility, typically ranging from 1,000 to 10,000 $cm^2V^{-1}s^{-1}$ due to strong scattering from optical phonon modes in the $SiO_2$. Pisana[22,23] and Lu[24] also reported mobilities of 2,500 and 5,000 $cm^2V^{-1}s^{-1}$ in graphene devices on $Si/SiO_2$ substrates respectively.

The electronic properties of exfoliated graphene can be exceptional under certain conditions, but the difficulty of their acquisition and the small size of obtained flakes makes the technique impractical for mass-production. Graphene can be grown over large areas using chemical vapor deposition (CVD), producing a material which is more amenable to serial device manufacturing, but the resultant films may contain many defects and grain boundaries[83] that can negatively affect electronic transport. Friedman[25] and El-Ahmar[24] both fabricated EMR devices using CVD graphene. Friedman produced a graphene on $Si/SiO_2$ device with an electron mobility of 1350 $cm^2V^{-1}s^{-1}$, similar to the 1200 $cm^2V^{-1}s^{-1}$ of the graphene on SiC device reported by El-Ahmar.

Recently, a significant improvement in the graphene device quality was achieved by encapsulating graphene in flakes of hexagonal boron nitride (hBN).[84] hBN is an ideal substrate for graphene for several reasons: First, it is an atomically flat, wide bandgap semiconductor where its surface optical phonon modes have high energies which prevents their interaction with electrons in the adjacent graphene. Second, there is a very small lattice mismatch between the two materials. Third, it is inert and possesses a low density of of charged impurities. The hBN protects graphene from extrinsic disorder from sources such as water, adsorbed hydrocarbons, dangling bonds and trap states, leading to exceptional transport properties in encapsulated graphene such as micron-scale ballistic transport at room temperature[82,85] and a 2 or 3 orders of magnitude larger mobility than that of graphene-on-oxide systems. In encapsulated graphene devices, edge contact is made between graphene and metal, offering a lower contact resistance than top contacts.[85] This method has been used in other magnetometery applications; for example Hall bars with magnetic field resolutions as low as 50 $nT/\sqrt{Hz}$.[86] For EMR devices, Zhou et al. produced encapsulated graphene devices with a mobility of approximately 80,000 $cm^2V^{-1}s^{-1}$.[27]

Encapsulation is one method to preserve the properties of graphene, but other strategies can be used to avoid impurities and prevent interactions with the substrate. Suspending graphene over the substrate is another technique which has been commonly employed to isolate graphene. Kamada et al. produced suspended graphene Corbino discs by etching away a sacrificial layer of resist and demonstrated mobility values of around 100,000 $cm^2V^{-1}s^{-1}$.[28]

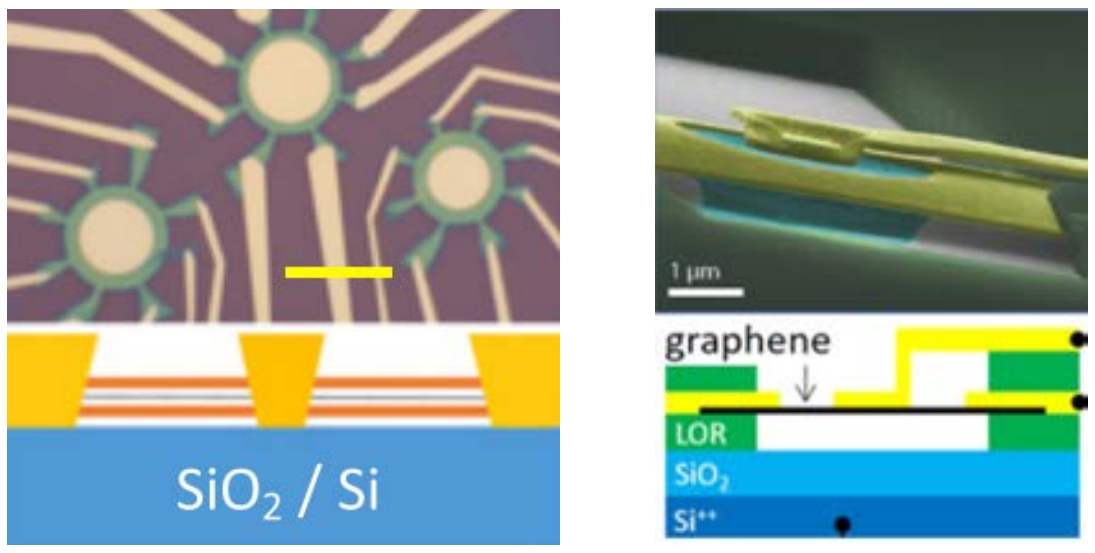

*Figure 3.24: SEM micrograph and schematic of shunted vdP device fabricated from encapsulated graphene (left).*[27] *False color image and schematic of a Corbino disc device made with suspended graphene (right).*[28]

The effect of the local environment on the mobility and subsequent magnetoresistance can be directly observed in the performance of the various devices. For example, Lu et al. reported the best graphene-on-oxide devices with a magnetoresistance on the order of $10^4$% at 8 T.[24] On the other hand, the encapsulated graphene device presented by Zhou with the same filling factor showed a magnetoresistance at the same magnetic field strength on the order of $10^7$%.[27] Designing sensors with vdW materials requires careful consideration of the device architecture and how the environment around the active graphene layer will affect electronic transport.

One of the key features of the electronic behavior of graphene is that the conduction and valance bands meet at Dirac points, thus making it a semi-metal. Both holes and electrons can be the majority carriers in graphene depending on the position of the Fermi energy relative to the Dirac point as shown in Figure 3.25. The Fermi energy can be modified through the application of a gate voltage and thus determine the number and sign of the dominant charge carriers. At the Dirac point itself the conductivity of graphene reaches a minimum and there are equal numbers of holes and electrons (charge-neutrality).[87] Natural defects and local potential fluctuations cause short-range disorder and result in the formation of electron and hole puddles which ensure that there is always a non-zero carrier density above 0 K.[59]

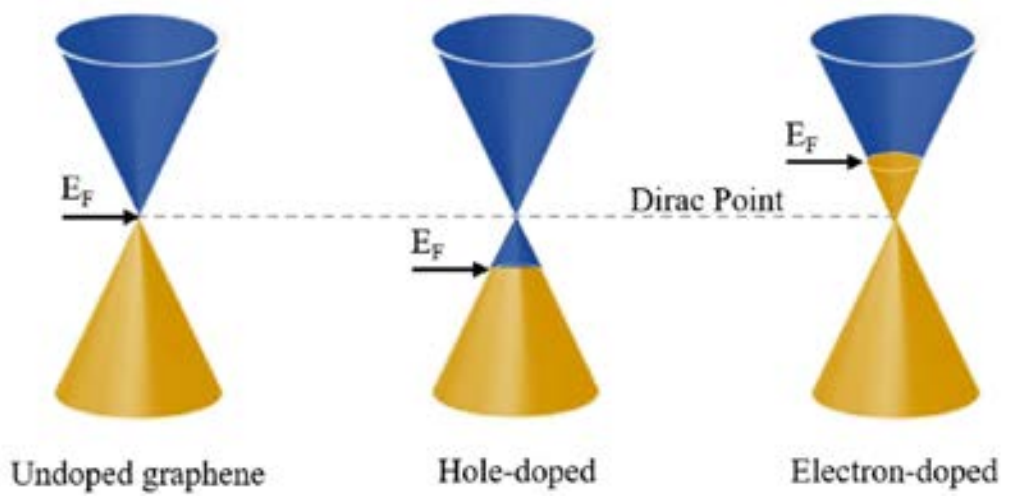

*Figure 3.25: Diagram of the linear dispersion with conduction and valence bands meeting at the Dirac point.*[87]

Graphene can show an intrinsic magnetoresistance by the presence of these electron-hole puddles.[88] Lu et al. demonstrated this effect by fabricating vdP discs made of pure metal or pure graphene. The metal disc showed a magnetoresistance of 5% at 9 T, which is consistent with the theoretical prediction for a pure metal. In contrast, the graphene device showed a magnetoresistance of 300-500% at 9 T, but only near the charge-neutrality point, otherwise no magnetoresistance was observed.[24] The intrinsic magnetoresistance that arises in graphene should be considered when operating sensors near the charge-neutrality point.

Charge-neutrality can be located by scanning the back gate voltage and identifying where the lowest conductivity occurs. If the Dirac point is found far away from a back gate voltage of 0 V it could indicate a strong presence of extrinsic dopants. This can be seen in Figure 3.26 which shows the results of the graphene-on-oxide devices reported by Pisana[14] and Friedman[25] where the minimum conductivity in shunted vdP EMR devices was found at back gate voltages in the range of 20-30V.

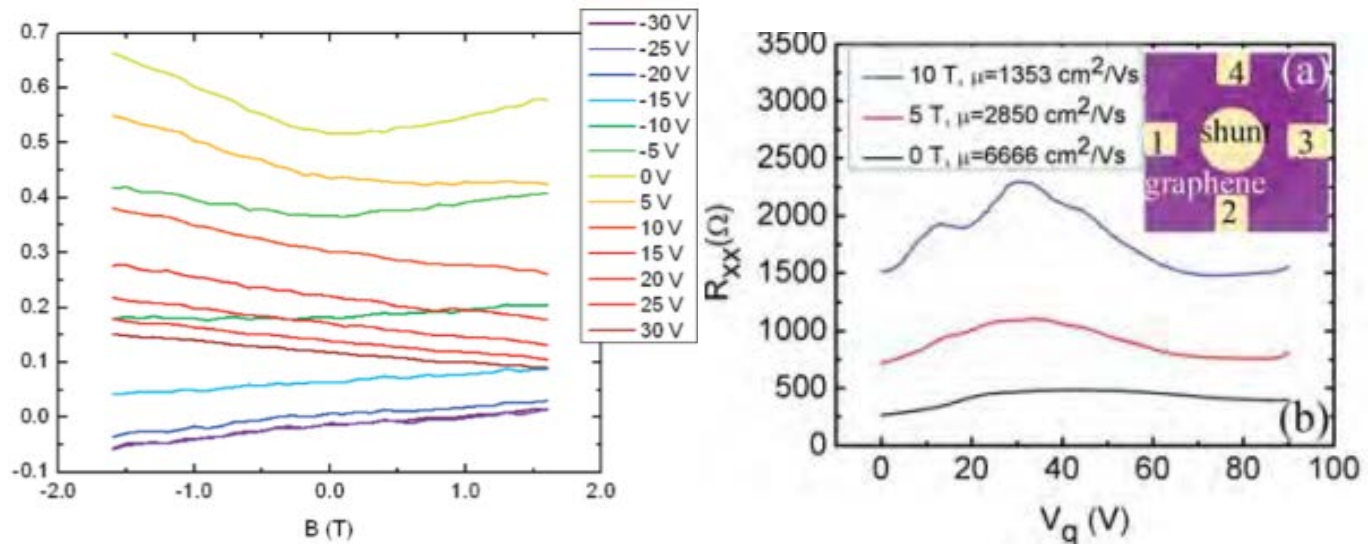

*Figure 3.26: The voltage drop in a shunted graphene vdP disc as a function of magnetic field for different back gate voltages. The backgate voltages are reported as the voltage difference from the Dirac point.*[14] *Resistance of the EMR device at three different magnetic fields as a function of back gate voltage. Here, the backgate voltages are not subtracted from the voltage resulting in charge neutrality.*[25]

Pisana et al. examined the effect of the gate voltage on the measured potential difference across the voltage electrodes in a shunted vdP EMR device. At large gate voltages and high carrier densities the measured potential difference varied linearly as a function of magnetic flux density. This linear behavior was attributed to the Hall effect dominating the response of the device in this regime. Close to charge neutrality, however, the response became quadratic (see Figure 3.26). The authors propose two possible mechanisms that explain this behavior: either that the presence of electron and hole puddles give rise to an intrinsic magnetoresistance in the graphene, or that the effectiveness of the shunt increases due to the higher resistivity of graphene at the charge neutrality point. Both of these mechanisms should produce a quadratic response according to theory.[14]

The Dirac point can also be observed in the high-field data from Friedman (see Figure 3.26 and Figure 3.30). At low magnetic fields, electronic transport occurs mostly through the gold resulting in an overall resistance which is low and has only a weak dependence of the applied gate voltage (Figure 3.26b). At higher fields, the current is expelled from the shunt and the overall behavior of the device is dominated by the properties of the graphene. In this regime, the resistance of the device varies significantly as a function of gate voltage, with the peak in resistance indicating the location of the charge neutrality point.[25]

Moktadir et al. developed simulations which included electron-hole puddles and indicated that homogeneous graphene is expected to yield a better EMR response.[59] This suggests better performance away from the charge-neutrality point. However this conflicts with experimental results reported by Lu[24] and Zhou[27] which demonstrated that graphene-based EMR devices function best when operated near charge-neutrality. Lu et al. showed that as the back-gate voltage is moved away from the charge-neutrality point, the carrier density of the graphene increases leading to a decrease in the ratio of $\sigma_m/\sigma_g$. The decreased conductivity contrast in the device results in a decrease in both the magnetoresistance and sensitivity (see Figure 3.27).[24]

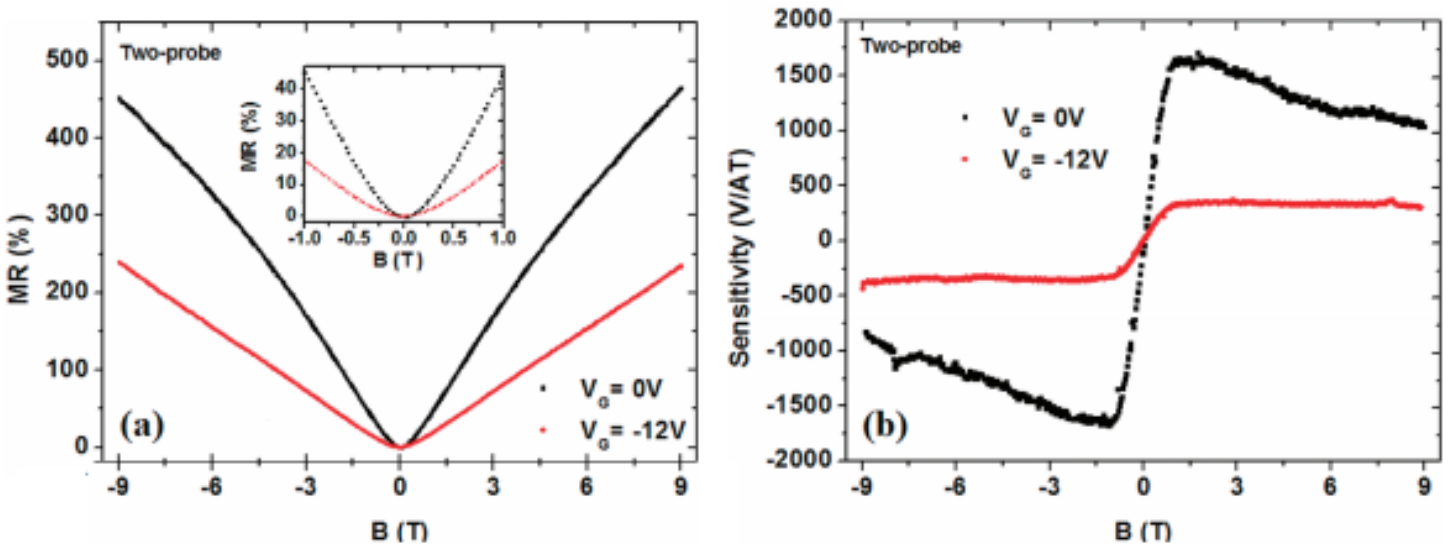

*Figure 3.27: Magnetoresistance (left) and sensitivity (right) for a graphene device tested in the two probe configuration at two different back gate voltages.*[24]

Zhou et al. showed how even small deviations in the back-gate voltage away from charge-neutrality can have large effects on the magnetoresistance (Figure 3.28a). Since the measured resistances for especially the positive magnetic fields for all three back gate voltages were similar (Figure 3.28b), the likely explanation is that the gate voltage can have a significant effect on the zero-field resistance, especially when operating near the charge-neutrality point of graphene.[27]

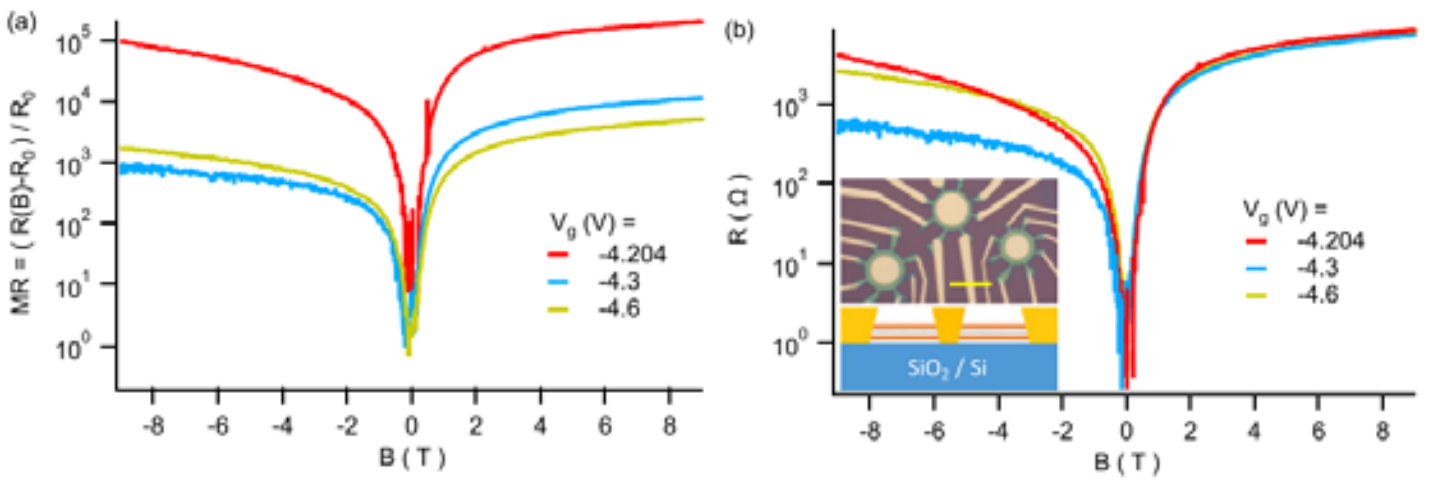

*Figure 3.28: a) Magnetoresistance ratio and b) resistance as a function of magnetic field for a graphene device at three back gate voltages near the charge neutrality point.*[27]

Contact resistance is also an important factor to consider when designing graphene devices. Since graphene is a single atomic sheet, the architecture of the device can determine the type of bonding that occurs between the metal contacts and the graphene. In the work by Pisana et al.[14] metal contacts were deposited directly onto the graphene flake in a top-contact configuration Lu et al.[24] etched the graphene using oxygen plasma into the desired shape and deposited the metal into the corresponding areas (see Figure 3.27). Converting the dimensions of Pisana's bar device into an equivalent filling factor[70] results in an $\alpha$ of 0.6. Although the electron mobility in the graphene used by Lu was only twice as high, the devices fabricated with the equivalent filling factor showed an order of magnitude improvement in the magnetoresistance; around 100% at 1 T compared to the 10% reported by Pisana.

Although results were not given to compare the contact resistivities between the two studies, the difference in performance between the two devices may be explained in part by the contact resistance. Side contacts to graphene have been shown to produce lower contact resistivities since they allow both the $p\pi$ and $p\sigma$ orbitals in carbon to contribute to electron transmission, compared to only the $p\pi$ orbitals when forming top contacts.[85,87] Pisana et al. reported contact resistances of $3.7\times10^{-6}$ $\Omega$cm$^2$, which is above the threshold calculated by Sun et al. (see Figure 3.12).[61] On the other hand, edge-contacted graphene devices in literature have been reported with contact resistances on the order of $10^{-9}$ $\Omega$cm$^2$.[85] While the contact resistance for Lu's devices is not reported, a lower contact resistivity could be formed by etching into the graphene and exposing the edge, which would lead toan improvement in the performance of the device. Graphene encapsulation with hBN also form edge contacts, which may also explain the low zero-field resistances and high magnetoresistances observed by Zhou et al.[27]

Although the magnetoresistance varied significantly between the devices described by Pisana and Lu, the reported sensitivity was 1000 Ω/T in both cases. This is also in good agreement with the simulations presented by Sun et al.[61] which suggests that sensitivity is more robust to changes in the contact resistance and can remain high even at values around $10^{-6}$ $\Omega$cm$^2$.

Resistance at the interface between graphene and metal depends not only on the type of bonding that occurs, but also on the relative Fermi energies of the two materials. Zhou et al. found that the zero-field resistance not only changes with gate voltage, but is asymmetric about 0 V (see Figure 3.29a). The authors posit that this phenomenon occurs due to Fermi-level pinning where the large density of states in the metal pins the Fermi level in the adjacent graphene, producing an interface-near region that is *n*-doped. However, in the bulk of the graphene the Fermi level is mostly controlled by the back gate voltage. Thus when the bulk of the graphene is *p*-doped, a *p/n*-junction is formed near the shunt, effectively increasing the zero-field resistance.[27] Computational modeling was used to estimate the length scale over which the Fermi level relaxes and was determined to be on the order of 100 nm (Figure 3.29b).[27] Kamada et al. determined that the formation of a *p/n*-junction affected the ability to accurately measure the mobility of holes and electrons in graphene Corbino discs. Only by gating the graphene with a strong bias voltage could the effect of the *p/n*-junction be eliminated.[28]

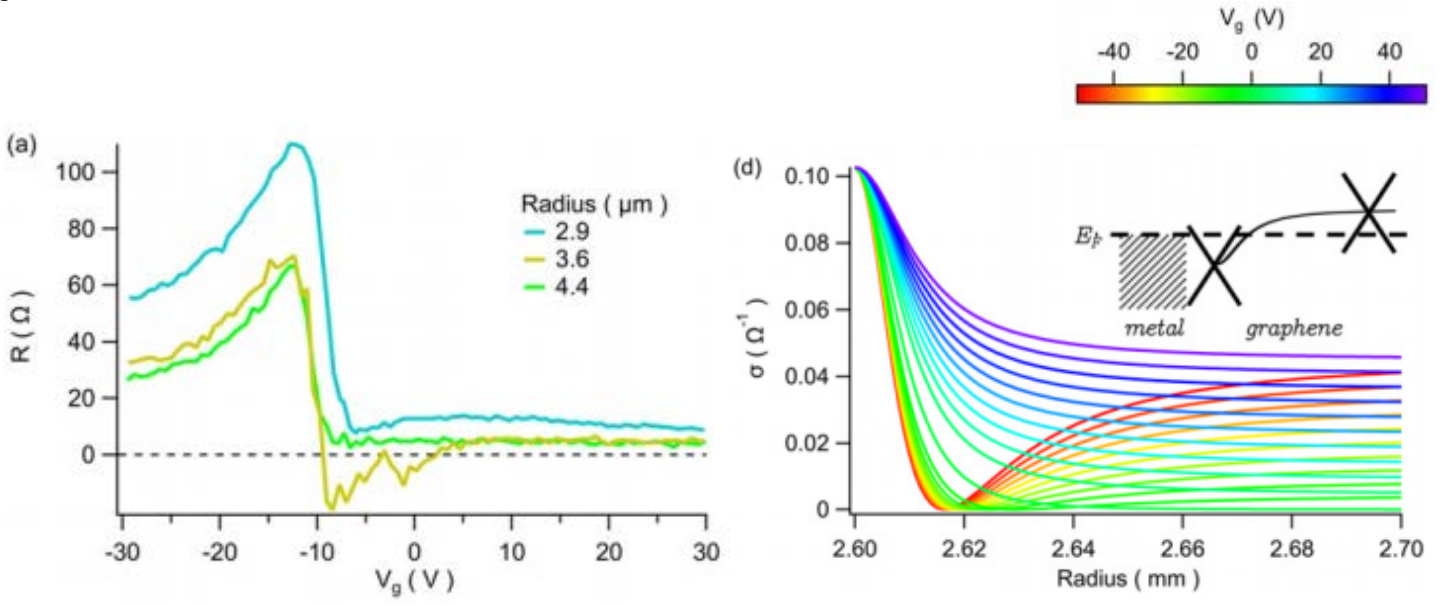

*Figure 3.29: a) Resistance as a function of back gate voltage for three hBN-encapsulated graphene devices with a circular shunt of varying inner radii. b) Simulated conductivity of the device near the metal-graphene interface at different back gate voltages.*[27]

The high electron mobility of graphene means that one has to consider quantum mechanical phenomena, particularly at low temperatures where phonon modes are suppressed and the mean free path of electrons becomes large. Zhou et al. observed negative resistances, which may be a signature of ballistic transport (see Figure 3.29a).[27] Friedman et al. observed oscillations in the resistance of their devices as the gate voltage was scanned at high magnetic fields and cryogenic temperatures, and posited that they are an example of the quantum transport. These oscillations are clearest for the case of an unshunted vdP sample (Figure 3.30) but can also be observed to a lesser extent in shunted vdP devices (Figure 3.26b).[25]

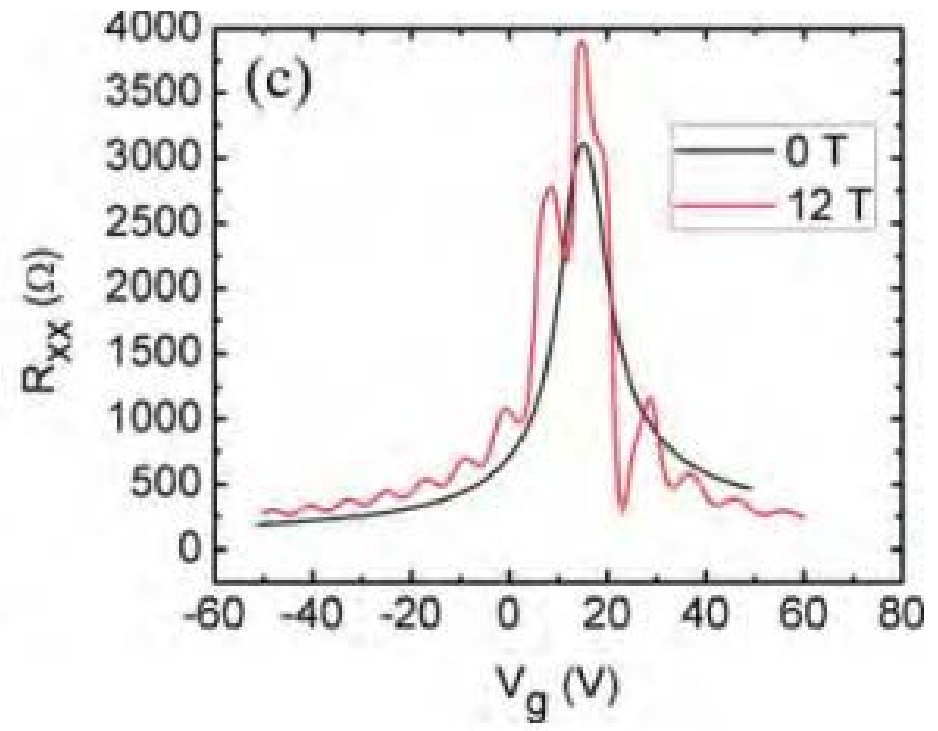

*Figure 3.30: Resistance of an unshunted vdP device at zero and high fields as a function of back gate voltage.*[25]

In addition to single-layer graphene, EMR devices made from bilayer graphene and $Bi_2Se_3$ have also been reported.[24] Bilayer graphene on SiC was found to have a higher mobility than monolayer (5,000 vs. 3,000 $cm^2V^{-1}s^{-1}$), likely due to electronic screening of charged impurities provided by the graphene layer in contact with the substrate. The device made with bilayer graphene showed an order of magnitude increase in magnetoresistance compared to the monolayer graphene device. No explanation is provided by the authors, and while the mobility of the bilayer graphene is higher, the difference in the carrier mobility of the two devices is not sufficiently large to fully account for the improvement. It is possible that factors such as a lower contact resistance in the bilayer device played a role. Bilayer graphene has been demonstrated to produce lower contact resistances than monolayer graphene by increasing the number of edge contacts.[87] $Bi_2Se_3$ is not strictly a vdW material but can be exfoliated in the same manner. The measured magnetoresistance was extremely low, likely due to its low electron mobility of 50 $cm^2V^{-1}s^{-1}$.[24]

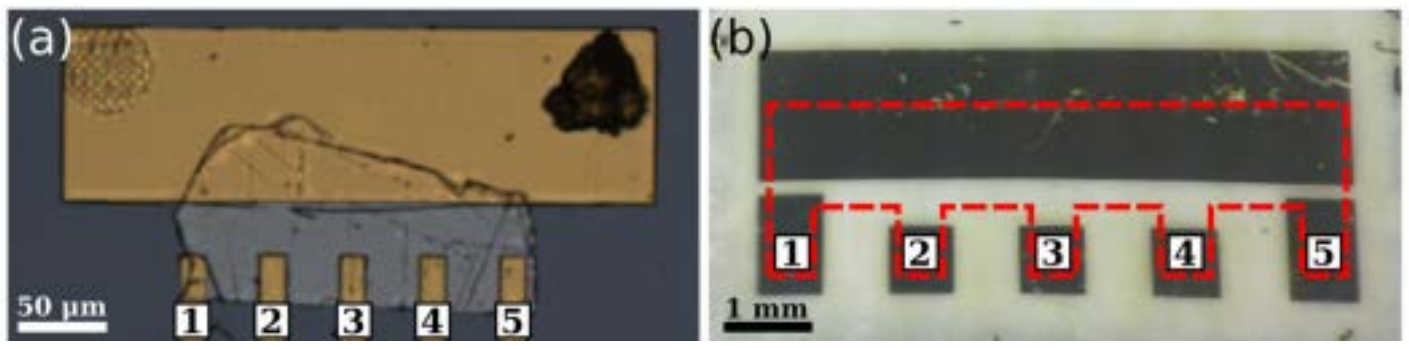

*Figure 3.31: Images of planar EMR devices made from a) $Bi_2Se_3$ flakes and b) CVD graphene.*[26]

# 4. Magnetometers

A key application of extraordinary mangetoresistance is to form the basis of magnetometers that can detect magnetic fields using either simple 2-terminal or 4-terminal electrical measurements. The use of EMR devices as magnetometers has been investigated in several studies and will be reviewed in this section. Important benefits of EMR magnetometers are the potentially large range of conditions that they can be operated under, including wide ranges of temperatures and magnetic fields. In addition, as displayed in the previous chapters, EMR devices possess extremely rich possibilities for tuning their performance as magnetometers by varying the device geometry and material properties

## 4.1. Fabrication of Devices

Bar-shaped geometries are often considered when fabricating EMR magnetometers as this geometry facilitates simple fabrication and even realization of small sensors with nanoscale spatial resolution.[3,5,70] To further improve the spatial resolution, the voltage contacts are closely spaced and the metal shunt only contacts the high-mobility semiconductor in the region between the voltage contacts (contact 2 and 3 in Figure 4.1). This region then becomes the active region, which determines the lateral resolution together with the distance between the sensor and the magnetic field source. The latter is improved by 1) placing the conductive layer close to the sensor surface and capping this with a nanometric insulating layer such as $Si_3N_4$ to prevent shorting and 2) by decreasing the thickness of the active layer in the EMR sensor, e.g. using quantum well structures with high mobility obtained through modulation doping and reduction of dislocations using epitaxial buffer layers.[70] With this strategy, Solin et al. fabricated EMR sensors from InSb/AlInSb quantum wells with an active volume of 35x30x20 $nm^3$.[3,5] As shown in Figure 4.2, this nanoscopic sensor produced an asymmetric magnetoresistance of up to 150% at -1 T. EMR magnetometers based on graphene, InAs, and GaAs/AlGaAs have also been considered,[5,13,16,19,23] as well as other nanoscopic sensors.[19] Of particular concern is the contact resistance between the semiconductor or graphene and the metal shunt, which can severely lower the magnetic field sensitivity. Here either poor electrical contacts, Schottky barriers, or *p/n* junctions may form,[5,27] which can limit the operation of magnetometers, particularly at low temperatures. Therefore, special care should be taken in the choice of materials and during fabrication as described by Solin et al.[5]

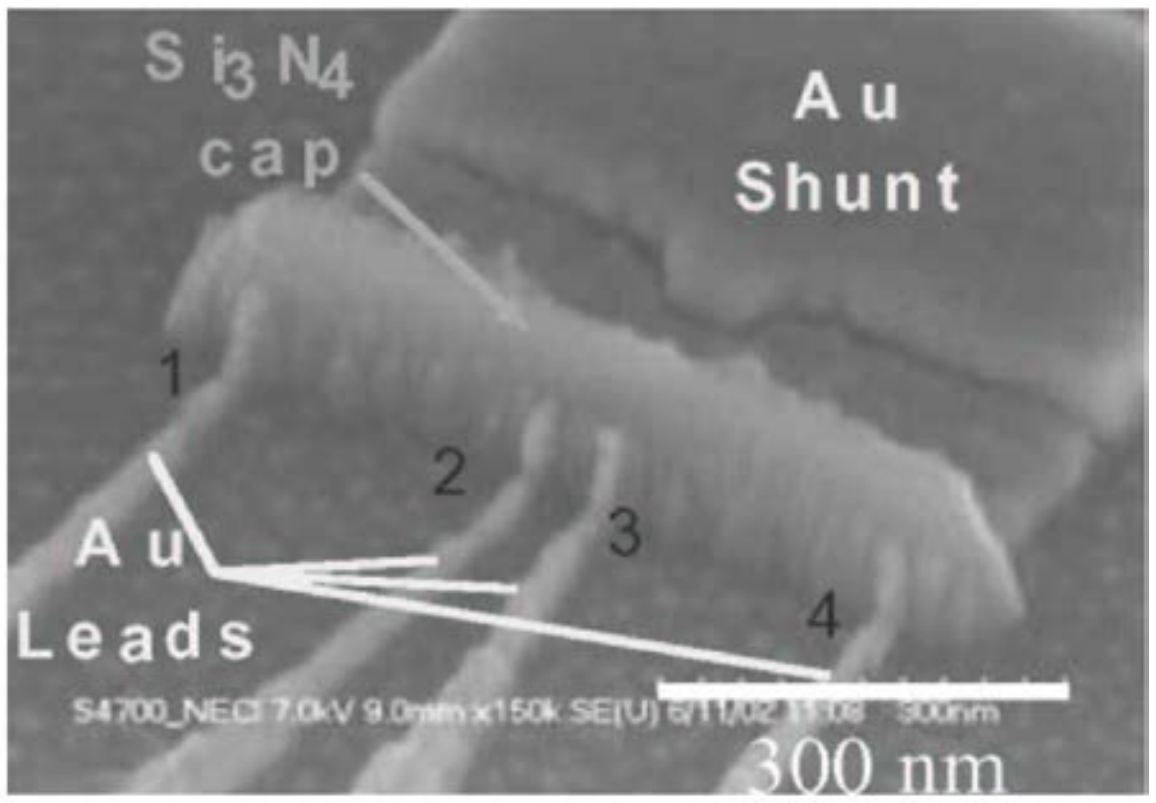

*Figure 4.1: Electron micrograph of a nanoscopic EMR device made from an InSb quantum well patterned using e-beam lithography. The $Si_3N_4$ capping layer was added on top of the structure to prevent shorting the leads and shunt.*[3]

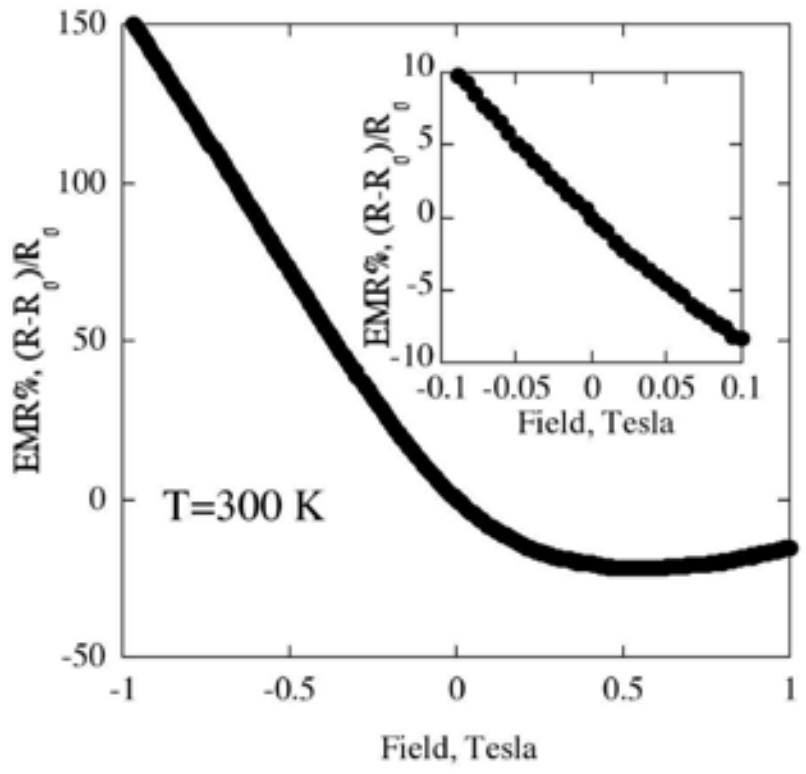

*Figure 4.2: Magnetoresistance as a function of magnetic field for a nanoscopic EMR device. The inset shows a zoom in on the low-field region.*[5]

## 4.2. Signal-to-Noise Ratio and Noise Equivalent Field

The two key figure of merits for EMR magnetometers are the signal-to-noise ratio (SNR) and noise equivalent field ($B_{NEF}$). The typical operation of EMR devices involves passing a constant current through the device while measuring the field-dependent voltage drop across either the current contacts in a 2-terminal configuration or separate voltage contacts in a 4-terminal configuration. The SNR is then defined as:

$$SNR = \frac{V_{signal}}{V_{noise}}$$

When measuring small magnetic fields differences $(B_{signal} - B_{bias})$ around a bias magnetic field $(B_{bias})$, the voltage signal can be approximated by[89]:

$$V_{signal} = I \left|\frac{dR}{dB}\right|_{B_{bias}} (B_{signal} - B_{bias})$$

Here, the bias magnetic field may be used to tune the EMR devices into the most sensitive region, or alternatively to provide a description for the case where a background magnetic field is present in the measurements.

In contrast to conventional giant magnetoresistance devices, the typical EMR devices do not contain any magnetic elements, which eliminates the magnetic noise contribution and stray fields from the magnetic component. Noise in the EMR sensors is generally characterized by only two contributions: A frequency-independent thermal noise contribution and a low-frequency $1/f$ noise contribution. Depending on the material platform used, other contributions may also contribute to the noise including generation-recombination noise[90], however these contributions have generally not been considered for EMR sensors. An example of the noise sources in an InSb EMR device is shown in Figure 4.3, which features both a low noise and a low frequency corner of 140 Hz separating the $1/f$ and white noise regions[16]. The frequency corner may however vary widely and has been reported to exceed 5 GHz in InAs EMR with Ta/Au as shunt metal.

The thermal noise $(V_{thermal})$ is well-established with an origin in thermal agitation of the charge carriers, which increases as the temperature and resistance is increased:

$$V_{thermal} = \sqrt{4k_B T R_{V,2pt}(B_{bias})\Delta f}$$

where $k_B$ is the Boltzmann constant, $R_{V,2pt}(B_{bias})$ is the two-terminal resistance of the voltage probes evaluated at the bias field and $\Delta f$ is the measurement bandwidth.[89] In the case of a two-terminal EMR device, $R = R_{V,2pt}$, with this resistance including the resistance of the leads as well as any contact resistance in the circuit. For four-terminal devices, the devices resistance $R$ may be significantly different from $R_{V,2pt}$, as $R$ is strongly affected by the relative placement of the current and voltage terminals and does not include the resistance of the leads or the contact resistance between the leads and the device. The $1/f$ noise contribution ($V_{1/f}$) is less rigorously understood, but is often described using:

$$V_{1/f} = \sqrt{E^2 \gamma \mu e R_{V,2pt} \frac{\Delta f}{f}}$$

where $E$ is the bias electric field applied to drive the constant current in the EMR device, $\gamma$ is the dimensionless Hooge parameter and $\mu$ is the electron mobility.[89] A key difference between the two noise contributions is that the thermal noise regime does not depend on the magnitude of the current whereas the $1/f$ voltage noise increases linearly with current ($E \propto I$).[19,91]

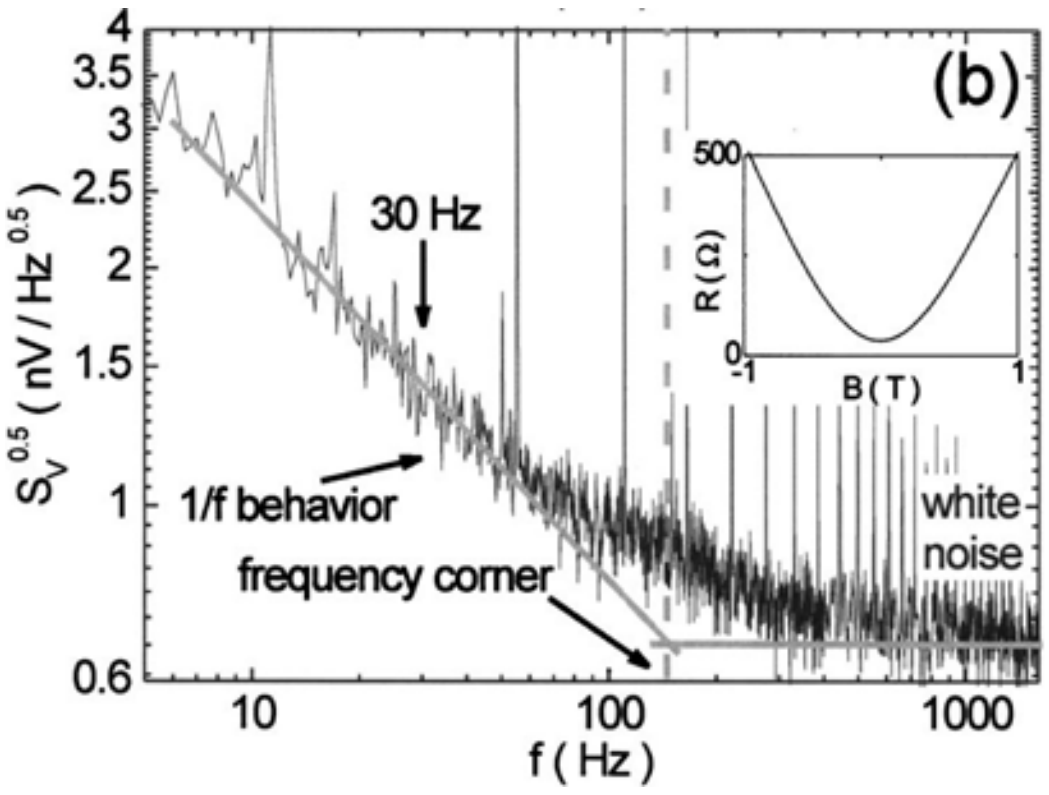

*Figure 4.3: Voltage noise ($S_V^{0.5}$) as a function of frequency (f) for a InSb/Au EMR device.*[16]

The SNR is defined as:

$$SNR = \frac{I \left|\frac{dR}{dB}\right|_{B_{bias}} (B_{signal} - B_{bias})}{\sqrt{4k_B T R_{V,2pt}(B_{bias})\Delta f} + \sqrt{E^2 \gamma \mu e R_{V,2pt} \frac{\Delta f}{f}}}$$

with a cross-over frequency between the $1/f$ noise and thermal noise occurring at $f_c = E^2 \gamma \mu e / (4 k_B T)$. When measuring weak magnetic fields without a bias ($B_{bias} = 0$),the SNR can be used to extract the noise equivalent field ($B_{NEF}$) in units of $T/\sqrt{Hz}$ representing the magnetic field strength which produces a signal strength equal to the noise value (SNR = 1). For the thermal noise regime this field detection limit is given by:

$$B_{NEF}^{thermal} = \frac{\sqrt{4k_B T R_{V,2pt}(B=0)}}{I \left|\frac{dR}{dB}\right|_{B=0T}}$$

Whereas the low-frequency noise-equivalent field becomes:

$$B_{NEF}^{1/f} = \frac{\sqrt{E^2 \gamma \mu e R_{V,2pt}/f}}{I \left|\frac{dR}{dB}\right|_{B=0T}}$$

In both noise regimes, a low magnetic field detection limit is obtained by increasing the sensitivity (*dR/dB*) while lowering the 2-terminal resistance at the voltage probes. In addition, it is favorable to increase the measurement frequency in order to operate in the thermal noise limited regime. In this regime, lowering the temperature and increasing the current further boost the sensitivity. The optimal current applied to the EMR magnetometer depends on the device geometry, constituent materials, and the application of the sensor. Solin argued that the maximum current in InSb-based EMR devices is limited by the onset of non-linear transport caused by a drop in the carrier mobility at high electric fields.[5] The optimal current is typically a trade-off between improved signal strength when increasing the current in the thermal noise regime and either current limitations of the sensor – such as the onset of non-linear transport – or the emergence of the current-dependent $1/f$ noise at the measurement frequency. With one exception,[16] the optimal current and the effect of current on the noise have not been identified for EMR devices. However, for an InAs device the white noise was found to be independent of the current up to approximately 0.6 mA, beyond which a moderate increase of 23% in the white noise was observed when increasing the current to 7.5 mA. In general, the noise equivalent field is typically estimated by only calculating the thermal noise using the resistance and temperature and hence excluding the impact of the current on the noise.

## 4.3. Detection of Homogenous Weak and Strong Fields

The noise equivalent field serves as the figure of merit for detecting magnetic fields that are homogeneous within the spatial extent of the EMR sensor. For 4-terminal devices, the expressions for the figure of merit as well as the optimization of the EMR magnetometers share several similarities with the more developed Hall magnetometers.[90,92,93] The magnetometers can be optimized in terms of the measurement frequency, operating temperature, probing current as well as the material parameters in the hybrid device. For the latter, the key parameters are the mobility and carrier density of the high-mobility material,[89] the contact resistance to the metal[61] as well as the conductivity contrast between the metal and high-mobility material.[78] In contrast to Hall-bars where the Hall signal is largely independent of the geometry, EMR sensors intrinsically have a large geometric influence as described in Section 2. An example is shown in Figure 4.4 for a bar-shaped InAs EMR device measured in 2- and 4-terminal configuration and compared to a corbino disk. Here, the sensitivities and resistances of the devices using a 1 mA applied current are measured as a function of an applied homogeneous magnetic field. The 2-terminal bar-shaped EMR device is found to have the largest sensitivity, exceeding the other two geometries by two orders of magnitude. The thermal noise was calculated from the resistance values, and used to extract the noise equivalent field, yielding values down to $B_{NEF}$ =10 nT/$\sqrt{\text{Hz}}$ for magnetic fields ranging from approximately 0.2 to 1 T. At weak magnetic fields the noise equivalent field is worsened significantly due to the reduced sensitivity in the symmetric devices. In contrast, a non-vanishing sensitivity can be obtained in asymmetric devices where the asymmetry is typically formed by *IVIV* contact configurations rather than the *IVVI* configuration as described in Section 2. Other studies report noise equivalent fields in EMR devices of 0.01-2200 nT/$\sqrt{\text{Hz}}$ where the noise is calculated in a similar manner[5,20] (See Table 1 for further details), with the lowest bound calculated using a high current of 100 mA. However, in a 20 µm x 8 µm InAs EMR device, the noise and its dependence of the current were measured, yielding a magnetic field sensitivity of around 1.3 nT/$\sqrt{\text{Hz}}$ at 0.8 T.

Of particular interest for magnetic field sensing is the micro- and nanoscopic EMR sensors such as the one shown in Figure 4.1 where the small sensor size yields a higher degree of homogeneity across the sensor as well as a great spatial resolution if employed in a scanning EMR magnetometer. The noise equivalent field is calculated for 35 x 30 x 25 $nm^3$ EMR sensors made from InSb/AlSnSb, which gives a value of $B_{NEF} = 4.1\ \mu T/\sqrt{\text{Hz}}$ at a field of 9 mT and probing currents of 2.2, respectively[5]. Predictions for similar sensors made from InAs gave $B_{NEF} = 2.4\ \mu T/\sqrt{\text{Hz}}$ at a field of 9 mT and probing currents of 3.8 µA. Larger microscopic sensors with dimensions of 1000 x 1000 x 25 $nm^3$ result in noise equivalent fields of 10-20 nT/$\sqrt{\text{Hz}}$, which was found to compare favorably to Hall sensors of identical size based on 2DEGs in GaAs/AlGaAs heterostructures.[5] The potential and challenges of using such nanoscale EMR sensors in scanning magnetometers with nanoscale resolution have been addressed by Solin.[5]

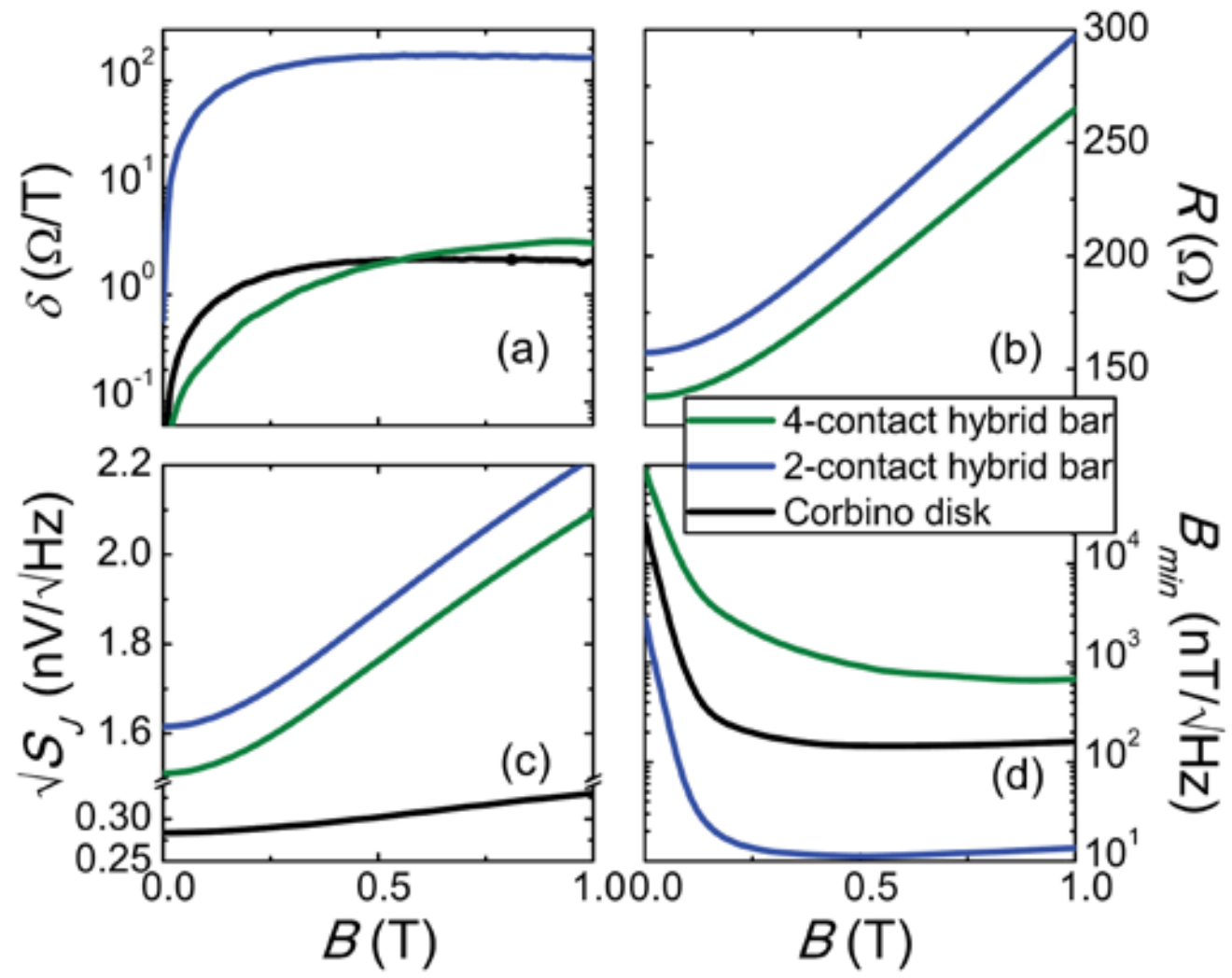

*Figure 4.4: Sensitivity ($\delta = dR/dB$), resistance (R), calculated thermal noise ($\sqrt{S_J}$) and noise equivalent field ($B_{min}$) as a function of magnetic field for three InAs/metal devices: 1) A bar-shaped EMR device measured in 4-terminal configuration, 2) a bar-shaped EMR device measured in 2-terminal configuration and 3) a corbino disk. Figure from Ref.* [13]

### 4.4. Inhomogeneous Magnetic Fields

It remains unknown how EMR sensors react to inhomogeneous magnetic fields that vary significantly across the EMR sensor. Only a single study exists that assesses the capabilities for EMR devices to sense local magnetic fields[52] i.e. fields smaller than the physical size of the sensor. The device investigated was a bar-shaped semiconductor/metal hybrid device intended for sensing extremely localized magnetic fields. Its characteristics were investigated using a numerical finite element model identical to that of Moussa et al.[32] The authors considered a localized magnetic dot with a spatial extent equal to the width of the semiconductor region in the device. The bar-shaped EMR device was long and thin with a length-to-width ratio of the semiconducting layer of 50 as shown in Figure 4.5. Using an asymmetric voltage probe configuration within the bar-shaped device, the finite element model revealed that for the ideal position of the magnetic dot, the magnetoresistance in a ±50 mT field is 18%. While this is much lower than the 134% obtained if the magnetic field was homogeneous across the entire device, it is nevertheless considerable considering that the magnetic dot only covers 1/60 of the semiconductor area.

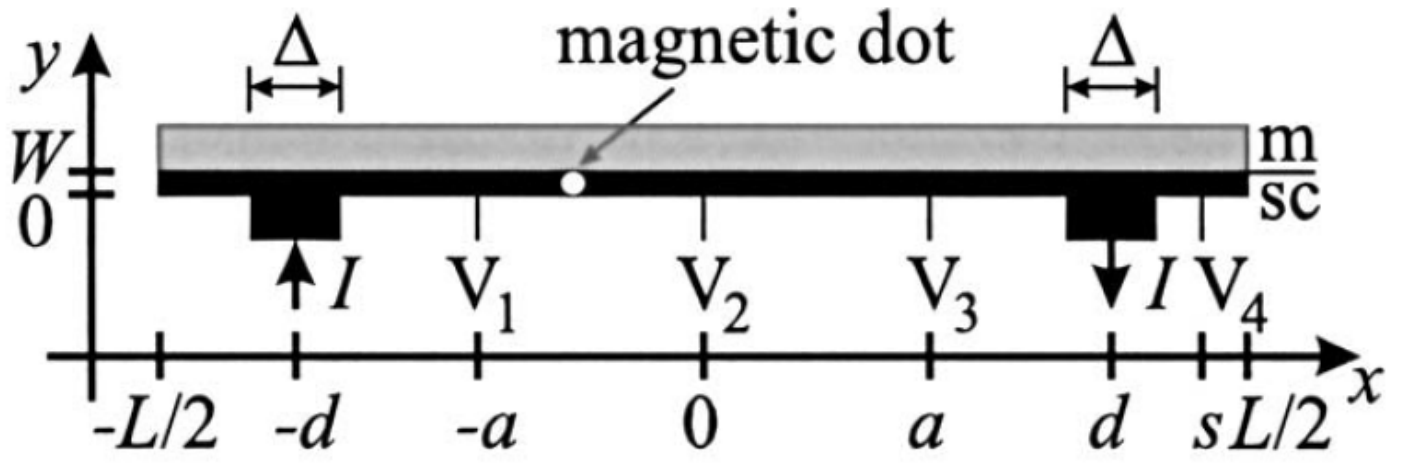

*Figure 4.5 A bar-shaped EMR sensor with a number of different voltage contacts. The magnetic field detected in localized in the small magnetic dot in the semiconductor layer of the device.*

# 5. Conclusion and Perspectives

EMR remains a relatively unknown species in the zoo of different magnetoresistance classes. In contrast to the majority of other magnetoresistance phenomena, EMR finds its root in a geometrically enhanced deflection of the current upon the application of a magnetic field. The effect has been realized in hybrid devices consisting of metal inclusions inserted into a high-mobility material. To date, the majority of EMR devices are based on graphene or III-V semiconductors with non-magnetic metal inclusions. The magnetoresistance in EMR devices has been shown to exceed $10^7$% at room temperature.[1,3,27] Although the current deflection has not yet been imaged experimentally, the origin of the EMR effect is considered to be well understood as justified by the good agreement between experimental magnetotransport data and both analytical and numerical models. Numerical modeling continues to be a valuable tool for investigating how variations in material parameters and geometry impact performance as well as to guide and understand experimental progress. Yet, most numerical and experimental studies published to date aim at describing the magnetoresistance in the device, and efforts to push the magnetic field detection limit in EMR magnetometers have been scarce.

Beyond magnetic field sensing, a range of other perspectives in EMR remains relatively untouched. Particularly interesting areas include the use of EMR devices for magnetic switches and magnetometers, the cross-over from the extensively studied diffusive devices to sparsely studied ballistic devices, as well as the family of other extraordinary phenomena that has developed in the wake of the discovery of the EMR effect. These are discussed briefly in the following subsections.

### 5.1. Magnetoresistive Switches and Magnetometers

The EMR effect can be harnessed for developing new switches and magnetometers with ample of degrees of freedom to tailor the performance through geometric and material optimization. For magnetic switches, the presence and absence of magnetic fields form a magnetically induced high on/off ratio in the electrical resistance of the device. This is of interest in, e.g., position sensing where the presence of magnetic objects can be electronically detected in the EMR devices, which may reveal positional information. Another application can be to magnetically induce a redirection of current as well as to close and open connections in a circuit which can be used for instance in solid-state switches. As observed by Solin et al. (ref) , magnetic switches can contain a low on-resistance, a high on/off ratio, contain no moving parts, and possess a fast switching speed provided that the magnetic field can be applied and removed quickly. Here, optimization of the geometry may be used to customize the performance as exemplified by Solin et al. (ref) where the onset field for switch operation is shifted from around 0.05 to 0.4 T by increasing the radius of the inner metal disk in concentric circular EMR devices. Another example is presented by Erlandsen where geometrically shifting of the inner metal disc in a shunted vdP device produce an EMR response where negative magnetic fields turn off the device and positive fields turn the device on (ref).

EMR devices can also be used as magnetic field sensors by sensing magnetically induced changes in the resistance similar to the extensively used Hall sensors and giant magnetoresistance sensors. This allows for a convenient all-electronic solid-state magnetometer with no moving parts that can operate in a wide temperature and magnetic field range. In contrast to giant magnetoresistance sensors, the general EMR sensor does not contain any magnetic elements, which eliminates magnetic noise, enables the use at higher magnetic fields and allows the sensors to be smaller without risking a spontaneous superparamagnetic spin reversal in the sensor. Despite the very immature state of EMR sensors, they still show a promising magnetic field sensitivity generally down to a few $nT/\sqrt{Hz}$[16] with ample of room for improvement through optimization of both materials and geometry. The possibility of using EMR sensors in 2-terminal mode as well as the large design space for sensor optimization sets EMR apart from Hall sensors. The geometry of Hall sensors has generally no impact on the Hall resistance, and the geometry primarily influences the noise level where scaling down Hall sensors results in an increasing noise.[90,93] Instead, optimization of Hall sensors is typically achieved either by lowering the sheet carrier density of the active area to increase the signal strength or increasing the mobility to decrease the noise level. This combination of material properties makes graphene and semiconducting 2DEGs the most promising material platforms for Hall sensors. As described in the previous section, EMR sensors have a higher ceiling for magnetic field resolution as they can be optimized to a very large extent by using both the material and geometric parameters of the sensor.

## 5.2. Diffusive vs. Ballistic Transport

While most of EMR studies were done in the diffusive transport regime, Zhou et al. found ballistic transport may significantly contribute to MR values as $R_0$ becomes extremely small and approaches zero.[27] This offers a new direction to study EMR in an almost entirely uncharted territory of EMR magnetometry. This experimental discovery is, however, contradictory to Solin's prediction in which they expect the MR value would be orders of magnitude lower in ballistic regime (ref).

The experimental transition from the diffusive transport regime to a quasi- or fully ballistic transport regime can be done through a variety of different means that increases the mean free path of the carriers beyond the characteristic lengths of the devices, including increasing the carrier mobility, reducing device sizes to a few micrometers or less, or decreasing the temperature. Finite element simulations only apply in cases where transport occurs in the diffusive regime and to study the EMR effect in the ballistic regime semiclassical trajectory simulations and multiscale tight binding calculations similar to Calogero et al.[94] could be used.

## 5.3. Other EXX Phenomena

The geometry of metal/semiconductor hybrid devices can have a profound impact on devices beyond magnetoresistive sensors. It is possible to change the resistance of hybrid metal/semiconductor devices by subjecting them to a variety of external stimuli. These phenomena are collectively known as the EXX family of effects, of which extraordinary magnetoresistance (EMR) is just one of the known members. Other documented effects include extraordinary optoconductance (EOC), extraordinary piezoconductance (EPC), and extraordinary electroconductance (EEC).[56,95] The extraordinary magnetoresistance effect as reviewed here remains the most studied phenomena but progress in this field may translate to the other EXX family members as well.

In extraordinary optoconductance, the semiconductor/metal hybrid device is irradiated with a focused laser, which excites carriers across the bandgap. Electrons and holes thus form and undergo diffusion. As the electron mobility in III-V semiconductors often exceed that of holes, the spatial distribution of photo-generated electrons is wider than the hole distribution, leading to a spatial fluctuation in the net charge and an associated electrostatic potential. The potential difference measured on two voltage contacts depends on the position of the laser beam with respect to the voltage contacts, leading to a light sensor with positional detection of the incoming light. The diffusion of electrons and hence the measured voltage can be modified by metallic inclusions with ohmic contacts to the semiconductor. This effectively constitutes an extraordinary light sensor as realized in the conventional bar-shaped geometry.[45] As in the case of EMR, the performance of extraordinary optoconductive sensors may be optimized through appropriate geometrical optimization.

In extraordinary piezoconductance[45] the interface between the semiconductor and the metal shunt forms a Schottky barrier that electrons can tunnel through. As the device is subjected to strain, the interatomic distances vary which changes the height of the Schottky barrier. Tunneling probabilities through the barrier have an exponential dependence on the barrier height, so the small variations induced by external strain result in a measurable change in resistance. The Schottky barrier between the semiconductor and metal inclusion is also at the heart of the extraordinary electro-conductance effect.[45] In contrast to extraordinary piezoconductance, variations in the Schottky barrier are induced by application of an electric field in electroconductive devices. As with the EMR and EOC, the ability of both piezoconductive and electronductive sensors to measure strain or electric fields may also be optimized through appropriate choice of geometry.

Overall, extraordinary magnetoresistance and the related phenomena constitute an exciting field where the geometric design of the sensor can be used as a versatile route for tailoring performance. In particular, this may be realized through the use of advanced numerical tools featuring inverse modelling to navigate through the countless possible geometric variations. To date however, the landscape is to a high degree uncharted territory which holds plenty of opportunities for future development within the field.

# 6. Acknowledgements

DVC, RB and BZ acknowledge funding from the Novo Nordic Foundation, grant no. NNF21OC0066526 (BioMag), and DVC further acknowledges support from the Novo Nordic Foundation Nerd program, Grant No. NNF21OC0068015 (Superior).

# 7. Mendeley References